# Nanocartography: Planning for success in analytical electron microscopy

Olszta, M.J.[1*], Fiedler, K.R.[2]

[1] Pacific Northwest National Laboratory, Richland, WA 99354, USA

[2] Washington State University, Tri-Cities, Richland, WA, 99354, USA

Phone: (509) 371-7217

Fax: (509) 375-3033

[*]Corresponding author email: matthew.olszta@pnnl.gov

## Abstract

With the increasing diversity in material systems, ever-expanding number of analysis techniques, and the large capital costs of next generation instruments the ability to quickly and efficiently collect data in the electron microscope has become paramount to successful data analysis. Therefore, this research proposes a methodology of nanocartography that combines predictive stage motion with crystallographic information to provide microscopists with a sample map that can both reduce analysis time and improve confidence in data collected. Having a road map of the stage positions linked to microstructural (e.g., interfaces and growing directions) and crystallographic orientation data (e.g., specific poles and planes) provides microscopists with the ability to solve orientation relationships, create oblique tilt series movies, and also solve complex crystallographic unknowns at extremely small scales with minimal information. Most importantly, it can convert any sample orientation relationships across microscopes to increase optimization and collaboration throughout the field.

# 1. Introduction

Prior to the invention of global positioning systems (GPS), humanity relied mainly on accurate cartography to optimize navigation (Council, 1995). Navigation by cartography was not only confined to terrestrial travel, but extraterrestrial travel as well through the knowledge of celestial orbits of the moon and planets (Gehrz et al., 2007). Yet, even the most detailed maps eventually became outdated and possibly misread or misinterpreted depending on the level of spatial awareness of the traveler. The advent of GPS in 1973 eventually brought about a revolution among travelers for ease of use, but more importantly for the confidence, it instilled in that no matter how complicated the route, a step-by-step guide was provided. A similar approach is needed for the nanoscopic world of electron microscopy. A guide that provides microscopists the ability to record where they have been in a sample, and to develop roadmaps for themselves and others to assist in future analysis. Those positions could be converted between microscopes if and when the sample needed to be re-analyzed. Just as a physical road map need be turned and flipped to compare to a landmark, fiduciary marker, or reference position, so too should electron microscopists have the ability to flip and rotate positional data when moving a sample to a different microscope.

Since the invention of electron microscopy (Mulvey, 1996, Knoll and Ruska, 1932), not only did the physical observation of microstructures become important (e.g., shape, size, and morphology), but so too did relating the crystallographic knowledge to those microstructures. The spatial resolution of transmission electron microscopes (TEM) opened an entirely new world as compared to X-ray and spectroscopic techniques. It lead to the discovery and confirmation of deoxyribonucleic acid (DNA) in the 1950s (Watson and Crick, 1953), and is currently providing materials researchers the ability to examine crystals atom-by-atom (Meyer et al., 2008) as well as to move atoms one by one to create larger structures at the atomic scale (Dyck et al., 2018). The push towards picometer resolution a short 80 years after its introduction has been rivaled by few technologies for volumetric analysis such as atom probe tomography (Blavette et al., 1993, Cerezo et al., 1988). Yet, the ever-expanding spectral, structural, and crystallographic techniques available in the TEM still make it the most versatile and attractive analysis technique for a wide range of research fields.

The ability to understand the crystallographic and microstructural orientations of any region of interest within a TEM sample in relation to the physical stage movements is crucial to extracting the most concise and relevant information possible in the shortest amount of time. The geometry and physics of extracting and understanding these data have long been understood and published (Duden et al., 2009, Klinger and Jäger, 2015, Liu, 1994, Liu, 1995, Qing, 1989, Qing et al., 1989, Zhang et al., 2018). Programs such as Desktop Microscopist[1], CrysTBox, ALPABETA, CrystalMaker, JEMS, τompas, SPICA, and K-space Navigator provided a variety of ways to understand crystallographic data (Cautaerts et al., 2018, Klinger and Jäger, 2015, Duden et al., 2009, Stadelmann, 1987, Palmer, 2015, De Graef and McHenry, 2012, Xie and Zhang, 2020, Li, 2016). In the conclusion of Liu's calculations on the prediction of cubic crystals a statement was made that, "If an interface between the microscope and the computer is developed, an automated on-line method can also be developed…" (Liu, 1994). Others have utilized stage positions and knowledge of crystalline poles to address grain orientations, and more importantly grain boundary misorientations (Jeong et al., 2010, Liu, 1994, Liu, 1995). This research has been widely available, but there is still not a concise, user-friendly manner in which to fully utilize this knowledge for mapping out an entire sample.

---

[1] Desktop Microscopist is a trademark of Desktop Microscopist

There is a need for increased speed and efficiency in electron microscopy due to a wider field of materials being analyzed, an increasing amount of analytical techniques being developed, higher capital costs associated with purchasing newer instrumentation, and decreased sources of funding (Maia Chagas, 2018). Current generation spectrometers can be as costly as the base microscope itself. With the revolution of aberration correction advancing resolution to the picometer scale (Yankovich et al., 2014), the inclusion of a corrector, whether image or probe, has increasingly commonplace on all new purchases. This increased technology has added to the steep costs of doing innovative microscopy. These factors have made it such that each minute spent in any microscopy session is precious. It has also made collaboration and user facilities an attractive option for researchers who do not have the capability to perform higher end research at their home institutions. All of this taken into consideration, the future of electron microscopy will be geared towards doing smarter microscopy and automation (Spurgeon et al., 2020), similar to what has been accomplished in the field of X-ray crystallography (Abola et al., 2000). The eventual progression into full automation presents the possibility of much of the underlying mathematics and physics being overlooked as microscopes will eventually perform much of the data collection.

Automation and machine learning, while first pioneered and developed in biological microscopy, is steadily being developed for materials science applications (Carter and Williams, 2019, Jansen et al., 2013). The genesis of automated detection and tomographic techniques within the framework of understanding biological materials was born out of a need for observing microstructural information over longer length scales, such as counting cells (Porter et al., 1945, Lidke and Lidke, 2012). The complex nature of the electron interaction physics of material science research such as crystallinity, defects, and variable Z contrast, makes automation more difficult and most likely why it has slowed the adoption and development in the field. This is not to mention the exceedingly smaller length scales that become crucial to understanding any number of atomic phenomena that control bulk materials properties.

As electron microscopy is a projection technique, there will always be a conundrum in analyzing material properties in that the thinner the sample becomes the more accurate the information collected (e.g., decreased multiple scattering); yet, the thinner the sample becomes the less representative the information is of the entire bulk sample (e.g., only a thin slice of a three dimensional object is being observed). Additionally, the thinner the sample the more questions of surface effects dominating the analysis arise (Carter and Williams, 2019, Findlay et al., 2010). Machine and smart learning algorithms require more demanding analytical image analysis techniques within the realm of materials science (Braidy et al., 2012, Jones et al., 2017, Jones et al., 2015, Jansen et al., 2013). To be able to position a sample to understand specific **g** vectors, contrast changes, and orientation effects requires more math than simple edge detection or shape recognition. Even when algorithms are developed to address this, the nature of relevant nanoscopic information within a finite sample thickness (e.g., even with a sample being 40-50 nm thick) may hinder their widespread acceptance. Therefore, there is still a need for the materials science microscopist to interact and guide the collection of data, and as such, there needs to be an intermediary that provides microscopists with tools to better analyze and understand their data.

More importantly, due to a more an ever-increasing reliance on metadata and digital capture, there has been less concentration on dictation and annotation of data. Electron microscopy is becoming more of a tool than a science, and although there are many programs to process and analyze data, there are few that serve as a digital notebook. While at first seemingly counter intuitive, current research into the human memory suggests that the brain is less likely to remember captured data than what is observed (Soares and Storm, 2018). This should seem familiar to any microscopist in discussing microscopy sessions with collaborators in that they "saw" additional features not apparent in the recorded data. There is a need to develop programs that act as a prediction tool, but as well a digital notebook.

Therefore, it is essential to have more accurate and directed electron microscopy to provide a pathway in alleviating the increased demand on current and future instrumentation. While there are inroads being made into automation and machine learning, there will be an unfortunate gap before the technology becomes available and even fiscally tenable (Maia Chagas, 2018). This paper provides a way to link together a long database of crystallographic data and double tilt stage mechanics that can be applied to any microscope, regardless of age or technological advancement. The framework of this research is based upon many papers and formulas of past electron microscopists, but it serves to combine all these data as a different manner of thinking to make microscopy more efficient and concise. This research will fully document how to best utilize crystallographic and microstructural information in combination with a double tilt stage to collect the most pertinent information, but also provide a roadmap for future analysis or plan for analysis on a different microscope. The protocol, which is being coined nanocartography, provides insight on how to travel within any given crystal system, quickly plot and solve the orientation of unknown crystals with as little information as one diffracting plane, create oblique tilt series, rapidly position interfaces on edge, relate interfaces to adjacent crystallographic information, quickly understand the tilt limits of each crystalline grain, and most importantly translate any microstructural or crystallographic information collected when reloading a sample or transferring the sample to a different microscope. This goes beyond the broader description of "nano-cartography" described in an editorial by Demming (Demming, 2015) describing instrument agnostic analysis at the nanoscale to understand materials systems.

The advent of digital capture (first with charge-coupled devices (CCDs), and more recently with direct electron detection), has provided microscopists with a double-edged sword in terms of data (Oxley et al., 2020, Ophus, 2019). More data is always preferential, but it has provided a false sense of information capture in the form of metadata. The latent information that users typically believe is embedded within each digital capture often means less meticulous note taking in the belief all information is being transferred. The ability to capture k-space, whether through diffraction or through the Ronchigram, affords microscopists with an advantage in terms of not only knowing that the data collected is correct (as say compared to oversaturation in film), but more importantly the ability to digitally measure that information.

The basis of nanocartography is understanding the control and predictive tilting of a double tilt stage in relation to the orientation and motion of all crystal systems, in addition to the physical constructs within a sample (such as grain boundaries and interfaces) and their relationship to crystallographic orientations. The development of these formulations have long been understood, but rarely, if ever, discussed with relationship to one another (Cautaerts et al., 2018, De Graef and McHenry, 2012, Duden et al., 2009, Hayashida et al., 2019, Hayashida and Malac, 2016, Klinger and Jäger, 2015, Li, 2016, Liu, 1994, Liu, 1995, Qing, 1989, Qing et al., 1989, Xie and Zhang, 2020, Moeck and Fraundorf, 2006). Unfortunately, the wide breadth of literature on this subject has failed to yield a complete picture that provides clear methodologies for understanding and controlling the motion of samples using a double tilt stage in a transmission electron microscope (TEM). The need to connect the theoretical and practical into a single document is long overdue, and this work serves this purpose but as well expands upon methodologies inherent to more experienced microscopists. These latent data collection techniques are most often considered best practices within individual labs but rarely published. Incorporation into the context of nanocartography became exceedingly relevant.

The following publication is divided into three major sections followed by discussion and conclusions. The first develops mathematical concepts of nanocartography, illustrating how vector analysis, TEM stage movement, and crystallography can be combined to accurately navigate sample analysis. The second portion utilizes the derivations from section one to develop practical derivations for nanocartography, including the use of digital capture to more accurately navigate a sample as well as identification of grain boundary type. Finally, the third section explores practical applications of nanocartography; how to correctly calibrate stage motion as well as apply derivations from sections one and two. Taken in whole, this Element serves as a rich summary of optimization of TEM data collection and analysis.

## 2. Mathematics of Navigation and Orientation in Three Dimensions

In order to most accurately utilize the TEM for crystallographic analysis it is first necessary to be able to express any crystal system as a geometrical concept. This approach serves to deconvolute the motion of a crystal as a solid object from the concepts of the crystal in reciprocal space, diffraction, or the physics of the electron beam interaction. These physics-based approaches are most often utilized in teaching materials science analysis using the TEM, but this adds additional complexity too early. By first treating any sample (which may or may not contain crystallographic material) as a solid three-dimensional object and how it can be manipulated using a double tilt stage, the additional concepts of diffraction and other electron beam interactions eventually becomes more intuitive.

Before any notion involving the addition of atomic positions to the discussion of crystallography and materials science, a general treatment of planes, plane normals, and other basic geometric constructs need be introduced and understood. This is necessary for a variety of reasons, most importantly of which is that it will provide a more solid foundation for both tilting throughout crystallographic space, as well as in general lay a solid groundwork for explaining Miller indices, Bravais lattices, and other such constructs. The extent of this discussion on geometry, while basic on some level is necessary to further build upon the knowledge base of crystallographic analysis.

The simplest geometrical concept in crystallography is the cube due to its high level of symmetry and uniformity. All three unit vectors [u,v,w] can be oriented along the [x,y,z] axes, respectively, and any combination of vectors can be constructed. The mathematics of a cube are dependent upon each of the three axes being of equal length and mutually orthogonal to one another. The angle between any two vectors can be described by the dot product, and the normal of any two vectors can be calculated through the cross product.

### 2.1 Unit Vectors – Real Space Map

The ability to travel through any crystal system is dependent upon understanding basic geometric principles of planes and directions. For example, the angle between the two cubic vectors [001] and [010] is 90°, between the [001] and the [011] is 45°, and lastly between the [001] and the [112] is 35.3°. These are all common low index (high symmetry poles) within any number of cubic crystals, and the understanding of how to move between these poles can be critical in proper microstructural analysis. Unit vectors are the fundamental descriptors of the cube and describe how to follow planes in the crystal to arrive at another location in the crystal. In this sense, the unit vectors function as the equivalent of roads on a map of a regularly laid out dense urban area.

Since there are six other crystallographic families by which atoms can be arranged, the importance of being able to travel between poles within each system is paramount to performing highly accurate crystallographic analysis. Similar vector notation can be used to describe any direction (i.e., vector) within each system (e.g., the [111] of a monoclinic crystal), yet the mathematical derivation of the angle between vectors or calculation of normal vectors is not as straight forward as applying the dot and cross-products due to either biases on the unit vectors or unit vectors not being orthogonal (or sometimes both depending on the crystal family). *Figure 1* provides a simple example illustrating how a naïve analysis fails to accurately predict the measured angle owing to the bias of one axis. By stacking two cubes on top of one another to create a tetragonal unit cell, the z-axis can be described as having a bias (in *Figure 1* is it biased by 2). In the schematic, the coordinate system is labeled in terms of what unit cell is being considered, either the biased cell or the cubic cell, and hence biased vectors in the tetragonal reference (or native format) are listed as [0,0,1] and [1,1,1], whereas in in the cubic reference they would be [0,0,1] and [1,1,2]. Calculating the angle between these vectors in both the biased and the cubic coordinates (through the dot product), it can be shown that the answers differ by 19.4°, with the calculations for the biased system being incorrect (as it will be shown later on in the paper, 54.7° is a common angle in the cubic system because it is the angle between the [001] and [111] poles). Only when the native vectors are converted to the cubic form is the correct answer, 35.3°, obtained. A demonstration of this conversion in a more complex, hexagonal system is provide in *Figure S1*.

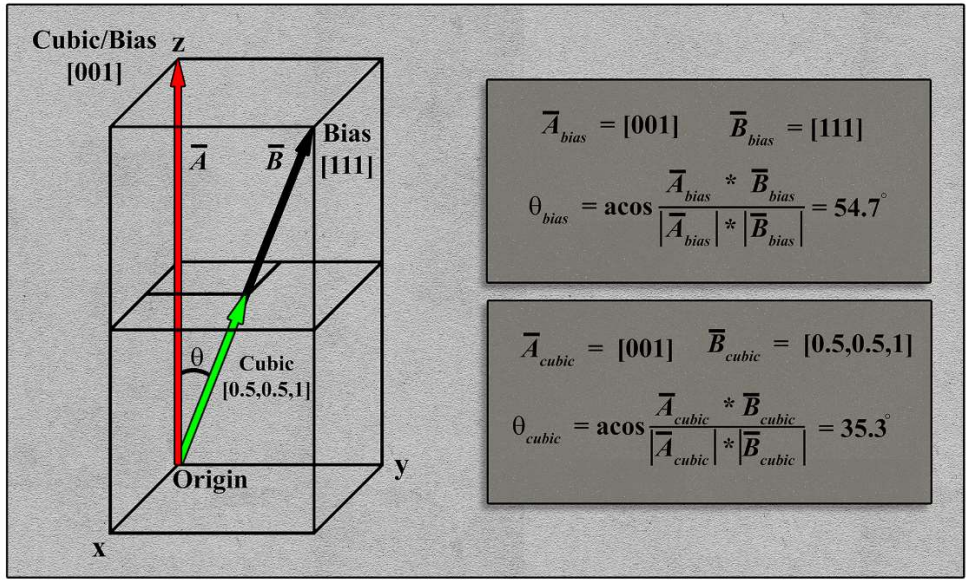

*Figure 1: Schematic illustrating how a bias on one of the axes (e.g., a tetragonal system) does not provide the correct angle between vectors. The bias of the z-axis being doubled incorrectly predicts the angle between the vectors. Note the positions are being listed in parenthesis and do not denote crystallographic planes.*

The importance of vector nomenclature as compared to the vector calculations can be further elucidated by comparing a single crystal system (e.g., hexagonal) where the ratio of the unit vectors are varied (*Figure 2*). Native vectors for hexagonal crystals with c/a ratios of 2.72 and 1.63 are described in the identical manner owing to the fact they are the same crystal system, but when converted to the cubic form the vectors are now noticeably different. Since the vectors within the basal plane (e.g., [100] and [110]) are not affected by the change in c/a ratio, the angle between the vectors does not change, but comparing the angle between the [001] and the [111] it can quickly be demonstrated that in the cubic system the [111] vector becomes [0.2 0.3 1] and [0.3 0.5 1] for the hexagonal crystals with c/a ratios of 2.72 and 1.63, respectively. Performing the dot product on the respective vectors illustrates how the angle between the [001] and [111] vectors can change by ~10.4° with a change ~66.8% c/a ratio (note that for ease of comparison to the cubic system, the three index notation is used for the hexagonal system). The ability to derive a conversion matrix for any given crystal system is necessary to perform these operations.

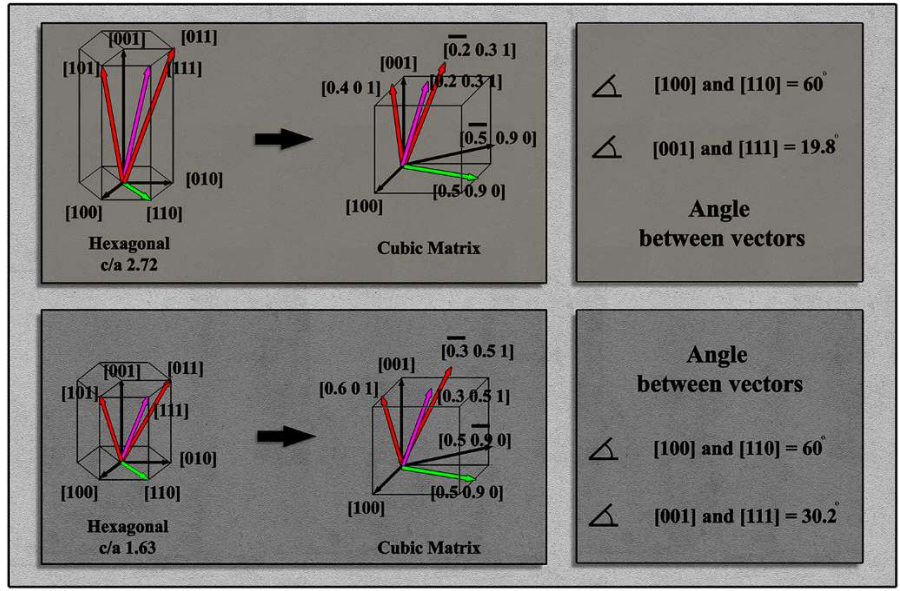

*Figure 2:* *Comparison of vector nomenclature in the native hexagonal system as compared to the cubic formulation.*

In order to travel about any of the other seven crystal systems (e.g., between poles) is it necessary to understand how to convert the axial and non-orthogonal biases into cubic/Cartesian form (note that for ease of comparison in the field of materials science, the term cubic will be utilized in the remainder of the paper). For the other orthogonal systems, the only consideration is the length of the bias, whereas for the remaining non-orthogonal systems there are angular dependencies on the axes in addition to length bias. In the overwhelming majority of materials science and electron microscopy textbooks, similar calculations are performed and presented as conversion formulas for each system. Again, these formulas are usually confined to describing angle between planes, and do not explicitly describe how to calculate the plane normals. This is typically performed because of the need to calculate the angle between diffracting planes and overlooks how to calculate the normals, which are needed to predict how to travel around each crystal system. The foundation for deriving all relevant crystallographic properties become available through the understanding of the pure geometric conversion of each crystallographic system to the cubic system. As an alternative formulation of this problem, these types of crystallographic computations can be calculated through the use of the metric tensor, as detailed by De Graef and McHenry (De Graef and McHenry, 2012) . While both approaches are mathematically correct, this work has chosen to preserve the intuitive sense of angles and distance in the cubic system, albeit with the requirement of a conversion matrix for non-cubic systems that will be described next.

The conversion of any non-cubic system (abc) to that of a cubic one (xyz) uses a conversion matrix (M) (Eqn. 1). A schematic of illustrating the two systems is shown in *Figure 3* along with the conversion matrix with the full derivation of this conversion matrix is provided in the Supplemental (Conversion to cubic in *Figures S1* and *S2*). Later, when discussing the microscope setup in more detail, the z-axis will be chosen to align with the electron beam. Consequently, we have chosen to align the c axis of the crystal that is to be converted with the z-axis for conceptual simplicity. This is certainly not the only choice, and the following derivations could be followed with a different convention, such as the one used by International Tables of Crystallography (Aroyo, 2016). However, the essence of the method is unchanged regardless of the specific convention used.

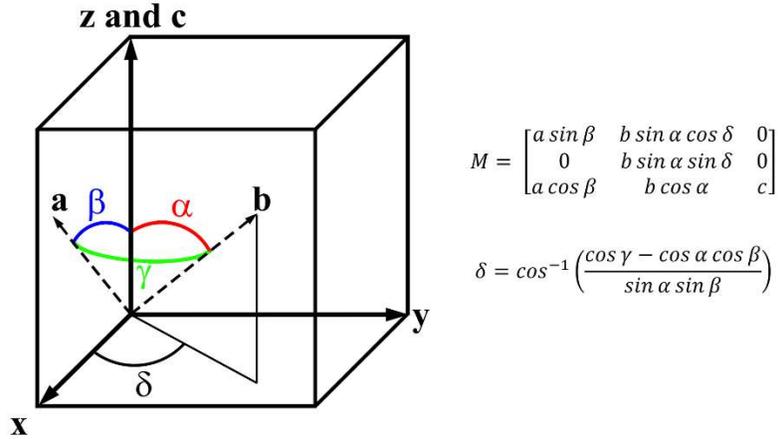

$$M = \begin{bmatrix} a \sin\beta & b \sin\alpha \cos\delta & 0 \\ 0 & b \sin\alpha \sin\delta & 0 \\ a \cos\beta & b \cos\alpha & c \end{bmatrix}$$

$$\delta = \cos^{-1}\left(\frac{\cos\gamma - \cos\alpha \cos\beta}{\sin\alpha \sin\beta}\right)$$

*Figure 3:* *Schematic illustrating how to convert any non-cubic vector coordinate system into a cubic system and the conversion matrix (M).*

The derivation of the conversion matrix (Eqn. 1), which includes the angle $\delta$ (Eqn. 2), can be calculated from the principal axis angles ($\alpha,\beta,\gamma$). The conversion matrix operates by setting one axis in the system to be converted equal to one axis in the cubic system (e.g., the c axis is first set commensurate with the z-axis in *Figure 3*). This is then followed by setting a second axis in the system to be converted and binding it within two of the axes in the cubic system (e.g., the a axis is restricted to the xz plane). The a axis can then be decomposed into the x and z components (i.e., there is no y component) through sine and cosine functions of the angle $\beta$, respectively. Lastly, the final axis of the system to be converted to the cubic system (in this case b to y) must be decomposed into all three axes of the cubic system (a full explanation is provided Supplemental). The introduction of a third angle, delta ($\delta$) (Eqn. 2) must be employed to account for the less symmetric crystals such as the monoclinic and triclinic where the angle $\gamma$ is not 90°. The conversion of any crystallographic vector in any crystal system can then be calculated by multiplying the vector by the conversion matrix (Eqn. 3).

Taking into consideration the [111] vector in the tetragonal unit cell as exhibited in *Figure 1*, when converted to the cubic system it can be described as a [0.5 0.5 1] (or [112]) vector, which is 35.3° from the [001] in the tetragonal system (the [001] converted by Eqn. 3 remains the [001]). These equations can then be utilized for any of the seven crystal systems for which the angle between vectors (i.e., poles) can be calculated with the understanding that the vectors are described in their native form but are calculated in the cubic form (an example of a hexagonal system conversion if presented in *Figure S3*). As such, all the remaining operations will be performed on cubic systems with the understanding that it can be generalized to any crystal system using the appropriate conversion matrix, or it's inverse.

*Equation 1*

$$M = \begin{bmatrix} a \sin \beta & b \sin \alpha \cos \delta & 0 \\ 0 & b \sin \alpha \sin \delta & 0 \\ a \cos \beta & b \cos \alpha & c \end{bmatrix}$$

*Equation 2*

$$\delta = \cos^{-1}\left(\frac{\cos \gamma - \cos \alpha \cos \beta}{\sin \alpha \sin \beta}\right)$$

*Equation 3*

$$Converted\ Vector = M * \begin{bmatrix} u \\ v \\ w \end{bmatrix} = \begin{bmatrix} ua \sin \beta + vb \sin \alpha \cos \delta \\ vb \sin \alpha \sin \delta \\ ua \cos \beta + vb \cos \alpha + wc \end{bmatrix}$$

While it has been described as the ability to travel throughout a crystal, a more elegant manner in which to describe these conversions is to envision all possible vectors within a cube. The "movement" around the system is the pathway between vectors. These pathways will eventually be considered traces of planes, and hence how to travel along these specific planes. The notion of predicting the location of all poles is important because it will subsequently be demonstrated that when considering electron beam interactions the structure factor will simply act as a filter to determine which of these poles are ultimately expressed for any given crystal.

## 2.2 Stereographic Projections – Rotation Maps

The stereographic projection is most often utilized by microscopists to navigate and understand crystalline sample orientations (***Figure 4***). As illustrated in ***Figure 4***a the stereographic projection is a calculation of the relative location of all vectors for a given crystal provided a specific normal orientation (e.g., the [001] in ***Figure 4***a). In the keeping with the context of this paper, the utilization of the stereographic projection keeps with the notion of first understanding the motion of the sample and crystal in real space and does not consider reciprocal space (i.e., Kikuchi bands).

There are different conventions for stereographic projections, typically whether the center of the sphere or the bottom of the sphere should be located at the origin, but they all contain the same fundamental information. Returning to the map analogy, this is analogous to different map projections (e.g., Mercator vs. Robinson) (Lapon et al., 2020). In this formulation, the sphere has a radius of 1 and is centered at the origin. To compute the location of these poles, consider a cubic crystal with one corner at the origin and [100] along the x-axis, [010] along the y-axis, and [001] along the z-axis. The intersection of the vector normal to a plane of atoms with the sphere and the top of the sphere define a line, as seen in ***Figure 4***a. The intersection of the line with the plane z = 0 defines the planar coordinates of the projection. Mathematically, if a pole is at [uvw], then it is normalized to have unit length (Eqn. 4) to find its intersection with the unit sphere. When observed in two dimensions (***Figure 4***b) the relative position of each vector and the trace between the vectors can be derived (Eqn. 5).

*Equation 4*

$$\begin{bmatrix} \dfrac{u}{\sqrt{u^2+v^2+w^2}} \\ \dfrac{v}{\sqrt{u^2+v^2+w^2}} \\ \dfrac{w}{\sqrt{u^2+v^2+w^2}} \end{bmatrix}$$

The line connecting the top of the sphere [0, 0, 1] and this point will intersect the z = 0 plane at

*Equation 5*

$$\begin{bmatrix} \dfrac{u}{\sqrt{u^2+v^2+w^2}-w} \\ \dfrac{v}{\sqrt{u^2+v^2+w^2}-w} \end{bmatrix}$$

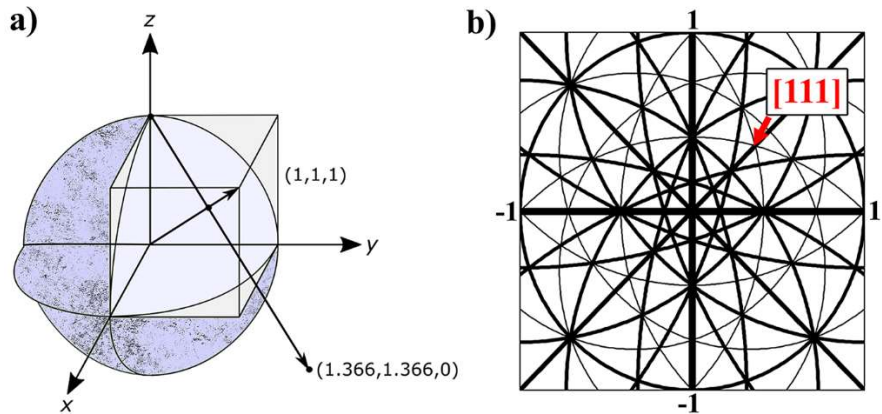

***Figure 4:*** *Geometric stereographic projection (in the [001] direction) in three dimensions (a) and the corresponding two-dimensional stereographic projection (b).*

It is important to note that when describing non-cubic pole figures, the native description of each pole is utilized such that when analyzing and reporting data there is an objective reference. Equally important to introduce here is the notion of the freedom of rotation within a pole figure. The stereographic projection can be considered as viewing a cube down a specific orientation and then determining how far and in which direction to rotate the crystal to align to another pole. The use of stereographic projections is useful when examining three-dimensional ball and stick models in computer programs designed to visualize orientations of crystals. Most programs will allow for the input of specific vectors, and it is worth pointing out that when a non-cubic system is visualized the poles/vectors are listed in a native coordinate system, but the mathematics are calculated by converting to a cubic system as shown above in Eqns. 1-3. In the electron microscope, the use of a double tilt stage adds an additional conversion that must be applied in order to travel throughout any crystal due to the limitation of the degrees of freedom.

## 2.3 Double Tilt Holder Coordinates – Tip/Tilt Map

As has been demonstrated by numerous other researchers, understanding the utilization of a double tilt stage in the analysis of solid materials and crystals is extremely important for accurate and reliable data collection (Cautaerts et al., 2018, Liu, 1994, Liu, 1995, Qing, 1989, Qing et al., 1989). While this has been reported on numerous occasions, the following derivation will be presented in a manner by which to convert three-dimensional rotations in simple geometric constructs to the double tilt stage and then demonstrate how this relates to reciprocal space and the physics of electron beam interactions. Given that different manufacturers utilize different terminology, as a matter of convention, the tilts of a double tilt stage will be denoted as α,β, and will be equivalent to X, Y tilts, respectively.

While stereographic projections are useful, they do not directly translate to the sample in the microscope due to the restrictions of the tip/tilt stage on which the sample is mounted. To accurately describe the position in terms of a double tilt holder, a tip/tilt map must be derived. This map plots the poles and planes of the crystal in terms of the tip and tilt coordinates of the double tilt holder and allows the prediction of all allowable poles within a specific orientation of a given sample. It should be noted that the derivations for converting from a stereographic projection to a tip/tilt map is reversible, and hence the tilt coordinates could and have been utilized overlaid on top of stereographic projections.

The importance of the motion of the double tilt holder as compared to a stereographic projection is the freedom of rotation considered for both. ***Figure 5*** provides the stereographic projection and tip/tilt map of a cube in the [001] orientation. In a stereographic projection the crystal can be rotated freely, hence oblique plane traces are straight (when emanating from the origin), whereas in the double tilt holder oblique plane traces will be curved (***Figure 5***). This curvature, as will be illustrated in the subsequent derivations, is a result of the β tilt axis changing as a function of the α tilt, and hence S-type curves are generated in a tip/tilt map. A trace of a single plane in the tip/tilt map (red line, ***Figure 5***b) has been overlaid on the stereographic projection (***Figure 5***a) to illustrate the difference. Additionally, when the position of three vector types (<103>, <114>, and <112>) in the tip/tilt map are transposed onto a stereographic projection, and it can be observed that the farther from the [001] orientation the larger the misorientation (e.g., the [114] vectors are located nearly in the same position, but the [112] are visibly misoriented).

This is also the exact reason why obliquely oriented g-vectors collected at high tilt angles in a double tilt stage (e.g., α,β:-25,-25) will rotate slightly in plane with relation to those collected at α,β:0,0. This is shown in the attached movie in ***Figure 5***c. Therefore, it is necessary to be able to convert a stereographic projection to a tip/tilt map for any given crystal system. This can be achieved through rotation matrices.

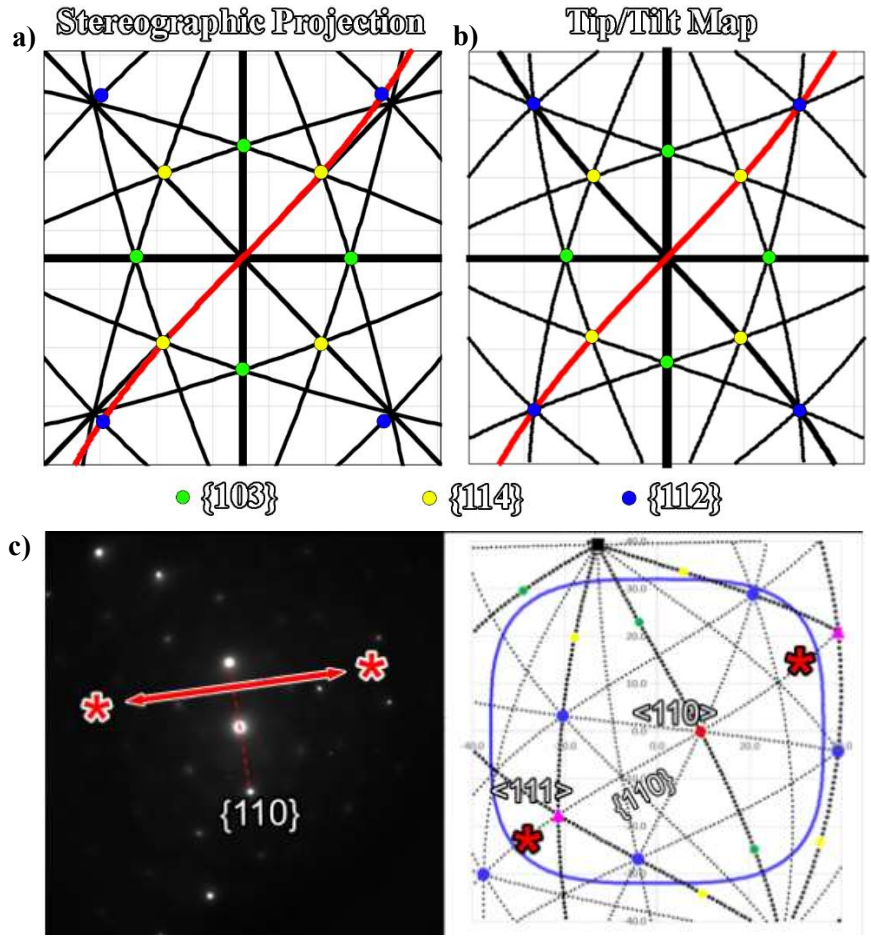

*Figure 5:* *Stereographic projection (a) versus a tip/tilt map (b) of a cube oriented in the [001] orientation illustrating how the oblique plane traces vary in path. A diffraction pattern collected along a {110} g-vector (between red stars) in an FCC crystal (c) Movie showing the tilt series along a {110} trace where the g-vector changes direction correlating to the angle of the line in the tip/tilt map.*

## 2.4 Rotation Matrices

While rudimentary, in terms of the overall development of the stage motion the fundamental mathematical operation being utilized is the rotation about a single axis. This rotation is typically described as a rotation about any of the primary axes (x,y,z), but more generally, any orientation can be described through successive rotations about these three axes (Eqns. 6-8). The rotation about any primary axis will be defined by the right-hand rule so that rotation matrices about the x-, y-, or z-axis through an angle θ are given by, respectively:

*Equation 6*

$$R_{\theta,x} = \begin{bmatrix} 1 & 0 & 0 \\ 0 & \cos\theta & -\sin\theta \\ 0 & \sin\theta & \cos\theta \end{bmatrix}$$

*Equation 7*

$$R_{\theta,y} = \begin{bmatrix} \cos\theta & 0 & \sin\theta \\ 0 & 1 & 0 \\ -\sin\theta & 0 & \cos\theta \end{bmatrix}$$

*Equation 8*

$$R_{\theta,z} = \begin{bmatrix} \cos\theta & -\sin\theta & 0 \\ \sin\theta & \cos\theta & 0 \\ 0 & 0 & 1 \end{bmatrix}$$

Note that for these rotation matrices they are defined by the angle of rotation and the axis about which the rotation occurs. In addition to rotations about a single axis (i.e., proper rotations), there are also improper rotations (i.e., reflections about an axis) that describe mirroring about a single axis (Eqns. 9-11). These can be illustrated in matrix form:

*Equation 9*

$$R_{-x} = \begin{bmatrix} -1 & 0 & 0 \\ 0 & 1 & 0 \\ 0 & 0 & 1 \end{bmatrix}$$

*Equation 10*

$$R_{-y} = \begin{bmatrix} 1 & 0 & 0 \\ 0 & -1 & 0 \\ 0 & 0 & 1 \end{bmatrix}$$

*Equation 11*

$$R_{-z} = \begin{bmatrix} 1 & 0 & 0 \\ 0 & 1 & 0 \\ 0 & 0 & -1 \end{bmatrix}$$

These are the *basic building blocks* upon which more general rotations can be built and will be referred to continuously when performing matrix operations to rotate the crystal and orient the sample. Whether rotating along a specific interface, tilting from pole to pole, or re-loading a sample and converting prior tilt conditions, these six formulae will be the basis set. In the following derivation of the tip/tilt map the utilization of diffraction and crystallographic terminology will be utilized but only as a frame of reference and not in terms of the electron beam interaction (i.e., traces of planes and not g vectors or Kikuchi bands).

In practice, the manner in which a microscopist interacts with a crystalline sample is at the most basic level through diffraction spots or Kikuchi lines. While the principle knowledge of what these optical markers represent goes to a fundamental understanding of electron beam interactions with samples, at the very core of electron microscopy as an observational tool, the understanding of these as fiduciary markers for roadmaps provides the basis for nanocartography. That is to say, regardless of whether one understands the physics of why a zone axis (ZA) appears, the observation and acknowledgement of a zone axis as a combination of a series of geometrically oriented Kikuchi lines (or order diffraction patterns) is necessary to be utilized as a map.

These traces along with the knowledge of the double tilt stage can be used to solve unknown crystals and also predict the motion of interfaces. In the current context of deriving a tip/tilt map, the knowledge of a specific pole type is assumed for the purpose of ease of explanation. This assumption is made as an initial means to connect vector motion to crystallographic analysis. For instance, as shown in *Figure 6*a, the six fold symmetry of the [111] in an FCC steel is presented along with a tilt position (e.g., α,β : 5,10) in which this orientation was discovered in the microscope (note that the correct nomenclature should be <111> as it is a description of the family of poles, but for ease of explanation a single vector will be used).

This "known pole" provides the first bit of knowledge as to a global position with respect to the remainder of the crystal. This could also be a diffraction pattern of the [111] pole, but for ease of understanding a convergent beam electron diffraction pattern (CBED) is presented. The in plane orientation of the crystal is described by the angle ($\varphi_c$) that can be used to describe the rotation of the crystal about the known pole. As will be subsequently described, the angle $\varphi_c$ can be used to freely rotate the crystal (*Figure 6*b-d), or it can be assigned as a specific fiducial marker (such as the (1-10) Kikuchi line) with known relation to the calibration of the α tilt axis. The tilt positions observed (e.g., α,β : 5,10) can then be used to mentally envision the relationship of the found ZA to the tilt stage (*Figure 6*b, where the stage has been returned to α,β:0,0). This crystal rotation can best be considered by illustrating a vector in a cube (e.g., [111] in *Figure 6*) in both the standard projection (*Figure 6*c) and a projection normal to the vector (*Figure 6*d), and the $\varphi_c$ being the relative rotation by which the entire cube rotates. Again, while the CBED pattern of the [111] is presented it is only used to orient the practical aspect of the microscope to developing a tip/tilt map. From these initial data, the tip/tilt map can be derived as follows.

There are two orientations that will be considered: the orientation of the crystal with respect to the probe, and the stage with respect to the probe. The probe will be considered an objective frame of reference oriented at [001]. The orientation of the sample can be described for any crystallographic orientation within the sample, whether a single grain or the orientation of two crystals across a boundary. *Figure 6*b shows an observed pole (red and blue) for two grains, of which the unit vectors for each crystal in black. The knowledge of the location of the observed poles and the unit vectors will be important for obtaining the most information from a sample (e.g., the local misorienation between two grains).

In order to create a tip/tilt map, the known vector will be rotated through a sequence of rotations to align it with the [001] probe orientation, subsequently rotated with respect to the crystal orientation ($\varphi_c$) using Eqn. 8, and finally the stage will be tilted from the probe position to the found conditions (observed tilts) (*Figure 6*e-g). The known pole will be utilized to derive the entire rotation matrix representing the crystal orientation, and subsequently any other possible vector can then be plotted accordingly through that matrix.

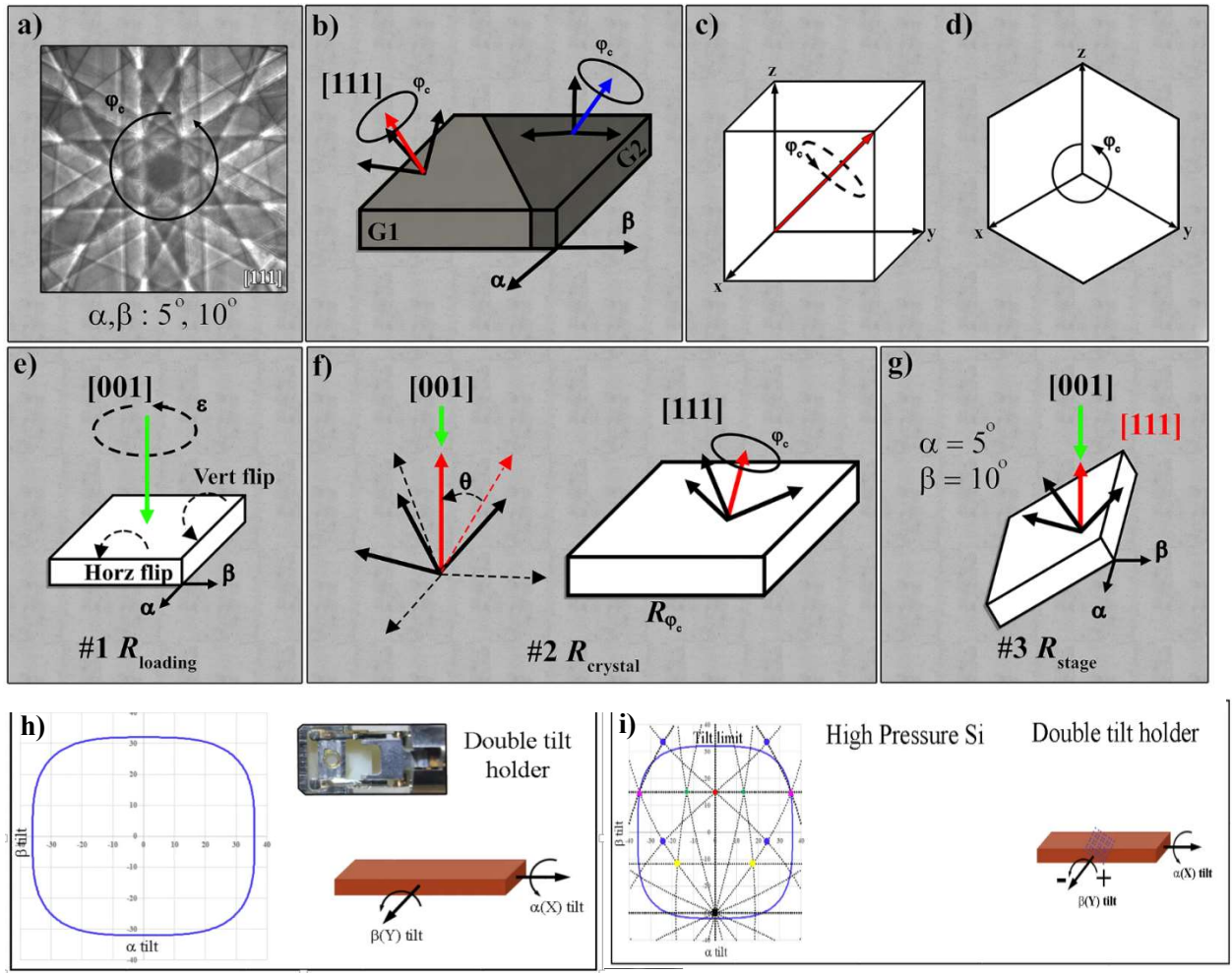

*Figure 6:* *Schematic and movies illustrating crystal rotation around a single vector/pole in relation to the stage tilt axes and their relationship to the rotation matrices detailed in the text. a) Kikuchi pattern of a [111] pole at* ***α****,:5,10. b) Crystallographic orientations of given grains (G1, G2) and angle of rotation **φ**$_c$ about each pole. c and d) Rotation about an arbitrary vector through a given angle **φ**$_c$ in two different projections. e) Matrix rotation R$_{loading}$ that describes sample rotation (**ε**)about the beam normal [001], and vertical and horizontal flips about the **α** and **β** axes, respectively. f)Matrix rotation R$_{crystal}$ about a given angle (**φ**$_c$). g) Matrix rotation R$_{stage}$ describing the found pole back to the stage tilts. h and i) Movies showing stage motion.*

The mathematical derivations of the full rotation matrix are divided into three steps. The first step, $R_{loading}$, represents the orientation of the sample with respect to the holder as it is inserted into the microscope (*Figure 6*e). The second step, $R_{crystal}$, aligns the mathematical description of the crystal with the one found in the microscope (i.e., α,β of the known pole and rotation of the crystal φ$_c$), but as shown in *Figure 6*f it does not consider the tilt conditions. The last step, $R_{stage}$, puts the pole at the location corresponding to the known α,β coordinates observed (*Figure 6*g). Taken together, the multiplication of these matrices yields an overall rotation matrix, $R_{total} = R_{stage}R_{crystal}R_{loading}$ that contains all the orientation information about the crystal as it is situated in the microscope. Again, it is important to note that the only necessary functions are combinations of the rotations provided in Eqns. 6-11.

One of the major advantages of these derivations with regards to previous calculations on stage motion ((Cautaerts et al., 2018, Klinger and Jäger, 2015, Liu, 1994, Liu, 1995, Qing, 1989, Qing et al., 1989)) is in the power of creating a sample map which can be utilized in subsequent analyses whether on the same microscope or at other institutions. This allows for rapid re-analysis of samples without losing previous crystallographic orientation data. Three terms are required that will allow the sample to be reloaded into any microscope in any orientation and convert any previously recorded tilt coordinates to the current sample loading. The application of these matrices account for the sample being flipped in the holder either horizontally ($R_{horz}$, Eqn. 12) or vertically ($R_{vert}$, Eqn. 13) to the long axis of the holder, and as well if the sample had been rotated in-plane by any angle $\varepsilon$ ($R_{position_{\varepsilon,z}}$, Eqn. 14) (***Figure 6***e).

The combination of these sample reloading matrices can be combined into one matrix $R_{loading}$ as shown in Eqn. 15. Note, the angle $\varepsilon$ is measured and recorded through a global fiduciary marker (e.g., the surface of a FIB lamella) during each analysis. In the instance where there is no horizontal or vertical flip (i.e., the initial analysis of the sample), then these matrices are be replaced with the identity matrix (i.e., no rotation is applied).

*Equation 12*

$$R_{horz} = R_{-x}R_{-z} = \begin{bmatrix} 1 & 0 & 0 \\ 0 & -1 & 0 \\ 0 & 0 & -1 \end{bmatrix}$$

*Equation 13*

$$R_{vert} = R_{-y}R_{-z} = \begin{bmatrix} -1 & 0 & 0 \\ 0 & 1 & 0 \\ 0 & 0 & -1 \end{bmatrix}$$

*Equation 14*

$$R_{position_{\varepsilon,z}} = \begin{bmatrix} \cos\varepsilon & -\sin\varepsilon & 0 \\ \sin\varepsilon & \cos\varepsilon & 0 \\ 0 & 0 & 1 \end{bmatrix}$$

*Equation 15*

$$R_{loading} = R_{position}R_{horz}R_{vert}$$

Once the sample has been loaded into the microscope and a known pole has been identified, the mathematical model of the orientation of the crystal is matched to the orientation of the sample. The first step in developing this correspondence is rotating a known vector (e.g., the [111]) to the probe direction [001] through a rotation matrix ($R_{\hat{r},\theta}$ ***Figure 6***f). It is important to note that this is setting the orientation of the crystal to the probe, and thus the tilt conditions of the stage are not considered. For ease of explanation, the use of vector terminology instead of crystallographic designations (such as ZA or crystal pole) will be utilized to describe the motion of a crystal in a stage.

The rotation of this vector to the probe direction can be achieved through a number of pathways (e.g., combination of rotation matrices), but the most direct is a rotation about an arbitrary axis by an angle ($\theta$) (*Figure 6*f). As the vector will always be rotated to the [001] direction to be aligned with the probe, the axis of rotation will always lie in the xy plane and will take the form of [uv0] because it is calculated through the cross-product of the known vector and [001] (see *Figure S4*). It should be noted that this is special to this case, and a more general formulation needs to be derived for a general operation. This will be subsequently utilized to describe the trace of planes.

The known pole need first be normalized to create a unit vector. The dot product is used to compute the angle required to move this unit vector in the direction of the pole from its standard orientation (i.e., at $\frac{1}{\sqrt{u^2+v^2+w^2}}(u,v,w)$ in Cartesian coordinates):

*Equation 16*

$$\theta = \cos^{-1}\left(\frac{w}{\sqrt{u^2+v^2+w^2}}\right)$$

The axis of rotation is determined from the cross product of the normalized known pole and the beam direction:

*Equation 17*

$$\hat{r} = \begin{vmatrix} \hat{x} & \hat{y} & \hat{z} \\ 0 & 0 & 1 \\ \frac{u}{\sqrt{u^2+v^2+w^2}} & \frac{v}{\sqrt{u^2+v^2+w^2}} & \frac{w}{\sqrt{u^2+v^2+w^2}} \end{vmatrix} = -\frac{v}{\sqrt{u^2+v^2+w^2}}\hat{x} + \frac{u}{\sqrt{u^2+v^2+w^2}}\hat{y}$$

For these specific axes of rotation that have no z-component (i.e., in the derivation of tip/tilt maps), and the general result simplifies to (where $r_x$ and $r_y$ are derived from Eqn. 17, and $\theta$ from Eqn. 16):

*Equation 18*

$$R_{\hat{r},\theta} = \begin{bmatrix} r_x^2 + r_y^2 \cos\theta & r_x r_y (1-\cos\theta) & r_y \sin\theta \\ r_x r_y (1-\cos\theta) & r_y^2 + r_x^2 \cos\theta & -r_x \sin\theta \\ -r_y \sin\theta & r_x \sin\theta & (r_x^2 + r_y^2)\cos\theta \end{bmatrix}$$

The mathematical derivation of the rotation matrix ($R_{\hat{r},\theta}$) of an angle $\theta$ about an arbitrary axis is presented in full in the Supplemental (*Figures S4-S5*). As an aside, it should be noted that with respect to crystallographic tip/tilt maps, the rotation about an arbitrary axis is not necessary. Two rotations (and subsequent inverse rotations) can be utilized that will accomplish the same rotation, but in subsequent utilization of these derivations for calculation of the local misorientation angle and axis between two adjacent grains there will arise a misalignment depending on the order of rotation. This is discussed in further detail in the Supplemental section (*Figure S5*).

As previously described, in order to orient the crystal with respect to the known pole (*Figure 6*a) an additional rotation is required. Since the crystal has been rotated to the z-axis, the rotation of the crystal through the angle $\varphi_c$ about the z-axis (Eqn. 8, $R_{\varphi_c,z}$) will rotate the crystal about the known pole. Combining the rotation of the known pole to, and about, the z-axis provides the full definition of $R_{crystal}$.

*Equation 19*

$$R_{crystal} = R_{\varphi_{c,z}} R_{\hat{r},\theta}$$

Whereas the rotation of the known vector to the probe direction was accomplished through a direct rotation from one position to another, the majority of double tilt stages do not operate in this manner and are performed through a two-step process with one axis beholden to the other. As can be illustrated in **Figure S4**, the order of rotation in a two-step process can affect the outcome of the final position, and hence order of tilt is a necessary consideration. The rotation of any vector to the final tip/tilt location $\alpha/\beta$ is accomplished by multiplication by $R_{\alpha,x}$ followed by $R_{\beta,y}$. This combination is called the rotation matrix of the stage $R_{stage}$ (Eqn. 20) where the tilt conditions for the known vector $\alpha,\beta$ are substituted in to Eqns. 6 and 7, respectively.

*Equation 20*

$$R_{stage} = R_{\beta,y} R_{\alpha,x}$$

The order of these rotations is important in the sense that the tip/tilt stage rotations are not interchangeable. Since the rotation of one holder axis ($\alpha$ in this case) does not change the axis of rotation of the second tilt ($\beta$), the first rotation is about the $\alpha$ axis. This allows the use of an active rotation framework presented above without difficulty where the axes are considered fixed. While a passive rotation formulation would be logically equivalent, mixing the two would lead to incorrect results. The most striking example of the fact that the order of rotation matters is that the tip/tilt diagram is not symmetric in the location of poles as previously illustrated in the stereographic projection as compared to the tip/tilt map (**Figure 5**).

To summarize this set of operations, the action of all these rotations in concert can be summarized in the total rotation matrix:

*Equation 21*

$$R_{total} = R_{stage} R_{crystal} R_{loading}$$

Finally, it is important to understand that this matrix operation provides the Cartesian coordinates of the poles (i.e., a 3x1 matrix), and hence these final values must be converted to $\alpha/\beta$ coordinates. An intuitive way to understand this conversion is to consider the rotation of a vector from (0, 0, 1) to (X, Y, Z) in Cartesian coordinates, where $X^2 + Y^2 + Z^2 = 1$. This conversion amounts to solving for the angles $\alpha/\beta$ that satisfy:

*Equation 22*

$$\begin{bmatrix} \cos\beta & 0 & \sin\beta \\ 0 & 1 & 0 \\ -\sin\beta & 0 & \cos\beta \end{bmatrix} \begin{bmatrix} 1 & 0 & 0 \\ 0 & \cos\alpha & -\sin\alpha \\ 0 & \sin\alpha & \cos\alpha \end{bmatrix} \begin{bmatrix} 0 \\ 0 \\ 1 \end{bmatrix} = \begin{bmatrix} X \\ Y \\ Z \end{bmatrix}$$

After multiplying and solving the individual equations, the final tilt angles are:

*Equation 23*

$$\alpha = \tan^{-1}\left(-\frac{Y}{\sqrt{X^2 + Z^2}}\right)$$

*Equation 24*

$$\beta = tan^{-1}\left(\frac{X}{Z}\right)$$

The X,Y,Z terms are not the vector describing the known vector or any starting vectors, but the final converted vectors through $R_{total}$ (e.g., if the known vector was [111], XYZ would not be defined by [111]). Note that these can be expressed in terms of other trigonometric functions that are equivalent mathematically, but it is most convenient to use the inverse tangent function in practice because it accepts signed inputs for both its inputs which allows angles to range anywhere from –π to π. This removes the requirement to adjust the quadrant of $\alpha/\beta$ explicitly. Once $\alpha/\beta$ have been computed for every pole of interest, then the poles can be plotted as a function of $\alpha/\beta$ to create the tip/tilt diagram detailed above. A demonstration of stage movement is shown in *Figure 6* h and i that shows simple sample tilt in beta (*Figure 6*h) as well as with a representative BCC Si phase ball and stick model to illustrate how the crystal would rotate with the stage (*Figure 6*i).

Examples of plotting of various poles and tilt conditions is shown in for both cubic and hexagonal systems. Cubic vectors [001] and [111] oriented at the (α,β:0,0) condition with a variety of other vectors are shown (**Figure 7**a and b, respectively). The asymmetry of the double tilt stage movement presented in *Figure 5* and again in *Figure 7*a becomes apparent, with the [112] not being located at equal α,β conditions when the (-110) is oriented 45° to the α tilt axis (it is observed at (α,β:24.1,26.6)). This is due to the β tilt dependency on the initial α tilt. Subsequent analysis of directions between poles will elucidate this in greater detail.

The vectors presented in these representations are arbitrarily based on what would represent lower index crystallographic poles, and as it were, any vector could be plotted. The plot of the [110] (*Figure 7*c) at tilts (α,β:10,20) is shown to illustrate how a discovered pole at a non α,β:0,0 tilt condition would appear to provide a more representative scenario of what would be observed in the microscope. In this orientation the <111>, <100>, and <112> low index poles are in the field of view. Additionally, the bounds of the tilt stage limits can be overlaid upon these maps to further discriminate the allowable poles within a specific grain.

In order to demonstrate how the conversion of non-cubic systems are handled, hexagonal plots are presented (*Figure 7*d-e). These figures illustrate a hexagonal system with a c/a ratio of 1.63 in both the basal [001] and primary prism [210] orientations at tilts (α,β:0,0). In the basal [001] orientation the [111] pyramidal poles are plotted, and in the primary prism [210] the secondary prism [100] are observed at (30,0) and (-30,0). As a demonstration of how the vector projections change with a change in c/a ratio, the [001] projection at (α,β:0,0) for a hexagonal system with a c/a ratio of 2.72 is presented in *Figure 7*f. The elongation of the c axis draws the [111] type vectors closer towards the (0,0) tilt position and as well the [1-11] type vectors are now within the applicable 40° tilt range. This change in c/a ratio can also be observed in *Figure 2*. Additionally, with the change in c/a ratio it is also noted that the angle between the primary and secondary prism poles do not change because they are orthogonal to the c axis, and hence are unaffected. The reader is guided to the online code (insert inline documentation here) to create basic tip/tilt maps for any system at their leisure.

Systems that are more complex could also be illustrated (see Supplemental *Figure S6* for examples), but it should again be mentioned that a) that while the vectors are described in their native format, the math is done in a cubic form, and b) any vector possible may be plotted because these are vector representations. The plots in *Figure 7* represent generic crystals/maps for the given crystal system (i.e., cubic and hexagonal), and do not represent real crystals. The presentation of these maps are solely meant to illustrate the tilt parameters for solid objects in real space. This sets the basis for the derivation of crystals in reciprocal space to explain the travel of planes of atoms within a crystal.

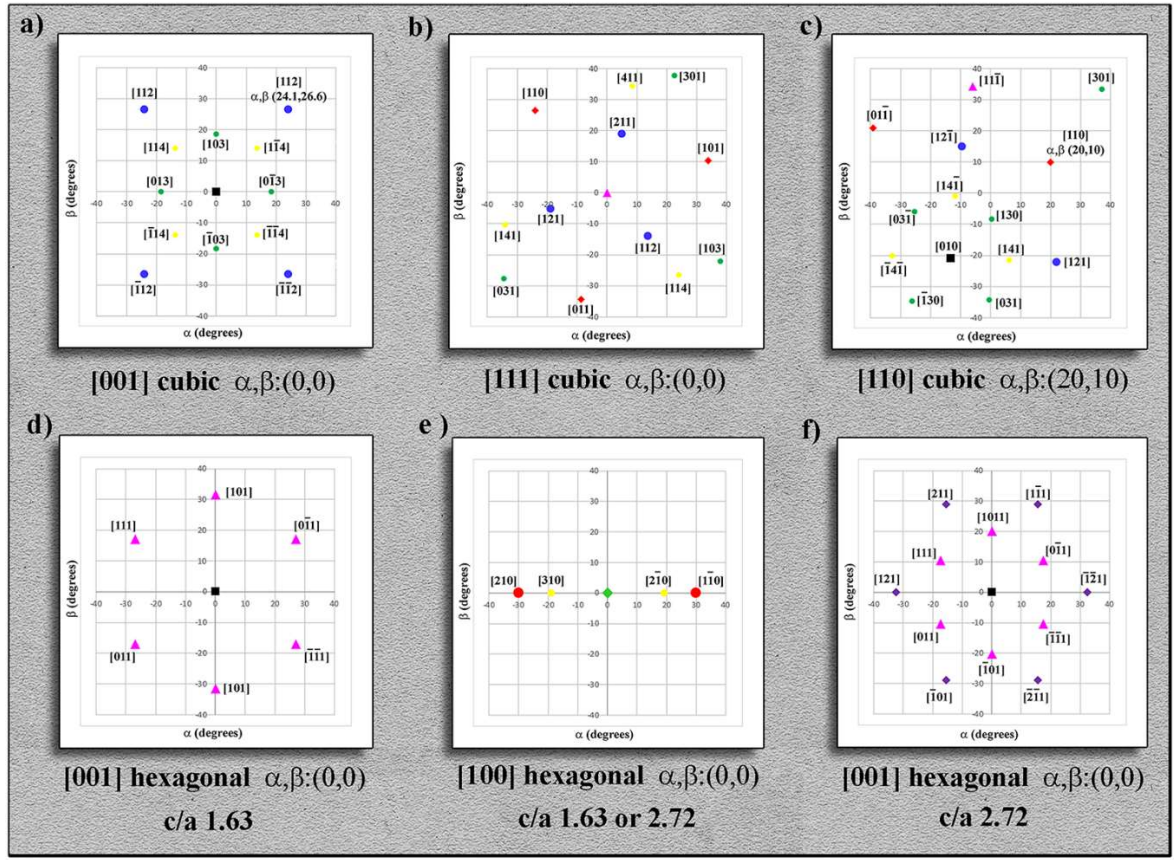

*Figure 7*: Tip/Tilt plots of the cubic system with [001],[111] at the α,β:0,0 (a and b, respectively) and the [110] at α,β:20,10 (c), and the hexagonal unit cell (d-f) with c/a ratios of 1.63 (d,e) and 2.72 (f) with either the [001] (d,f) or [100] (e) at α,β:0,0

## 2.5 Calculation of Planes in a Tip Tilt Map

The development of a tip/tilt map for any given crystal system provided a manner in which to predict the tilt motion of any possible vector within each system. These tip/tilt maps are most relevant to stereographic projections or poles figures that indicate the motion between poles within a freely rotating system. In order to make a more complete comparison, it is necessary to add a description of the travel between poles. This will also facilitate the transition from real space to reciprocal space when discussing crystallographic planes. The understanding of the real space calculations is not only imperative for crystallographic motion, but as will be demonstrated, the identical formulations can be utilized to define the pathways of other physical constructs, such as interfaces and free surfaces within the sample.

This discussion must be prefaced with the explicit understanding of these motions with respect to crystallographic terminology as to not further confound the already difficult task of differentiating real space and reciprocal space. Kikuchi lines are a representation of inelastic scattering that diffracts from crystallographic planes at the Bragg angle, and while the derivation and presence of allowed diffracting planes will be considered in subsequent sections, their introduction here is used as a manner by which to suggest that just as plotted poles can be represented as in both stereographic projections and tip/tilt maps, so too can the travel between any of these poles be calculated or mapped. Again, the presentation of these will be discussed in simple geometric terms and then later elaborated upon in terms of crystallography and electron beam interaction.

Concerning plotting actual Kikuchi lines as compared to plotting the tilt coordinates between various poles, a standard convention must be adopted. While Kikuchi lines are formed in pairs corresponding to both the positive and negative g vectors, within the accuracy of any double tilt stage given possible errors such as motor backlash and machining tolerance it is more convenient to plot a single set of directions for the trace of any given plane whose vector has been normalized (i.e., the normal of the (222) can be described as [111]). This is not to say that the mathematics could not be derived for the exact tilt coordinates for each specific allowed plane for any crystal, but in terms of practical analysis, the normalized vector for each family will be considered. *Figure S7* illustrates a tilt map for an FCC austenitic stainless steel (unit cell ~3.86 Å) oriented in the [111] orientation with the {440} planes expressed, and a CBED pattern in the same orientation. These Kikuchi bands represent a major plane that would be expected to be oriented farther out within k-space, and still the tilt angle is ~1°. Therefore, the proposed method is a conversion of a stereographic projection into tip/tilt space more than it is a conversion of a crystalline stereographic projection.

To plot the trace of any given crystallographic plane, especially for non-cubic systems, the normal to the plane must first be calculated and the subsequently converted to the cubic form. This formulation is modulated by the crystal structure and structure factor that is discussed in detail in the subsequent section. Only the cubic form will be discussed herein since the description of the normal to the plane is the same as the plane itself. The trace of the plane can be considered as the plot of all possible vectors within the plane, and therefore a rotation matrix with the plane normal substituting for the arbitrary axis of rotation is necessary (see *Figure 8*). As will be shown, this rotation matrix is nearly identical to the rotation about an arbitrary axis ($R_{\hat{r},\theta}$) derived in Eqn. 18 with the caveat that a more generic derivation can be developed that is not required to tilt to the beam direction.

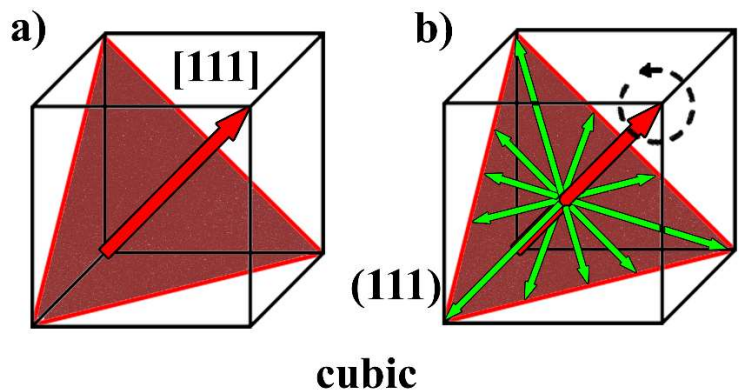

**cubic**

*Figure 8:* Schematic illustrating the derivation of traces of vectors along a crystallographic plane (green arrows) given the plane normal (red arrow). a) [111] vector and (111) plane. b) Plotting vectors along the (111) plane rotating about the [111] vector.

The rotation matrix can be calculated by first determining any normal vector (green arrows *Figure 8*b) to the plane normal (red arrow *Figure 8*a,b), where this vector lies in the desired crystallographic plane. This truly arbitrary rotation matrix is found by evaluating Eqn. 18 with the arbitrary axis of rotation being the plane normal and the angle of rotation becoming the desired step size of the line to be plotted. A set of vectors is created by applying the rotation matrix to one of the poles repeatedly until it returns to its original location (in the case of 1-degree steps this will yield 360 total vectors). Because these vectors are in standard orientation, they must be rotated as the poles were above through the multiplication of $R_{tot}$ (Eqn. 21). This will yield the Cartesian vector sequence which then is required to be converted to $\alpha/\beta$ coordinates using Eqns. 22-24. Computing these sequences for various low index planes of interest yields in each system the complete tip/tilt diagram that can be seen in *Figure 9*. The attached python module allows the reader to create tip/tilt diagrams for generic cubic and hexagonal constructs. Variation of crystal parameters, starting poles, tilt conditions, and stage limits are allowable. Whereas the normals can be described by the Miller indices in the cubic system, for the hexagonal planes the plane normal first needed to be calculated, subsequently converted to cubic, and then plotted. As has been previously mentioned (*Figure 5*), due to the motion of the double tilt stage the traces of the planes can exhibit S-curves and are not always straight. This motion is exactly how the planes of atoms within the microscope behave across the entire tilt space, and the *reason why g-vectors in diffraction patterns collected at different ZA can appear to rotate in relation to one another*. The calculation of the vector normal, or g-vector, to these hexagonal planes will be discussed in the next section.

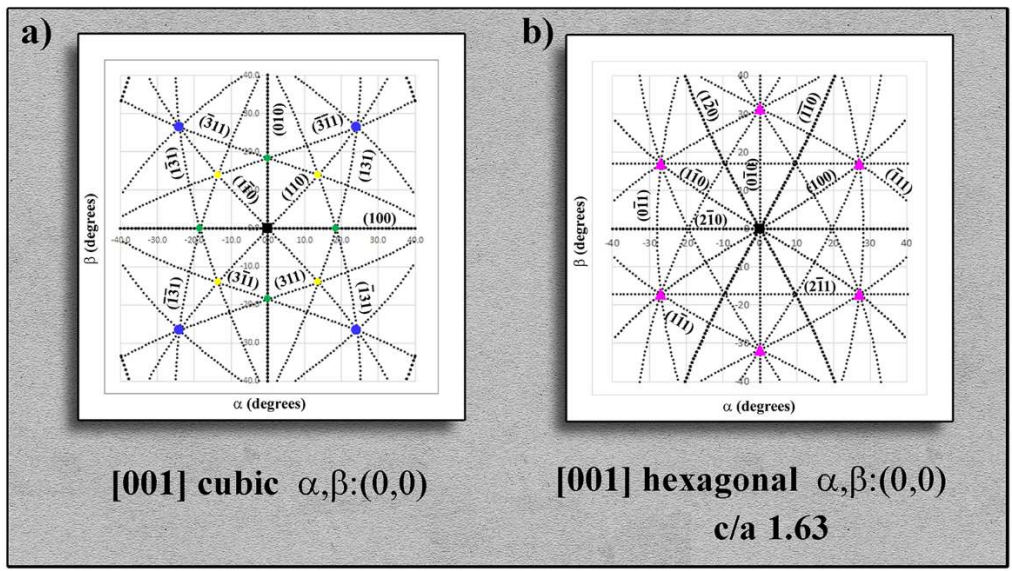

*Figure 9:* Tip/tilt maps of cubic and hexagonal crystals oriented at with the [001] and [001](a and b, respectively) at the (α,β:0,0) positions.

These derivations combine many different aspects of previously published research but have been presented in the manner of simple geometric considerations for the purpose of deconvoluting the physical nature of electron microscopy samples from the physics of electron beam interaction. The ability to understand the motion of and samples within a double tilt stage is imperative, *and* then subsequently being able to connect that knowledge to the physics of electron beam interaction can elevate any research whether the crystal structure of the desired sample is known, or more importantly *if it is not*. As an example, the description of interface motion can be modeled after the motion of the trace of crystallographic planes, thus allowing for the microscopist to orient crystals and physical objects such as grain boundaries or surfaces.

## 2.6 Reciprocal Lattice Vectors – Reciprocal Space Maps

The study of nanoscale electron beam interactions with solid materials (most importantly crystals) in the electron microscope has been a hallmark of the technique and has drastically expanded any number of scientific fields. The previous sections treated samples only as geometric objects in order to more easily orient the reader as to how vector and matrix mathematics can be utilized to travel through a sample. This knowledge is extremely useful even when the material is unknown. When crystalline orientations are known, this information can then be mapped onto the strict geometrical derivations previously discussed. While there are countless topics (over a century's worth of research) surrounding electron beam interactions with crystals, this paper will concentrate solely on the structure factor due to its role in which planes of atoms (and hence poles) are exhibited for any given crystal. It is beyond the scope of the work to go beyond this, and with respects to the topic of nanocartography it would not be relevant.

With the considerations of any crystal as a simple geometric construct or unit cell (e.g., a cube or hexagon), the introduction of the atomic packing within these cells will dictate when a crystallographic plane will be expressed via diffraction. By first illustrating how any infinite number of planes or vectors can be plotted and manipulated, the discussion of which planes can be observed for a given crystallographic sample becomes clearer than first introducing reciprocal space and then demonstrating how it can be manipulated through three-dimensional space.

The description of the travel between poles has been previously demonstrated, and these derivations can also be considered to travel along a specific plane within that body. In the derivation, the tilt coordinates for every normal to a desired pole were calculated, hence forming the directions along the plane. In order to relate these derivations to electron beam diffraction it is necessary to convert the crystal to reciprocal space and determine the normal of any plane.

It should be noted that this is slightly different than how this topic is typically presented with the angle between planes being derived for each crystallographic system in addition to the d-spacing within a crystal (Carter et al., 1996). While this is important for the analysis of diffraction patterns, it does not account for calculating the normal for any plane within any crystal. It is well known (and is often the basis of materials science education) that the description of the normal to a plane of atoms in a cubic crystal is the same description as the plane (i.e., [uvw] = (hkl)). While the mathematical analysis of the angle between poles is straightforward, more often than not when the discussion of non-cubic systems is broached the introduction of the angle between planes is introduced without further explanation (e.g., hexagonal systems).

The conversion of any pole within any crystal was demonstrated in Eqns. 1-3, but this was considered in the realm of real space. Crystallographic analysis with regards to diffraction is always considered in reciprocal space, and hence the derivation of the normal to any crystallographic plane is necessary. The unit vectors in reciprocal space must be derived by first considering the unit vectors in real space (i.e., the [100], [010] and [001]) (Eqns. 25-27). For a cubic system this simply becomes [a00], [0b0], and [00c] because the orthogonal nature of the crystal precludes any of the trigonometric operators in the conversion matrix from being anything other than 1 or 0. As shown in *Figure 1*, introducing the unit cell bias of a tetragonal system where a equals b but does not equal c, the unit vectors are still the similar description as the cubic system because of the orthogonality of α, β, and γ, save for the magnitude of c. When these three angles are not mutually orthogonal, the length of the unit vectors is a combination of the lengths of the unit cell and angles describing the cell as calculated by the conversion matrix (Eqn. 1) multiplied by the unit axes vectors:

*Equation 25*

$$M \begin{bmatrix} 1 \\ 0 \\ 0 \end{bmatrix} = \hat{a} = \begin{bmatrix} a \sin \beta \\ 0 \\ a \cos \beta \end{bmatrix}$$

*Equation 26*

$$M \begin{bmatrix} 0 \\ 1 \\ 0 \end{bmatrix} = \hat{b} = \begin{bmatrix} b \sin \alpha \cos \delta \\ b \sin \alpha \sin \delta \\ b \cos \alpha \end{bmatrix}$$

*Equation 27*

$$M \begin{bmatrix} 0 \\ 0 \\ 1 \end{bmatrix} = \hat{c} = \begin{bmatrix} 0 \\ 0 \\ c \end{bmatrix}$$

Once the unit vectors in real space have been derived, the unit vectors in reciprocal space are then formulated by crossing the opposite unit vectors in real space and then dividing by the volume of the cell (Eqn. 28) to gain the lengths of the unit vectors in reciprocal space (Eqns. 29-31). The volume of any parallelepiped can be calculated by taking the cross product of two of the unit vectors dotted by the third. In condensed form it appears in Eqn. 28.

*Equation 28*

$$V = (axb) \cdot c = abc\sqrt{1 - \cos \alpha^2 - \cos \beta^2 - \cos \gamma^2 + 2 \cos \alpha \cos \beta \cos \gamma}$$

*Equation 29*

$$\underline{a} = \frac{\hat{b} \times \hat{c}}{V} = \begin{bmatrix} \frac{bc \sin \alpha \sin \delta}{V} \\ \frac{-bc \sin \alpha \cos \delta}{V} \\ 0 \end{bmatrix}$$

*Equation 30*

$$\underline{b} = \frac{\hat{c} \times \hat{a}}{V} = \begin{bmatrix} 0 \\ \frac{ac \sin \beta}{V} \\ 0 \end{bmatrix}$$

*Equation 31*

$$\underline{c} = \frac{\hat{a} \times \hat{b}}{V} = \begin{bmatrix} \frac{-ab \cos \beta \sin}{V} \\ \frac{ab(\sin \alpha \cos \beta \cos \delta - \sin \beta \cos \alpha)}{V} \\ \frac{ab \sin \alpha \sin \beta \sin}{V} \end{bmatrix}$$

The cubic unit cell can then be calculated by combining the reciprocal unit vectors (Eqns. 29-31) into a 3x3 matrix which can then be used to calculate the g-vector (Eqn. 32) in the cubic form for any plane. Derivation of the inverse of this matrix multiplied by a given native normal will provide the plane associated with that pole. It should be noted that while the native description of the plane of atoms (hkl) (e.g., (111) tetragonal c/a =2) is utilized for this calculation, the resultant g-vector is in cubic form. As previously noted, the cubic form is necessary to plot planes of atoms in a tip/tilt map, as well calculate the angle between planes (Eqn. 33) and determine the d-spacing of plane (Eqn. 34, the distance between any plane is then is the length of the normal vector in cubic form). It should be stressed that when plotting or representing the planes, the nomenclature for the <u>native planes</u> are still used. The description of the native normals can also be calculated for demonstration purposes (Eqns. 35 and 36) by multiplying the cubic description of the normal by the inverse of the conversion matrix ($M^{-1}$). The initial example provided in this article (***Figure 1***) utilized a tetragonal cell with a c/a ratio of 2 to demonstrate the calculation of the angle between two vectors. A similar schematic illustrated in ***Figure 10*** for a similar tetragonal crystal where the plane normals for (111) and (212) planes are shown both in their native and cubic forms. Whereas in (***Figure 1***) the [111] native normal was listed, it does not describe the normal for the (111) plane. ***Figure 10*** illustrates that the native normal for the (111) is actually the [441] (which converts to [221] in the cubic form).

*Equation 32*

$$g_{(hkl)} = \begin{bmatrix} \dfrac{hbc \sin \alpha \sin \delta - lab \cos \beta \sin \delta}{V} \\ \dfrac{-hbc \sin \alpha \cos \delta + kac \sin \beta + l(\sin \alpha \cos \beta \cos \delta - \sin \beta \cos \alpha)}{V} \\ \dfrac{lab \sin \alpha \sin \beta \sin \delta}{V} \end{bmatrix}$$

*Equation 33*

$$\cos \theta = \frac{g_{(hkl),1} * g_{(hkl),2}}{|g_{(hkl),1}| * |g_{(hkl),2}|}$$

*Equation 34*

$$d_{hkl} = \frac{1}{\sqrt{g_{(hkl)}^2}}$$

*Equation 35*

$$g_{native(hkl)} = M^{-1} * g_{(hkl)}$$

*Equation 36*

$$M^{-1} = \begin{bmatrix} \dfrac{1}{a \sin \beta} & \dfrac{-\cos \delta}{a \sin \beta \sin \delta} & 0 \\ 0 & \dfrac{1}{b \sin \alpha \sin \delta} & 0 \\ \dfrac{-\cos}{c \sin} & \dfrac{\cos \beta \sin \alpha \cos \delta - \sin \beta \cos \alpha}{c \sin \alpha \sin \beta \sin \delta} & \dfrac{1}{c} \end{bmatrix}$$

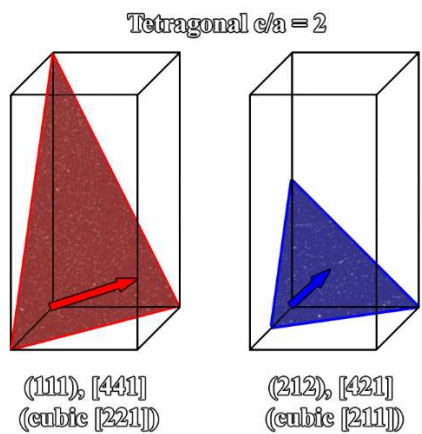

*Figure 10:* *Schematic of two planes in a tetragonal crystal showing the relationship of the plane normal in the native and cubic form.*

## 2.7 Structure Factor – Tip/Tilt Filter

The organization of this research was designed to begin in real space, define how to take any solid geometrical object such as a cube or hexagonal prism, and then rotate that object. Next, these Cartesian coordinates were converted to double tilt stage coordinates and it was demonstrated how confining the degrees of freedom required a second set of coordinates. The conversion of real space into reciprocal space was then examined in order to best introduce the idea of how the description of real crystals could be mapped on top of the tip/tilt calculations. Just as the real space calculations considered any pole/vector normal possible, so did the description of crystals in reciprocal space provide any possible set of poles/planes/vectors based on the simple geometry of the crystal. The last portion of this discussion goes further into examining real crystals and how they interact with an electron beam, more specifically the structure factor and how it acts as a simple filtering function for the aforementioned calculations. That is to say, all possible combinations of vectors, crystal systems, planes, and normals were provided, and the structure factor provides a way to determine which of all of those combinations are exhibited in any crystal.

As noted prior, countless other electron beam interactions relate to diffraction and scattering contrast that could be discussed. In the context of this paper the structure factor is most relevant, and even then only a cursory explanation will be provided to illustrate the power of understanding the connections between the real space and reciprocal space in regards to materials analysis. More detailed descriptions of the physics of these interactions can be found in any number of electron microscopy texts (Carter et al., 1996, De Graef and McHenry, 2012, Thomas, 1962).

The structure factor as it pertains to this discussion is a means to determine which planes of atoms within any given crystal will diffract. In terms of diffraction and the TEM, tallying the combinations of allowable diffracted planes can then be used to create a list of allowable expressed poles. The use of these lists can then be utilized to create tip/tilt maps by which to travel throughout any crystal given provided recognition of specific planes and poles is possible (*Figure 11*).

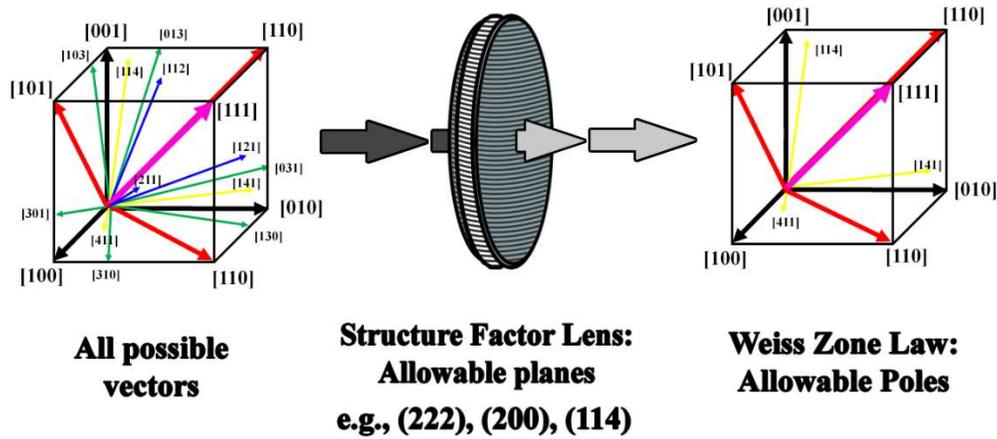

*Figure 11:* Schematic showing the use of the structure factor as a filter to determine which of the infinite number of vectors in real space are expressed in reciprocal space (ZA).

Whereas the conversion of crystal systems in real space into reciprocal space considered all possible combinations within each structure, the physics of real crystals are defined by the arrangement and packing of any number of atoms within the unit cells of the 7 different crystal systems. The complexity and variation of this packing is evidenced by the 230 possible space groups within these systems, not to mention the increased complexity of quasi-crystals and quasi-crystal approximants. In order to distinguish which planes within each crystal will diffract, the position and scattering power of each atom is considered. The equation for the structure factor (Eqn. 37) is provided below for any given plane of atoms described by (hkl), and depending on whether the solution is 1 or 0 dictates whether or not the plane will diffract, respectively. This can be further expanded to account for more complex crystals with any number of atoms each at any position within the unit cell. Note that since the atomic positions of each atom are used, there need not be any conversion from non-cubic systems.

*Equation 37*

$$F_{hkl} = \sum_{j=1}^{N} \sum f_j \, e^{[-2\pi i(hx_j + ky_j + lz_j)]}$$

where $f_j$ is the scattering factor of the j-th atom, $x_j, y_j, z_j$ are the atomic coordinates, and hkl defines a reciprocal lattice point corresponding to real space planes defined by the Miller indices. After a list of allowable planes is calculated, the trace of each of these planes could be plotted in either a stereographic projection or tip/tilt map, of which they would automatically intersect at the possible poles expressed for each crystal. Moreover, a combination of allowable poles could be derived by determining only those planes that satisfy the Weiss Zone law (Eqn. 38). Depending on the definition of applicable poles (i.e., which poles exhibited appreciable Bragg diffraction spots), the positions of those poles could be calculated using Eqn. 38 and plotted.

*Equation 38*

$$hu * kv * lw = 0 \; given \; [uvw] \; and \; (hkl)$$

These tip/tilt maps of well-defined crystals are only a small part of what can be accomplished utilizing the information contained herein.

# 3. Practical Derivations of Nanocartography

The methodologies and protocols derived subsequently in this paper build off the derivations in section 2, and serve to better interface crystallographic and stage motion in a practical manner. Transmission and scanning transmission electron microscopy (S/TEM) data presented in this paper were collected on an aberration, $C_s$ corrected JEOL ARM200CF and include diffraction, convergent beam electron diffraction (CBED), bright field (BF), and STEM high angle annular darkfield (HAADF). The data are generic examples used for demonstration purposes to elucidate various protocols and will not be described further than identification of the imaging mode and/or base crystal type.

## 3.1 K-space Calibration, Small Angle Tilting

In the pursuit of analyzing beam sensitive samples or smaller volumes within a polycrystalline field it is often necessary to have a guide by which to be able to blindly drive the stage while either blanking the beam, lowering the magnification, or defocusing the beam (in STEM) such that the area of interest is not accumulating dose or is lost amongst a field of other adjacent crystals. The approaches provided in later sections presupposes that one has some knowledge of the crystal, but often if the sample is only tilted some observable distance from a desired ZA or plane of atoms, it is not necessary to understand the overall orientation, just that a specific ZA is within a small tiling angle. Therefore, once the location of the α and β axes have been identified, it is conceivable to create a small angle tilt template such that if a sample is, for example, less than 5° off a ZA, one can rapidly calculate the tilt coordinates without further observation of the sample past the initial collection of the current pattern.

This is important for beam sensitive samples and small samples within a polycrystalline matrix where either the beam can destroy the sample or the non-eucentricity of the stage translates the sample away from the field of view during tilting. In order to calibrate a tilt map, one of two methodologies can be utilized depending on level of programing expertise (for example in Gatan Microscopy Suite).

Calibration of the digital capture of k-space is first necessary such that a subsequent point and click on the computer screen to tilt any desired pole/plane to the center position could be accomplished (*Figure 12*). At any point within the double tilt stage the immediate motion of the stage transverses in a linear fashion out to ~7-10°, at which point due to the motion of the β tilt in relation to the α any trace begins to rotate and nonlinear effects become noticeable. Since most local digital fields of view illuminate ~5-6° of tilt (~90-100 mrad), the calibration will be considered linear. Only the α need be considered, as the entire relationship of tip/tilt map can be deduced from this measurement.

The calibration of the α tilt is required for the specific TEM approach and should be performed with a crystalline sample with a ZA fiduciary marker close to α,β: 0,0. The location of the probe (red dot/circle in *Figure 12*) or transmitted beam should first be identified digitally (i.e., the pixel location on the screen, x/y, should be correlated with the center of the beam) and noted as the origin ($x_0,y_0$). Next, the crystal can be tilted in the pure negative or positive α direction ~4-5° (or to the edge of the field of view) such that the digital position ($x_{ref},y_{ref}$) can be calibrated to the tilt (green dot/circle in *Figure 12*). This position will be denoted as the calibration, and a calibration vector can be produced by subtracting the reference position from the origin. This vector, shown in Eqn. 39, is normalized to produce a unit vector in the direction of α tilt. The direction of β tilt is perpendicular to this, and the unit vector $\hat{y}$ is shown in Eqn. 40.

*Equation 39*

$$\vec{x} = \begin{bmatrix} x_{ref} - x_0 \\ y_{ref} - y_0 \end{bmatrix} \Rightarrow \hat{x} = \frac{\vec{x}}{|\vec{x}|} = \frac{1}{\sqrt{(x_{ref} - x_0)^2 + (y_{re} - y_0)^2}} \begin{bmatrix} x_{ref} - x_0 \\ y_{ref} - y_0 \end{bmatrix}$$

*Equation 40*

$$\hat{y} = \frac{1}{\sqrt{(x_{ref} - x_0)^2 + (y_{ref} - y_0)^2}} \begin{bmatrix} -(y_{ref} - y_0) \\ x_{ref} - x_0 \end{bmatrix}$$

The tip/tilt coordinates for any position (x,y) in the field of view (blue dot/circle in **Figure 12**) can be calculated to align the feature of interest (e.g., zone axis) with the probe by decomposing this location into components along $\hat{x}$ and $\hat{y}$. The decomposition must be solved for the amount along $\hat{x}$ ($c_1$) and the amount along $\hat{y}$ ($c_2$) in Eqn. 41.

*Equation 41*

$$c_1\hat{x} + c_2\hat{y} = \begin{bmatrix} x - x_0 \\ y - y_0 \end{bmatrix} \Rightarrow \frac{c_1}{\sqrt{(x_{ref} - x_0)^2 + (y_{ref} - y_0)^2}} \begin{bmatrix} x_{ref} - x_0 \\ y_{ref} - y_0 \end{bmatrix} + \frac{c_2}{\sqrt{(x_{ref} - x_0)^2 + (y_{ref} - y_0)^2}} \begin{bmatrix} -(y_{ref} - y_0) \\ x_{ref} - x_0 \end{bmatrix} = \begin{bmatrix} x - x_0 \\ y - y_0 \end{bmatrix}$$

Solving this system of equations for the weights yields:

*Equation 42*

$$c_1 = \frac{(x - x_0)(x_{ref} - x_0) + (y - y_0)(y_{ref} - y_0)}{\sqrt{(x_{ref} - x_0)^2 + (y_{ref} - y_0)^2}}$$

*Equation 43*

$$c_2 = \frac{(y - y_0)(x_{ref} - x_0) - (x - x_0)(y_{ref} - y_0)}{\sqrt{(x_{ref} - x_0)^2 + (y_{ref} - y_0)^2}}$$

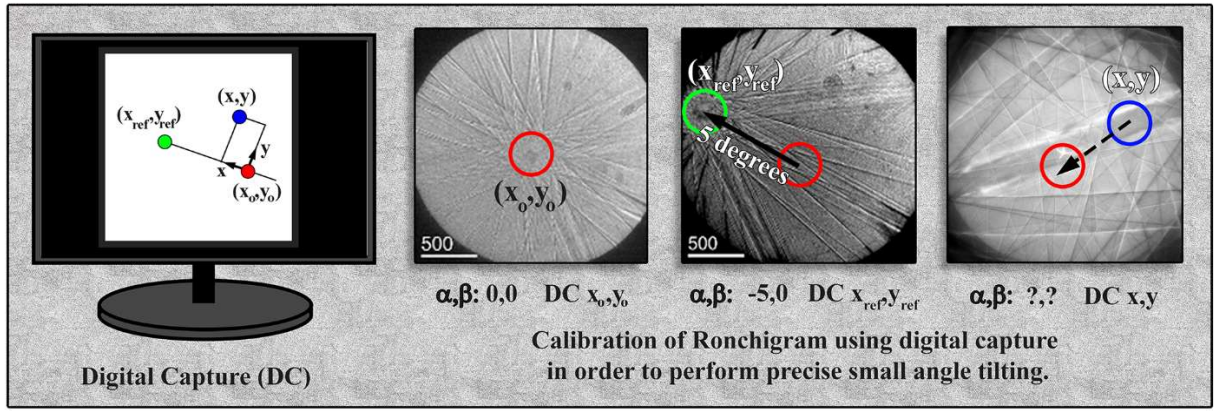

*Figure 12: Calibration of digital capture for precise, small angle sample tilting.*

Once the weights are known, they are converted to tip/tilt coordinates using a scaling factor that was derived from the initial α calibration tilt divided by the length of the vector. The equation for scaling factor is:

*Equation 44*

$$Scaling\ Factor = \frac{\alpha}{\sqrt{(x_{ref} - x_0)^2 + (y_{ref} - y_0)^2}}$$

Utilization of these formulae allows for precise small angle tilting using CBED or a Ronchigram where there could be a high density of additional Kikuchi lines from adjacent, smaller crystals. More importantly, if the sample is beam sensitive only a single image capture need be collected to predict how to tilt within the field of view in k-space.

## 3.2 Center-beam Darkfield Tilting

Precise off axis tilting of the electron probe using the condenser lens deflector coil system has long been utilized to examine the location of specific diffracted beams (center beam darkfield (Carter et al., 1996)) and also to perform techniques such as hollow cone diffraction (Kondo et al., 1984) and precession electron diffraction (PED) ((Vincent and Midgley, 1994, Midgley and Eggeman, 2015)). Tilting the beam can be considered a conjugate of tilting the sample, and hence the use of digital capture can be utilized to dictate the tilt of the beam.

If the same mathematical calculations are completed with the tilt conditions replaced with condenser lens deflector outputs, the precise beam deflections can be utilized (Eqns. 39-44). The difference is that the beam deflections are most often read as hexadecimal, and therefore the calculations need to utilize the hexadecimal outputs instead of stage tilt positions.

This technique can be utilized in a wide variety of methods to examine darkfield tilting and is important for analysis of beam sensitive samples. Once a single diffraction pattern is collected, the beam can be blanked by appropriate methods, and the beam tilt conditions can be performed by pointing and clicking on the viewing screen to obtain the correct deflections (e.g., *Figure 13*a where the (-1-31) spot is deflected to the central beam). Similarly, when there is no visible diffraction spot but there is a crystalline phase suspected (or possibly a weak superlattice reflection), digital alignment can be performed blindly (white circle in *Figure 13*a). Additionally, the beam could be deflected in a circular manner by which to explore all possible g-vectors in k-space during a long, darkfield exposure (*Figure 13*b). This technique could then be utilized to program in all desired g-vectors of a given crystal system at once and compare the resulting image to a second set of deflections corresponding to a different crystal (e.g., FCC versus BCC). It is beyond the scope of this paper to go into more detail, but precise digital control of the beam deflectors in TEM mode could be highly beneficial for a wide range of materials analyses.

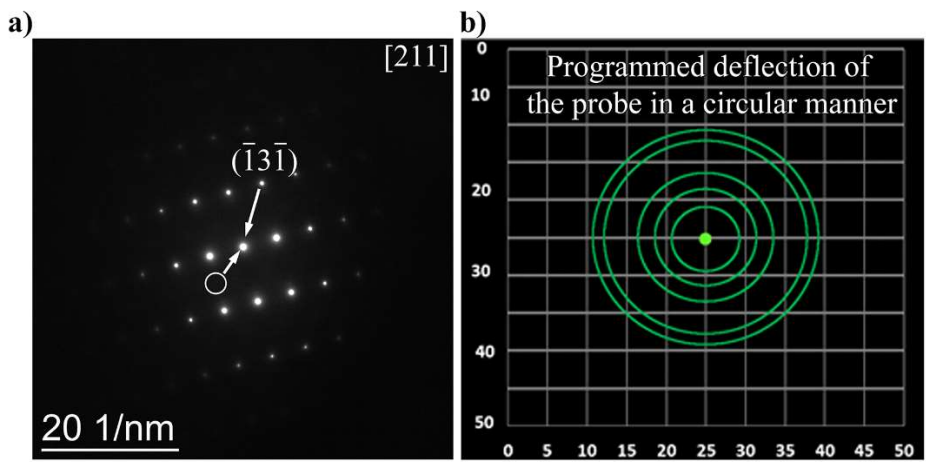

*Figure 13:* Example of digital capture darkfield tilting using a diffraction pattern of an FCC crystal in the [211] orientation (a), and a schematic illustrating a theoretical example of deflecting the probe in a circular manner (b).

## 3.3 Image Montaging

Although not directly related to the stage tilt movement and additional protocol that has been proven extremely productive in scanning electron microscopy (SEM) is the notion of montaging images at a specific magnification/resolution to create a larger image. The increased resolution of the higher magnification maps provides for richer, more meaningful data sets as compared to a single, low magnification overview image. Although in principle an increased resolution could be used at lower magnifications, there is a physical limit to the camera/detector size that makes it prohibitive. Most often on the TEM montaging is performed manually by the microscopist because a small number of maps are required to cover an area of interest, and as well because of the time prohibitive nature of limited scope availability. The ability to automatically montage data has not been a necessary feature on most microscopes, but with the coming age of automation, there will be a need to perform overnight montaging of samples for data triaging in subsequent sessions. As has been demonstrated throughout this work, the ability to have a map or specific list of commands provides a sense of direction for the microscopist. While a montage could be done manually, gauging where the previous region of interest overlaps with the current image, a table of stage tilts would be more beneficial to the user. Given a desired distance, X, image dimension, Y, and necessary image overlap (p, as a fraction) the number of maps necessary to create a montaged image is given in Eqn. 45. The table of stage positions can then be calculated through Eqn. 46 where the addition or subtraction of the (X-Xp) or (Y-Yp) terms are a function of the numbering logic of the stage and are added to the previous image coordinates (movie showing montaging in *Figure 14*). This of course takes into account the fact that the sample is flat, and a Z term has not been introduced. This could be addressed for highly titled samples by observing the height at either end of the desired montage range and scaling each position accordingly. This also does not consider the movement of backlash within the stage, and is only meant as a starting point for creating montaged images.

*Equation 45*

$$N_{maps} = \frac{(X - pY)}{(1 - p)Y}$$

*Equation 46*

$$Image_{N+1} = X_0 \pm (X - Xp), Y_0 \pm (Y - Yp)$$

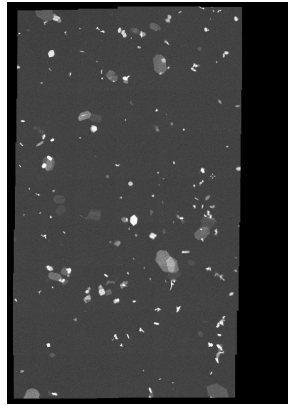

*Figure 14: Movie illustrating montaging of MoO$_3$ particles on a carbon film.*

## 3.4 Pure Tilt Between Tip/Tilt Conditions

It is often necessary to calculate the angle between specific tip/tilt conditions, as will be demonstrated later when solving unknown crystal structures. The derivation for computing the angle between stage tilt coordinates (α,β) is based off of the tip/tilt convention, where the stage tilt is first rotated about the α tilt axis (Eqn. 47) and then subsequently about the β tilt axis (Eqn. 48) to the to the beam normal [001] (Eqn. 49). That is, given a normalized vector at any tip/tilt position (α,β) can be rotated to the [001] position (Eqn. 50).

*Equation 47*

$$R_{-\alpha,x} = \begin{bmatrix} 1 & 0 & 0 \\ 0 & \cos\alpha & \sin\alpha \\ 0 & -\sin\alpha & \cos\alpha \end{bmatrix}$$

*Equation 48*

$$R_{-\beta,y} = \begin{bmatrix} \cos\beta & 0 & -\sin\beta \\ 0 & 1 & 0 \\ \sin\beta & 0 & \cos\beta \end{bmatrix}$$

*Equation 49*

$$R_{\theta,total} = R_{-\beta,y}R_{-\alpha,x} = \begin{bmatrix} \cos\beta & 0 & -\sin\beta \\ \sin\alpha\sin\beta & \cos\alpha & \sin\alpha\cos\beta \\ \cos\alpha\sin\beta & -\sin\beta & \cos\alpha\cos\beta \end{bmatrix}$$

*Equation 50*

$$R_{\theta,total}\begin{bmatrix}0\\0\\1\end{bmatrix} = \begin{bmatrix}-\sin\beta\cos\alpha\\ \sin\alpha\\ \cos\alpha\cos\beta\end{bmatrix}$$

This rotation can be performed for any two sets of tilt conditions, $(\alpha_1, \beta_1)$ and $(\alpha_2, \beta_2)$, and therefore the dot product between two vectors (Eqn. 51) will provide the angle between the two tip/tilt conditions (Eqn. 52). It should be again be noted that the order of rotation, α then β, is important, and reversal of the order will provide erroneous results.

*Equation 51*

$$\cos\theta = \sin\beta_1\sin\beta_2\cos\alpha_1\cos\alpha_2 + \sin\alpha_1\sin\alpha_2 + \cos\alpha_1\cos\alpha_2\cos\beta_1\cos\beta_2$$

*Equation 52*

$$\theta = \cos^{-1}(\sin\alpha_1\sin\alpha_2 + \cos\alpha_1\cos\alpha_2\cos(\beta_1-\beta_2))$$

## 3.5 Grain Boundary Misorientation

Possessing the crystallographic solution for two adjacent crystals (*Figures 15*a and b) of the same crystal system provides additional information, namely the grain boundary misorientation angle and axis of rotation (Chesser et al., 2020). This ability to calculate and report this additional sample descriptor can be a powerful tool where the only additional analysis that must be performed is the calculation (i.e., only the two crystal orientations are necessary). There are a number of methods by which to derive the local misorientation (Jeong et al., 2010, Liu, 1994, Liu, 1995), but in all cases the crystal orientation of two adjacent crystals are utilized to determine the directions of the unit vectors (*Figure 15*c). The comparison of the unit vectors of each crystal are used to calculate the misorientation angle about a shared misorientation axis ([uvw]) through a misorientation matrix (*Figure 15*c). The first step in developing this matrix is to solve for the location of each of the unit vectors in each crystal.

In *Figure 15*, Crystals A (a) and B (b) are observed in a given orientation, and the given tip/tilt positions of any three vectors within each can be determined through the crystallographic solution (e.g., Crystal A [110], [111] and [201], and Crystal B [110], [112], and [111]). Note that the choice of these three is arbitrary, but that they must be linearly independent and contain three distinct directions (i.e., [111] and [222] would not be distinct directions). These three vectors can be used to determine the location of their respective unit vectors (*Figure 15*d).

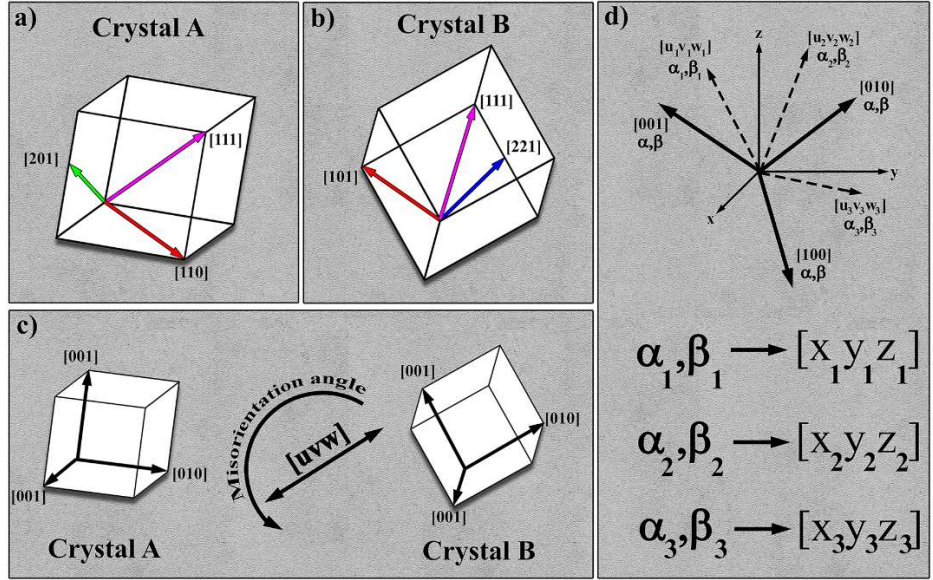

***Figure 15:*** *Schematic illustrating how the local misorientation between two crystals is formulated. a and b) Crystals A and B in a given orientation. c) Misorientation angle and axis between the two crystals. d) Conversion to primary axes coordinate system.*

The development of a misorientation matrix will describe the pure angle required to rotate the unit vectors of Crystal A to align with the unit vectors of Crystal B (***Figure 15***c) and the shared axis between the two crystals about which this rotation can be accomplished. In Cartesian space the orientation of the vectors can be utilized in developing this matrix, but in the microscope the description of the vectors are defined by the coordinates of the double tilt stage.

The three observed or known vectors for one crystal (e.g., $[u_{A1}, v_{A1}, w_{A1}]$) are observed at a given tip/tilt position (e.g., $\alpha_{A1}, \beta_{A1}$) (***Figure 15***d). These tip/tilt positions need to be converted into Cartesian space similar to the operation performed in Eqns. 47-52 during the development of the angle between tilt positions (e.g., $[x_{A1}, y_{A1}, z_{A1}]$). In order to derive the Cartesian vector form of the unit vectors ([100], [010], [001]) for the crystal, the tilt position vector in Cartesian form ($[x_{A1}, y_{A1}, z_{A1}]$) is first required to have the same magnitude as the crystallographic vectors ($[u_{A1}, v_{A1}, w_{A1}]$). This can be accomplished by multiplying the tilt vector by the length of the crystallographic vector (Eqn. 53).

*Equation 53*

$$\vec{x}_{A1} = \begin{bmatrix} x_{A1} \\ y_{A1} \\ z_{A1} \end{bmatrix} = \sqrt{u_{A1}^2 + v_{A1}^2 + w_{A1}^2} \begin{bmatrix} -\sin \beta_{A1} \cos \alpha_{A1} \\ \sin \alpha_{A1} \\ \cos \alpha_{A1} \cos \beta_{A1} \end{bmatrix}$$

Similar expressions for the other two poles can be calculated, where the subscript 1 has been replaced with either 2 or 3. It is necessary to find three Cartesian vectors which add up to the three known vectors given the linear combination weights determined by the crystallographic poles.

*Equation 54*

$$\vec{p} = \begin{bmatrix} p_{Ax} \\ p_{Ay} \\ p_{Az} \end{bmatrix} \sim [100]$$

*Equation 55*

$$\vec{q} = \begin{bmatrix} q_{Ax} \\ q_{Ay} \\ q_{Az} \end{bmatrix} \sim [010]$$

*Equation 56*

$$\vec{t} = \begin{bmatrix} t_{Ax} \\ t_{Ay} \\ t_{Az} \end{bmatrix} \sim [001]$$

These three unknown Cartesian vectors are the unit vectors that describe the orientation of the crystal. In the microscope, regardless of the sample orientation (e.g., [111] at α,β:5,10) the location of the unit vectors (i.e., [001],[010] and [100]) are calculated. These vectors are subsequently utilized to describe how to translate from one crystal orientation to another (e.g., [100] of Crystal A to [100] of Crystal B). The linear combinations that connect these sets of vectors are:

*Equation 57*

$$u_{A1}\vec{p} + v_{A1}\vec{q} + w_{A1}\vec{t} = \vec{x}_{A1}$$

*Equation 58*

$$u_{A2}\vec{p} + v_{A2}\vec{q} + w_{A2}\vec{t} = \vec{x}_{A2}$$

*Equation 59*

$$u_{A3}\vec{p} + v_{A3}\vec{q} + w_{A3}\vec{t} = \vec{x}_{A3}$$

These equations can be solved for the components of the vectors $\vec{p}$, $\vec{q}$, and $\vec{t}$ since there nine equations and nine unknowns. The details of how these equations are rearranged are in the Supplemental, but after gathering like terms, it is equivalent to the augmented matrix:

*Equation 60*

$$\begin{bmatrix} u_{A1} & v_{A1} & w_{A1} & | & x_{A1} & y_{A1} & z_{A1} \\ u_{A2} & v_{A2} & w_{A2} & | & x_{A2} & y_{A2} & z_{A2} \\ u_{A3} & v_{A3} & w_{A3} & | & x_{A3} & y_{A3} & z_{A3} \end{bmatrix}$$

After row reducing this augmented matrix to reduced row echelon form:

*Equation 61*

$$\begin{bmatrix} 1 & 0 & 0 & | & p_{Ax} & p_{Ay} & p_{Az} \\ 0 & 1 & 0 & | & q_{Ax} & q_{Ay} & q_{Az} \\ 0 & 0 & 1 & | & t_{Ax} & t_{Ay} & t_{Az} \end{bmatrix}$$

The right-hand side of the row reduced matrix produces the elements of the vectors $\vec{p}, \vec{q}$, and $\vec{t}$ that give the axes of the crystal in Cartesian vector form. The reason these three vectors are necessary is because they give the rotation matrix that describes the orientation of the crystal from the standard orientation where the crystallographic axes ([100], [010], [001]) align with the coordinate axes ($\hat{x}, \hat{y}, \hat{z}$). This standard orientation alignment is the common feature that connects any crystal orientation to another. The rotation matrix is found from the transpose of the right-hand side of the row reduced augmented matrix. This can also be described as the unit vector matrix in that it can be used to describe the location of the unit vectors for a specific crystal.

*Equation 62*

$$R_A = \begin{bmatrix} p_{Ax} & q_{Ax} & t_{Ax} \\ p_{Ay} & q_{Ay} & t_{Ay} \\ p_{Az} & q_{Az} & t_{Az} \end{bmatrix}$$

The previous rotation matrix was derived for Crystal A (Eqn. 62), but the exact procedure applies to Crystal B without any modifications beyond the subscript, and is denoted by:

*Equation 63*

$$R_B = \begin{bmatrix} p_{Bx} & q_{Bx} & t_{Bx} \\ p_{By} & q_{By} & t_{By} \\ p_{Bz} & q_{Bz} & t_{Bz} \end{bmatrix}$$

Both of these rotation matrices convert from the standard orientation to the current rotation of their respective crystal. To get from one orientation to the other requires going from the current orientation back to the standard orientation and then to other crystal orientation. Mathematically, this is done by using the inverse of the rotation matrix to get back to the standard orientation.

*Equation 64*

$$M_{A \to B} = R_B R_A^{-1}$$

*Equation 65*

$$M_{B \to A} = R_A R_B^{-1}$$

These two overall rotation matrices are the misorientation matrices that describe the relative orientation of one crystal to another. Most frequently, the desired information is how far apart two crystals are misaligned and about which axis they must be rotated so that they would become aligned. This is called the axis-angle representation of the rotation matrix. If the general form a misorientation matrix is:

*Equation 66*

$$M = \begin{bmatrix} M_{11} & M_{12} & M_{13} \\ M_{21} & M_{22} & M_{23} \\ M_{31} & M_{32} & M_{33} \end{bmatrix}$$

Then the axis-angle representation is, where $\theta_M$ is the misorientation angle and $\vec{r}_M$ is the misorientation axis:

*Equation 67*

$$\theta_M = \cos^{-1}\left(\frac{M_{11} + M_{22} + M_{33} - 1}{2}\right)$$

*Equation 68*

$$\vec{r}_M = \begin{bmatrix} \dfrac{M_{32} - M_{23}}{2 \sin \theta_M} \\ \dfrac{M_{13} - M_{31}}{2 \sin \theta_M} \\ \dfrac{M_{21} - M_{12}}{2 \sin \theta_M} \end{bmatrix}$$

It should be noted that in that the relative orientation solution the adjacent crystals is important in the calculation of the misorientation angles depending on the crystal system chosen. This is especially true for higher symmetry systems (e.g., cubic) where redundant vector normals allow accurate prediction of tilting from one pole to another, but the relative orientations will become important when performing mathematical calculations such as the misorientation angle. It is beyond the scope of this paper to include a full discussion of all of the possible symmetrical operators, but the reader must be aware of this when utilizing stage tilts in the TEM to perform these calculations.

Lastly, in order to complete the full description of the grain boundary, the interface must be further characterized to describe the orientation of the plane of atoms with respect to the boundary and the adjacent grain. To access this data, the physical orientation of the grain boundary as a physical plane with respect to the stage is required. Once the tip/tilt conditions are determined, then these coordinates can be used to calculate the description of the plane normals.

## 3.6 Interface/Boundary Tilting

The ability to correctly and accurately predict the motion of crystals in an electron microscope using a double tilt stage is crucial to collecting the optimal data over a wide range of fields of study. In section 2, a full explanation of how to derive these calculations was conducted; first the crystal was treated as a physical object and then subsequently a physics based filter was applied through the structure factor. The advantage of this approach is that the motion of non-crystalline samples can be treated in the same way as the derivation of directions for planes or tilts between poles as interfaces are physical planes. Therefore, the ability to identify the orientation of the long axis with respect to the α tilt axis can be predicted similar to identifying the orientation of a plane of atoms.

As will be demonstrated in subsequent sections, this can be further utilized in a variety of techniques from creating oblique tilt series to rapid analysis of grain boundaries edge on. More importantly, prediction of the interface movement allows for more accurate data collection as it can also be related to adjacent crystalline material. For instance, if an edge on boundary condition is determined, then the boundary can be subsequently tilted along its long axis to any tilt condition that may be favorable to the adjacent crystal, such as a pole or specific plane of atoms. Additionally, when the interface is edge on, the adjacent crystallographic normal(s) can be calculated if the crystallographic solution of the crystal(s) has been measured.

The approach for predicting interface motion is similar to that of calculating planes of atoms, except that for planes of atoms there is an explicit normal previously defined by the plane of atoms in question. For an interface, the starting tip/tilt conditions are the only information available (e.g., α,β:-5,10 in **Figure 16**) as well as the measure of the long axis to the α tilt axis (e.g., θ in **Figure 16** of which the grain boundary is measured at ~75°). Note that the sign of rotation of the boundary to the α tilt axis is reversed between **Figure 16**a and b because of the sign convention of how the α tilt axis is calibrated. The relationship of the current tilt conditions to the fiduciary angle to the α axis is important because, as was demonstrated in the section 2 for planes of atoms, in the double tilt stage linear features will rotate when tilted to higher angles. That is to say, θ in *Figure 16*a will vary slightly based on the given α,β tilt conditions.

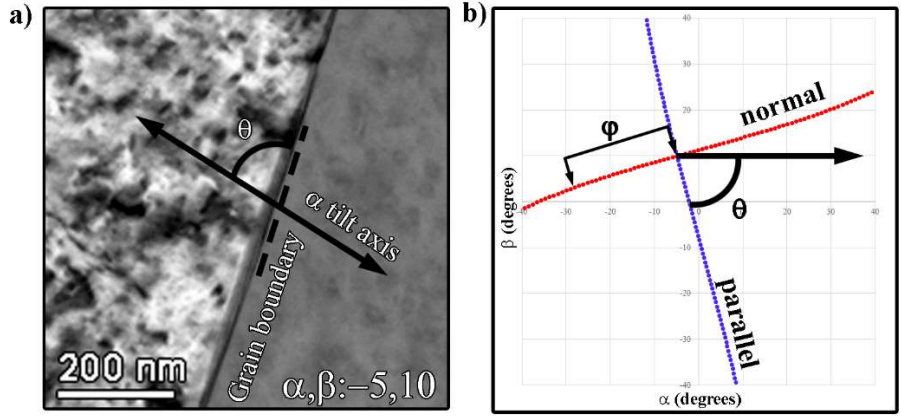

***Figure 16:*** *Plotting tip/tilt coordinates for interface analysis. A TEM (BF) image of a grain boundary is shown in with the angle to the α tilt axis (θ) highlighted (a). The trace of the boundary on a tip/tilt diagram (b) illustrates the angular movement (φ) normal to the boundary conditions shown in (a).*

As there is no crystallographic information utilized, the current tilt position (e.g., α,β: -5,10 in *Figure 16*a and b) is required to be converted into a Cartesian vector similar to what was performed in the calculation of the misorientation matrix. In order to calculate the vectors parallel and perpendicular to the direction of the interface Eqns. 69 and 70 need be derived, respectively. These two vectors lie in the xy plane and are determined solely by the angle θ.

*Equation 69*

$$\hat{a}_{parallel} = \begin{bmatrix} \cos\theta \\ \sin\theta \\ 0 \end{bmatrix}$$

*Equation 70*

$$\hat{b}_{perpendicular} = \begin{bmatrix} -\sin\theta \\ \cos\theta \\ 0 \end{bmatrix}$$

Section 2 details a rotation about an arbitrary axis (see Supplemental), and this will be used to rotate about both the vectors $\hat{a}_{parallel}$ and $\hat{b}_{perpendicular}$ (Eqn. 18). The general formula for rotation of angle $\varphi$ about an axis of rotation $\hat{u}$ (with length equal to one) is:

*Equation 71*

$$R_{\hat{u},\varphi} = \begin{bmatrix} u_x^2 + (u_y^2 + u_z^2)\cos\varphi & u_xu_y(1-\cos\varphi) - u_z\sin\varphi & u_xu_z(1-\cos\varphi) + u_y\sin\varphi \\ u_xu_y(1-\cos\varphi) + u_z\sin\varphi & u_y^2 + (u_x^2 + u_z^2)\cos\varphi & u_yu_z(1-\cos\varphi) - u_x\sin\varphi \\ u_xu_z(1-\cos\varphi) - u_y\sin\varphi & u_yu_z(1-\cos\varphi) + u_x\sin\varphi & u_z^2 + (u_x^2 + u_y^2)\cos\varphi \end{bmatrix}$$

In the case where $\hat{a}_{parallel}$ is the axis of rotation, the tilt series is perpendicular to the interface. Conversely, in the case where $\hat{b}_{perpendicular}$ is the axis of rotation, the tilt series is parallel to the interface. In either case, the angle of rotation ($\varphi$) determines how many steps will be in the tilt series (*Figure 16*b). Typically for a full rotation $\varphi = 1°$, hence there will be 360 steps in the series before returning to the original orientation. The simplified rotation matrices are:

*Equation 72*

$$R_{\hat{a},\varphi} = \begin{bmatrix} \cos^2\theta + \sin^2\theta \cos\varphi & \cos\theta \sin\theta (1-\cos\varphi) & \sin\theta \sin\varphi \\ \cos\theta \sin\theta (1-\cos\varphi) & \sin^2\theta + \cos^2\theta \cos\varphi & -\cos\theta \sin\varphi \\ -\sin\theta \sin\varphi & \cos\theta \sin\varphi & (\cos^2\theta + \sin^2\theta)\cos\varphi \end{bmatrix}$$

*Equation 73*

$$R_{\hat{b},\varphi} = \begin{bmatrix} \sin^2\theta + \cos^2\theta \cos\varphi & -\sin\theta \cos\theta (1-\cos\varphi) & \cos\theta \sin\varphi \\ -\sin\theta \cos\theta (1-\cos\varphi) & \cos^2\theta + \sin^2\theta \cos\varphi & \sin\theta \sin\varphi \\ -\cos\theta \sin\varphi & -\sin\theta \sin\varphi & (\sin^2\theta + \cos^2\theta)\cos\varphi \end{bmatrix}$$

As with all other tip/tilt conversions, the Cartesian vectors calculated in Eqns. 72 and 73 need to be converted to tilts through Eqns. 74 and 75 (note these are the same as Eqns. 23 and 24). These derivations will subsequently be utilized to create oblique tilt series and perform precise interface orientation calculations.

*Equation 74*

$$\alpha_{final} = \tan^{-1}\left(-\frac{Y}{\sqrt{X^2 + Z^2}}\right)$$

*Equation 75*

$$\beta_{final} = \tan^{-1}\left(\frac{X}{Z}\right)$$

## 3.7 Calculating Interface Orientations/Rapid Grain Boundary Analysis

Before the advent of atom probe tomography (APT), TEM had long been the most advanced technique for understanding materials properties at the highest chemical resolution (Blavette et al., 1993, Carter et al., 1996). Even with the ability to more precisely analyze interface chemistry by APT, S/TEM still provides a manner by which to analyze chemistry in addition to relating it to crystallography and other microstructural features such as dislocations and defects. More importantly, whereas the region of interest in APT is highly localized and is dependent on precise sample preparation (i.e., it is possible that only a small portion of an interface is captured within one tip), S/TEM allows a more global perspective for any given sample. One sample may contain tens of grain boundaries with lengths on the order of micrometers, thereby allowing the user to probe and provide a more representative analysis of the microstructure and microchemistry.

The ability to harness this much information for any given microstructure/sample is predicated on the speed of analysis combined with the correct orientation. In terms of interfaces, it is absolutely necessary that a boundary be analyzed edge on in order to best assess chemical gradients. For example, the depletion in sensitized stainless steels due to irradiation can be measured (Simonen and Bruemmer, 1998) and subsequently used to model the behavior of a material as a result of various external stimuli (e.g., heating, irradiation, chemical diffusion). Therefore, rapid and accurate alignment of interfaces on edge (i.e., completely aligned with the electron beam direction) is tantamount to performing the best possible analysis.

The utilization of a double tilt stage makes this entirely possible based on a number of tilting techniques. First, each tilt axis can be rocked independently in a positive and negative manner while the width of the desired interface is minimized. This requires that the same interface region be kept in the field of view during each tilting step. If the boundary has any deviation along its long axis (e.g., grain boundary curvature), determination of exactly when the boundary width is minimized can be challenging. Even more difficult is the assessment of the width of the boundary during tilting of the non-eucentric tilt axis (β), as the sample location often drifts even in piezo controlled stages. More importantly, as the length scale of the boundaries decrease towards the sub micrometer level, tracking the exact position of the boundary can be tedious and time consuming even on the eucentric tilt axis (α). In order to minimize the difficulty of tilting on the non-eucentric axis, a second method that can be employed is to measure the rotation of the long axis of the interface to the α tilt axis, remove the sample from the microscope, and then physically rotate the sample to such that it can be tilted solely on the eucentric axis (i.e., the long axis of the boundary is physically aligned along the non-eucentric axis so it can be tilted against the interface). The problem with this method is the amount of time for each analysis and lack of fidelity in physically rotating the sample. There do exist double tilt rotate holders specifically for this type of alignment, but they are most often not optimized for chemical analysis and are typically not standard to most microscopes.

Using the tilt methodologies developed in previously, a grain boundary in <u>any orientation can be tilted on edge by collecting two tilt conditions</u> where an interface width is measured at each tilt condition (provided reasonable assumptions of sample thickness and tilt range of a given holder). When the boundary is tilted against its long axis in a purely orthogonal manner in a known quantity (φ in ***Figure 16***b) the geometry of an inclined sample can be measured to determine not only the necessary tilt conditions to be aligned on edge, but as well provide a reasonable measure of the sample thickness (***Figure 17***).

For any given interface, the projected width (***Figure 17***a, $w_1$) can be measured at the current tip/tilt conditions through the difference in contrast where the interface intersects the top and bottom of the foil. While initially unknown, the interface has an inclination angle to the foil normal (***Figure 17***, $\theta_1$). The projected width ($w_1$) can be related to the interface length (IL) by the cosine of the inclination angle ($\theta_1$, Eqn. 76), and the foil width (FW) can be calculated by the sine of the inclination angle (Eqn. 77). Additionally, while the directionality of the boundary (top left to bottom right, or top right to bottom left) is inherently unknown, the manner in which the calculations are performed make this orientation irrelevant.

*Equation 76*

$$\cos(\theta_1) = \frac{w_1}{IL}$$

The angle of the interface's long axis to the α axis can be measured (i.e., θ in ***Figure 16***a) and using Eqns. 30-35, the tilt conditions for a pure orthogonal tilt normal to the boundary can be calculated (***Figure 17***b and c, $\theta_2$). This angle is commensurate with φ in ***Figure 16***b, and these tilt conditions can be either positive or negative (***Figure 17***b and c) based on the stage reference. For example, at a starting condition of α,β:0,0 with the boundary at 45° to the α axis and a tilt of 10° ($\theta_2$ or φ) puts the tilt stage at (α,β:-7.1, -7.1) or (α,β:7.1, 7.1).

At this tip/tilt condition, a second apparent grain boundary width ($w_2$) can be measured (***Figure 17***b). Depending on magnitude of $w_2$ as compared to $w_1$, and knowing the foil width (FW), Eqns. 77-79 can be used to calculate the angle necessary to tilt the boundary edge on, $\theta_3$, of which can be derived in the following manner.

Once tilted, the projected width of the boundary ($w_2$) can be related to the angle necessary to tilt the interface on edge ($\theta_3$) using trigonometry.

*Equation 77*

$$\sin(\theta_3) = \frac{w_2}{IL}$$

The interface length (*IL*) is initially unknown, but is constant between tilts and can be found using the Pythagorean Theorem once the initial width ($w_1$) and the foil width (*FW*) are known.

*Equation 78*

$$IL = \sqrt{FW^2 + w_1^2}$$

Substituting this above yields the final equation for $\theta_3$:

*Equation 79*

$$\theta_3 = \arcsin\left(\frac{w_2}{\sqrt{FW^2 + w_1^2}}\right)$$

In a double tilt stage the trace of a plane (or boundary in this instance) physically rotates due to the S-curve, and therefore there are two manners in which to calculate the final tilt conditions where the interface will be oriented edge on. These calculations consider the measure of the long axis of the boundary to the α tilt axis (θ in *Figure 16*), and this angle will change depending on the current tip/tilt conditions (i.e., in *Figure 16* θ at α,β:0,0 is 75°, but at α,β:-10,30 it would be change by ~2° due to the S-curve). Therefore, when performing the final tip/tilt calculations the choice of the appropriate measure of θ is imperative. One can either re-measure the angle of the boundary to the α tilt axis at the second tip/tilt conditions, or utilize the starting tilt conditions (where $w_1$ was measured) and the measure of the initial interface angle to the α tilt axis. The pure tilt (whether $\theta_3$ or the difference between $\theta_2$ and $\theta_3$) will depend on which conditions is chosen. Regardless, either can be performed by using Eqns. 76-79 to tilt the sample in a positive or negative fashion by $\theta_3$.

This can be visually demonstrated in *Figure 18*a where a theoretical grain boundary measured at 45° to the α tilt axis at starting tilts α,β:0,0 is measured at 40 nm wide. A positive tilt of 25° moves the stage to α,β:17.4,18.3 (along the black dotted line in *Figure 18*a) where the boundary is measured to be 60 nm wide (an increase of 20 nm), meaning the boundary needs to be tilted -35.4° from the starting tilt conditions (α,β:0,0) to a final tilt condition of α,β:-24.2,-26.7. Under these conditions, the sample thickness would be ~56 nm thick. If the final, edge on, tilt calculations were performed from the second tilt condition (α,β:17.4,18.3, red dotted line in *Figure 18*a) without re-measuring the grain boundary angle to the α tilt axis (which has rotated by ~2.5°) the final tilt conditions would be (α,β:-26.1,25.0) which is ~2.4° from the correct position. While still within the precision of most double tilt stages, it would not be correct.

The initial input tilt (*Figure 17*, $\theta_2$) will be dictated by the initial interface width due to the possibility of the tilting the sample through the edge on condition, thereby invalidating the calculations. The wider the initial boundary width, the more the initial tilt can be applied. To fully account for the possibility of tilting through the edge on condition, a third tilt and a third width, $w_3$ would be necessary, but by assessing the approximate width of the sample (e.g., ~100 nm) and the initial interface projected width ($w_1$) a protocol can be developed by which to determine the initial normal tilt ($\theta_2$).

*Figure 18*b provides a calculated guide for applicable tilt angles provided a starting apparent boundary width ($w_1$) for a number of approximate sample thicknesses as to not invalidate the calculations put forward in Eqns. 76-79. In the theoretical example shown in *Figure 18*a where the sample was on the order of ~56 nm thick and the starting grain boundary width of 40 nm, from *Figure 18*b the starting tilt of 25° was appropriate. If the assumed starting thickness of the sample was ~100 nm, 25° would not have guaranteed that this tilt would not tilt past the edge on condition.

As illustrated in *Figure 18*b, if the starting apparent interface width is on the order of 10 nm, the boundary is nearly edge on already, and a small angle calculation can be utilized (Eqns. 76-79). Instead of developing a tilt series, these equations can be utilized for small angles (1-5°) to tilt a boundary or interface normal to its long direction to close the apparent width of the boundary. The utilization of this rapid interface calculation methodology allows for successive, rapid analysis of any number of grain boundaries regardless of orientation to one another. After one boundary has been tilted edge on, an adjacent boundary width can be measured and then tilted edge on.

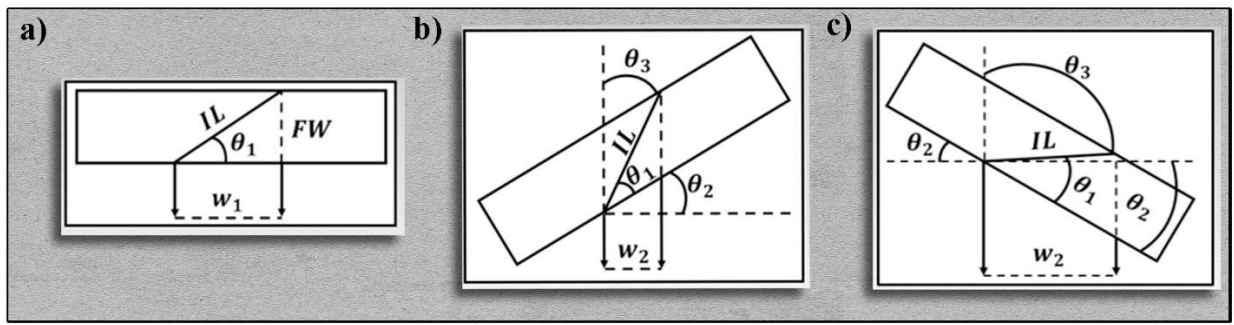

*Figure 17:* Schematics illustrating calculation of interface on edge conditions and sample thickness. a) Sample in original tilt, b) sample tilted negatively, and c) sample tilted positively.

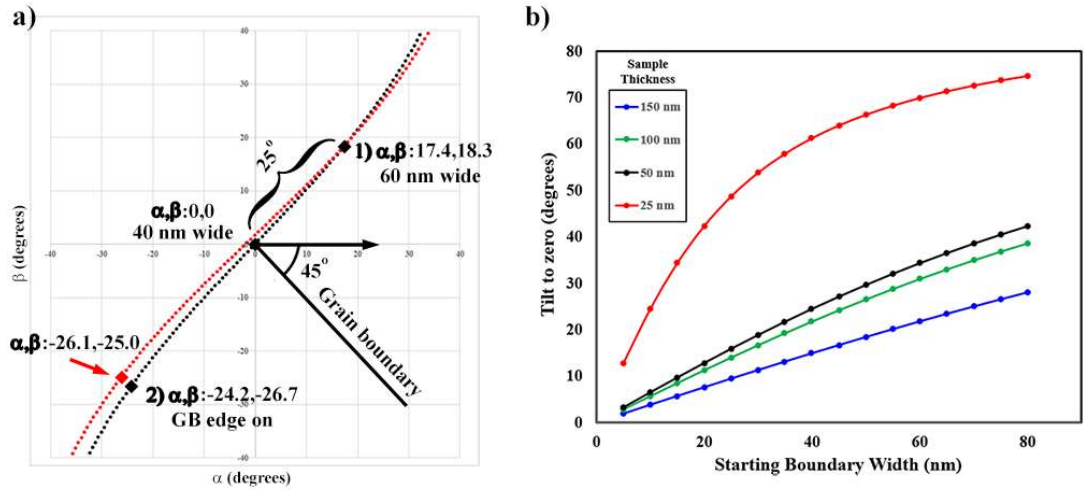

*Figure 18:* *Tip/tilt map of a theoretical grain boundary tilted edge on (a), and plot of normal tilt value versus initial projected interface width ($w_1$) for a number of sample thicknesses to gauge the applicable normal tilt angle ($\theta_2$) without crossing over the edge on condition (b).*

## 3.8 Interface/Crystallographic Normals

Predicting and controlling the motion of crystals and interfaces is important for grain boundary descriptions, determining growth directions of oxides from surfaces, and orientations of facets. In describing the relationship between two crystals, the first step is to determine the misorientation angle and axis (Eqns. 67 and 68). This data is commensurate with what is collected in electron backscatter diffraction (EBSD) in SEM, but TEM provides the advantage of further being able to describe the grain boundary orientation (*Figure 19*). The inclination of the grain boundary with respect to the stage, as shown in *Figure 19*a for grain 1 (G1), is a known variable and can be solved to orient the grain boundary edge on (Eqns. 76-79), as shown in *Figure 19*b. Given these tilt conditions, the angle of the grain boundary to the α axis (θ in *Figures 16*a and *19*b) and the crystallographic solution of the two adjacent crystal(s), the crystallographic normals and planes to the interface can be calculated (solid red arrow in *Figure 19*b).

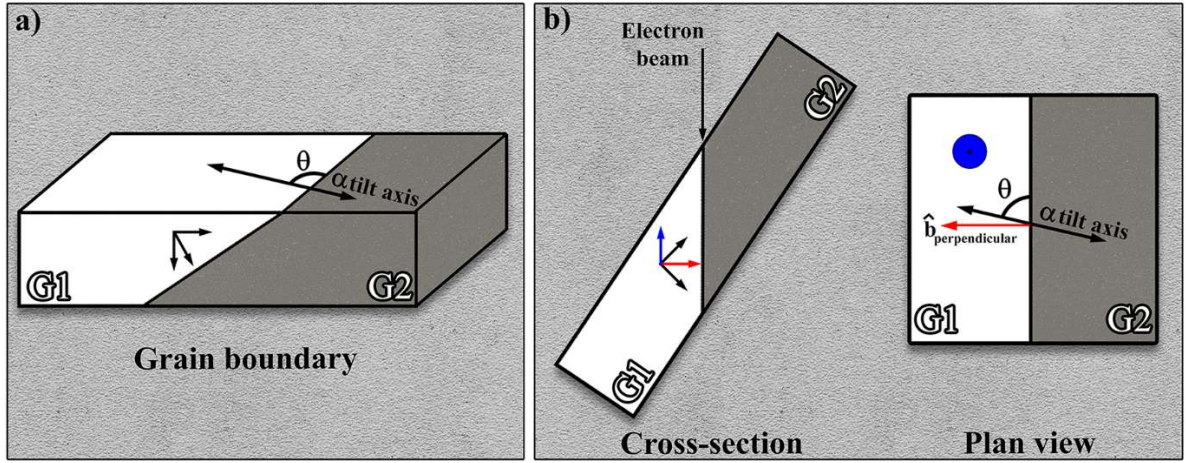

*Figure 19:* Schematic illustrating a grain boundary between two grains (G1 and G2) in a standard orientation (a) and with the boundary oriented edge on to the electron beam (b) shown in both cross-section and plan views.

As has been demonstrated through Eqns. 53-62, provided three known vectors with specific α,β coordinates the unit vectors for the crystal (i.e., [001]) can be derived, and therefore the vector describing any tip/tilt position (α,β coordinates) within that crystal can be described (e.g., the blue arrow/dot in *Figure 19*b for G1 in the edge on condition). This vector can be calculated by multiplying the rotation matrix of the given α,β conditions (Eqns. 47 and 48) by the unit vector matrix (Eqn. 62) to form a rotation matrix at that specific α,β tilt condition (Eqn. 80). This in effect rotates the unit vector at that specific tilt condition to the beam direction [001], and as such multiplying the inverse of this matrix $M_{\alpha\beta}^{-1}$ by the beam direction will provide the vector at the given tilt conditions (Eqn. 81).

The unit vector matrix, which describes the location of the unit cell axes, can be subsequently utilized to calculate the vector normal orientation of either crystal (red solid arrow in the cross-section view of *Figure 24*b) since the grain boundary plane edge on is commensurate with the crystallographic plane. That is to say, along the interface the grain boundary can be envisioned to have a corresponding plane of atoms in a specific orientation with relationship to the adjacent grain. The vector normal (Eqn. 70) to the long axis can be calculated (dashed line in the plan view of *Figure 24*b) and multiplied by the inverse of the tilt condition matrix, $M^{-1}_{\alpha,\beta}$. This vector describes the current tilt conditions where the grain boundary is observed edge on (Eqn. 82). Instead of calculating the vector parallel with the beam direction ([001]) the vector normal is substituted ([-sinθ,cosθ,0], Eqn. 82). This can be corroborated using a generic cubic crystal with the [001] positioned at α,β:0,0 and with the [100] oriented along the α tilt axis (such as in *Figure 7*a) and by subsequently inputting any α,β coordinates the listed vector/pole/ZA will be returned (e.g., α,β:24.1,26.6 will be the [112]).

*Equation 80*

$$M_{\alpha,\beta} = R_{-\beta,y} R_{-\alpha,x} R_A$$

*Equation 81*

$$\vec{u}_{\alpha\beta} = M_{\alpha\beta}^{-1} \begin{bmatrix} 0 \\ 0 \\ 1 \end{bmatrix}$$

*Equation 82*

$$\vec{u}_{\alpha\beta,\theta\ normal} = M_{\alpha\beta}^{-1} \begin{bmatrix} -\sin\theta \\ \cos\theta \\ 0 \end{bmatrix}$$

These calculations can be visually corroborated through a tip/tilt map of a basic FCC crystal with the [110] pole plotted at α,β:5,10 (***Figure 20***) with an representative interface observed edge on at α,β:5,10 with the long axis measured 45° to the α tilt axis (blue line). Expansion of the crystallographic tip/tilt plot to ±90° indicates that 90° from the edge on condition (α,β:5,10), the [001] is observed (red circle) at α,β:44.8,-85. Using Eqn. 52, the angle between (α,β:5,10) and (α,β:44.8,-85) is 90°. If the interface were oriented 135° to the α tilt axis, the [-110] at (α,β:-44.8,-75) would be the vector describing the normal to the interface.

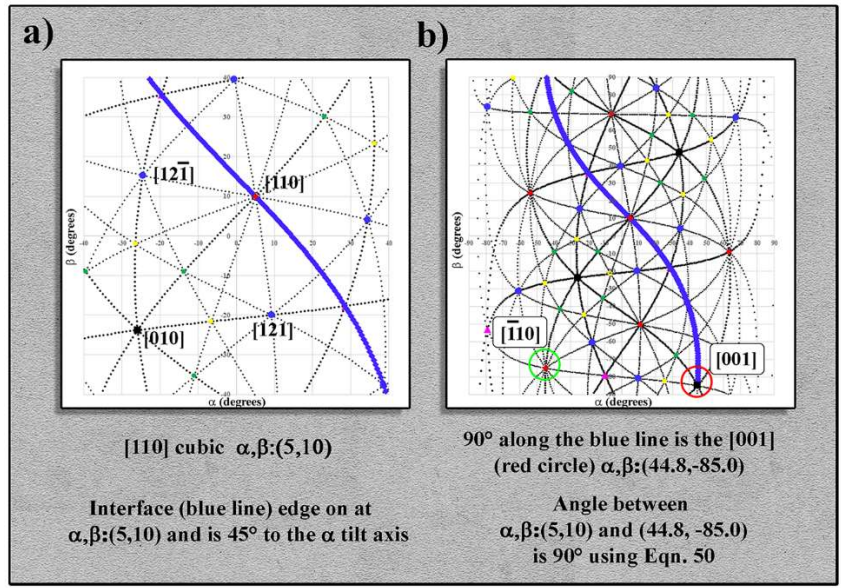

***Figure 20:*** *Plotting of a basic FCC crystal where the crystallographic solution is provided and the edge on condition is observed at α,β: 0,0 with the interface long axis 45° to the α tilt axis (a). The pole normal to the interface is demonstrated to be the [001] at α,β: 44.8, -85 (b).*

The mathematics utilized in these derivations are in the cubic or Cartesian form, and therefore these vectors need to be converted to the native form (for non-cubic systems) through multiplication with the inverse conversion matrix that was derived in the Eqns. 1-3. In order to calculate the plane of atoms in the native form, the reciprocal lattice matrix (Eqn. 83) must be utilized. Note that this is derived from Eqns. 28-31, where Eqns. 29-31 provide the reciprocal unit vectors and Eqn. 28 is the volume of the cell. The calculation of the native plane of atoms can be derived by multiplying the vector normal (Eqn. 84) by the inverse of the reciprocal lattice matrix (Eqn. 83).

*Equation 83*

$$M_{reciprocal\ lattice} = \begin{bmatrix} \dfrac{bc \sin\alpha \sin\delta}{V} & 0 & \dfrac{-ab \cos\beta \sin\delta}{V} \\ \dfrac{-bc \sin\alpha \cos\delta}{V} & \dfrac{ac \sin\beta}{V} & \dfrac{ab(\sin\alpha \cos\beta \cos\delta - \sin\beta \cos\alpha)}{V} \\ 0 & 0 & \dfrac{ab \sin\alpha \sin\beta \sin\delta}{V} \end{bmatrix}$$

*Equation 84*

$$\vec{u}_{interface\ normal} = \vec{u}_{\alpha\beta,\theta\ normal} M^{-1}_{reciprocal\ lattice}$$

## 3.9 Tilt Series

As previously described, the derivation of tilt motion of interfaces leads to a number of applications in the microscope. Small angle calculations (either normal or parallel with the long axis of the interface) have already been described, and expanding upon these calculations provides a methodology to perform oblique tilt series. While each are simple consequences of the interface motion, a brief discussion of each is necessary to build upon more complex protocols. Electron tomography has matured into a powerful tool for a wide variety of fields, from biology to materials science (Hayashida et al., 2019, Hayashida and Malac, 2016, Lidke and Lidke, 2012). The high tilt requirement often provides limiting factors for technique including special holders, larger pole pieces, and narrow sample geometries. When performed correctly, tomography can be a useful tool to circumvent the projection issues of TEM, but at the cost of losing the relationship with respect to the rest of the sample. This is considering small, needle shape regions of interest being utilized for tomographic analysis similar to APT analysis.

Tilt series for TEM foils can also be performed along a single orthogonal axis (whether it be α or β) using a double tilt stage, but there are limitations to this approach. Depending on the orientation of objects within the TEM foil, especially when they are oriented obliquely to the stage axes, single axis tilt series may not provide a clear picture. For instance, when a grain boundary decorated with precipitates is tilted in a non-logical manner (i.e., not against its long axis) it is difficult to observe the full distribution of precipitates or voids on the boundary (Badwe et al., 2018) (***Figure 21***). Yet, when the boundary is tilted against its long axis, the boundary moves in a logical fashion, and the distribution can be readily observed. Equally, if not more, important is the ability to create tilt series along specific planes of atoms that can be beneficial to demonstrate dislocation microstructures in three dimensions (Liu and Robertson, 2011, Hata et al., 2020, Yamasaki et al., 2015).

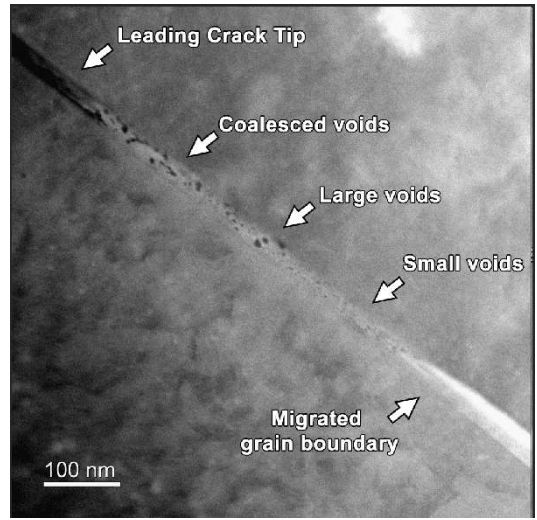

*Figure 21: Tilt series collected across a Ag-Au grain boundary exhibiting void evolution ahead of an stress corrosion cracking (SCC) tip in the binary alloy* (Badwe et al., 2018).

Using both the interface calculations combined with the crystallographic solution of any grain within the sample allows for rapid collection of tilt series. The tilt series calculations are performed by adjusting the tilt steps in the interface calculation (φ in **Figure 16**b), and most importantly since only small steps are being performed the lack of eucentricity on the β is minimized.

The mathematical derivations (Eqns. 69-75) for tilting along or against the long axis of an interface at discrete angular steps regardless of orientation (i.e., orthogonal or oblique to the major tilt axes) provides the ability to correctly orient interfaces for spectral analyses, and can also be utilized in combination with the solution of adjacent crystals to create useful tilt series data.

## 3.10 Kikuchi Bands and Diffraction Patterns

The ability to create tip/tilt maps of any crystal system is important for nanocartography at any length scale. Given a known pole and in plane orientation, microscopists can quickly and reproducibly tilt anywhere within the stage tilt limits. In section 2 it was noted that tip/tilt conjugates of crystalline stereographic projections (i.e., a stereographic projection including Kikuchi bands) could also be produced in a similar manner, except that the fidelity of such maps may not be practical given the accuracy of many double tilt stages. That is to say, only extremely small d-spacings would provide Kikuchi bands that could be tilted between with precision. With advancing technology, there may come a time when this could be possible, and therefore it will be discussed herein. The derivation that allows Kikuchi lines to be plotted can additionally be utilized to plot diffraction patterns as well.

Unlike a stereographic projection where free rotation is considered, the axes of tip/tilt motion are beholden to one another in that the β axis depends on the α rotation. Therefore, when plotting Kikuchi bands (which can be considered mirror images of the plane) it is not appropriate to calculate the scattering vector in order to map the Kikuchi band. The reason being is that the traces of ± g-vectors will eventually converge to a point, whereas the traces need be mirror images and not converge (**Figure 22**a). Within the first ~40° the two maps appear similar, but out past this point plotting the g-vectors will converge. This is because the trace of the plane normal to the g-vector will inevitably intersect the trace of the original point (since the g-vectors will be multiples of the trace of the plane). In order to plot Kikuchi lines correctly in a tip/tilt map (*Figure 22*b), g-vector calculations at each point along the trace of the plane must be performed in the following manner.

To be able to mirror the Kikuchi band, the specific orientation at the given α,β needs to be rotated by the desired ± Bragg angle about an axis (Eqn. 18) in the direction of the plane normal. In *Figure 23* the trace of a (111) plane in a cubic system is illustrated. While the remaining planes and poles are not shown, the crystal orientation is commensurate with *Figure 7*a where the [001] is plotted at α,β:0,0, with the [100] aligned along the α tilt axis. At any given point along this line a vector can be drawn that points directly to the [111] pole (inset of *Figure 23*a). Due to nature of the S-curve, the directionality of the vector will change depending on the position along the trace. As demonstrated in Eqns.80-82, the crystallographic vector can be derived at every α,β condition along the trace (e.g., at α,β:16.2,-33.1 the vector would be described as [-0.52,-0.28, 0.80]). Given that a single Kikuchi line scatters at the Bragg angle and **g**-vectors (diffraction spots) scatter at twice the Bragg angle, in order to plot either, a rotation about an arbitrary axis normal to the vector (Eqn. 81, i.e., a vector along the trace of the plane as shown in *Figure 23*b) and the plane normal at the desired angle (θ or 2θ) must be performed.

The arbitrary vector can be calculated by taking the cross of the normal to the trace of the plane (e.g., the [111]) and the vector describing the trace of the plane (e.g., [-0.52,-0.28, 0.80]). Using Eqn. 18 with this vector as the rotation axis (e.g., [0.62,-0.77,0.14]) the vector describing the trace of the plane (red circle in *Figure 23*) can be rotated about the desired multiple of plus or minus the Bragg angle (green circles in *Figure 23*c) to derive the position of the Kikuchi band or diffraction spot (note the Bragg angle of 5º is exaggerated for demonstration purposes). The trace of the Kikuchi line will then be a line plotted through each point, hence mirroring the trace of the plane.

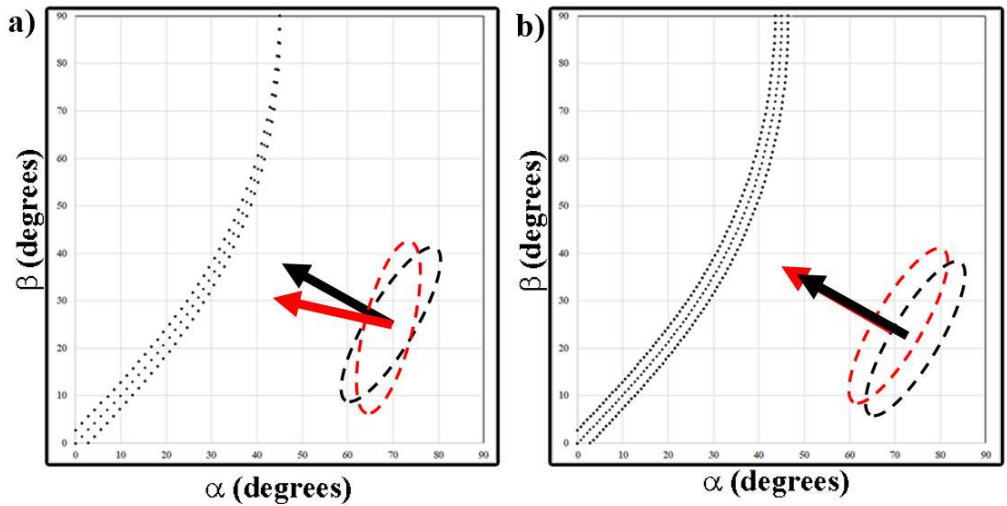

*Figure 22: Plotting Kikuchi lines as scattering vectors (a) as compared to plotting the Bragg angle at each point of the trace of the plane (b).*

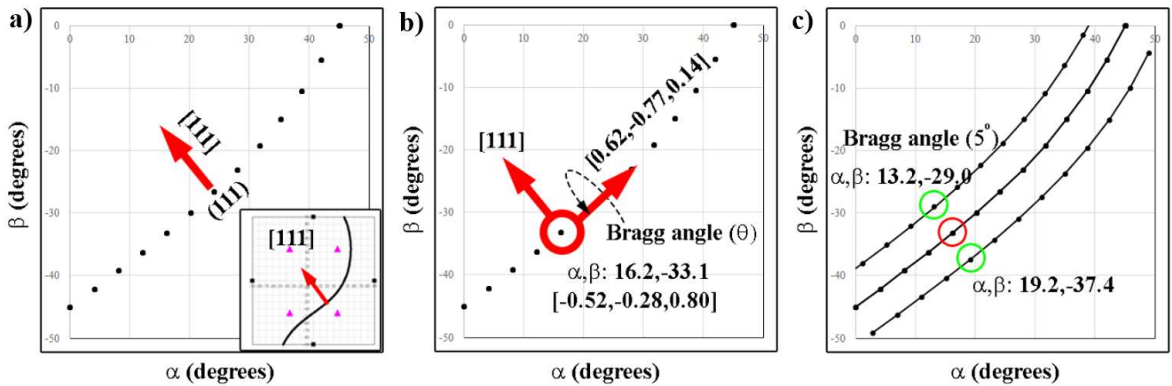

*Figure 23:* Plots of a basic cubic crystal and the trace of the (111) plane, and a Kikuchi bands at a Bragg angle of 5°.

As Kickuchi bands represent dynamical diffraction from crystallographic planes, so too can kinematical diffraction spots also be plotted in a similar fashion, except that instead of plotting lines at the Bragg angle, diffraction spots (±) could be plotted at twice the Bragg angle at any given tilt position within tip/tilt map. The diffraction spots will be dependent upon both the directionality of the plane normal (Eqn. 34) and the structure factor (Eqn. 37) as to whether the plane of atoms is expressed. As demonstrated by Cautaerts et al., if the accuracy of a double tilt stage is sufficient, tilting samples to a two beam condition could be utilized in this manner (Cautaerts et al., 2018).

## 3.11 Unknown Crystal Calculator

The development of crystalline stereographic projections, pole figures, and pole figure tip/tilt maps for both general crystals (i.e., for basic low index poles/planes and not derived from the structure factor) in addition to specific crystals (e.g., where the structure factor has filtered specific poles/planes) has been demonstrated to be a powerful tool for studying crystalline materials using a double tilt stage (Duden et al., 2009, Klinger and Jäger, 2015, Li, 2016, Liu, 1994, Liu, 1995, Qing, 1989, Qing et al., 1989, Xie and Zhang, 2020). Yet, when the crystal structure is unknown, especially for nanocrystalline materials, these calculations are ineffective because the poles/planes discovered can represent any particular crystal and many systems share similar diffraction patterns among various poles/planes. Therefore, it is necessary to develop a methodology to map unknown crystals using the aforementioned protocols, specifically the interface calculator.

Just as the interface calculator was based off the prediction of planes of atoms, so too can the inverse be applied for unknown crystals given the knowledge of what diffraction patterns and Kikuchi bands represent. Diffraction spots (g-vectors) and Kikuchi lines represent the directionality of planes of atoms within a crystal, and therefore can be treated as interfaces. The interface calculator was derived to predict motion of the traces of interfaces, and therefore it can be extended to unknowns. At first, the concept may seem trivial, but when extended to tracing the motion of a single diffraction spot or Kikuchi band in combination with the knowledge that planes of atoms intersect at crystallographic poles, the complexity of the protocol becomes more relevant (*Figure 24*). Regardless of whether the g-vector of a plane is deduced from an atomic column image/fast Fourier transform (*Figure 24*a) or a diffraction pattern (*Figure 24*b), the trace of the plane of atoms can be calculated with relation to the calibrated α tilt axis (given that the plane of atoms is normal to the g-vector) and the tilt conditions at which the g-vector was observed. Similarly, using CBED or Ronchigram mode where a Kikuchi pattern can be collected (*Figure 24*c), the orientation of the trace of the plane at the given tilt conditions can also be plotted.

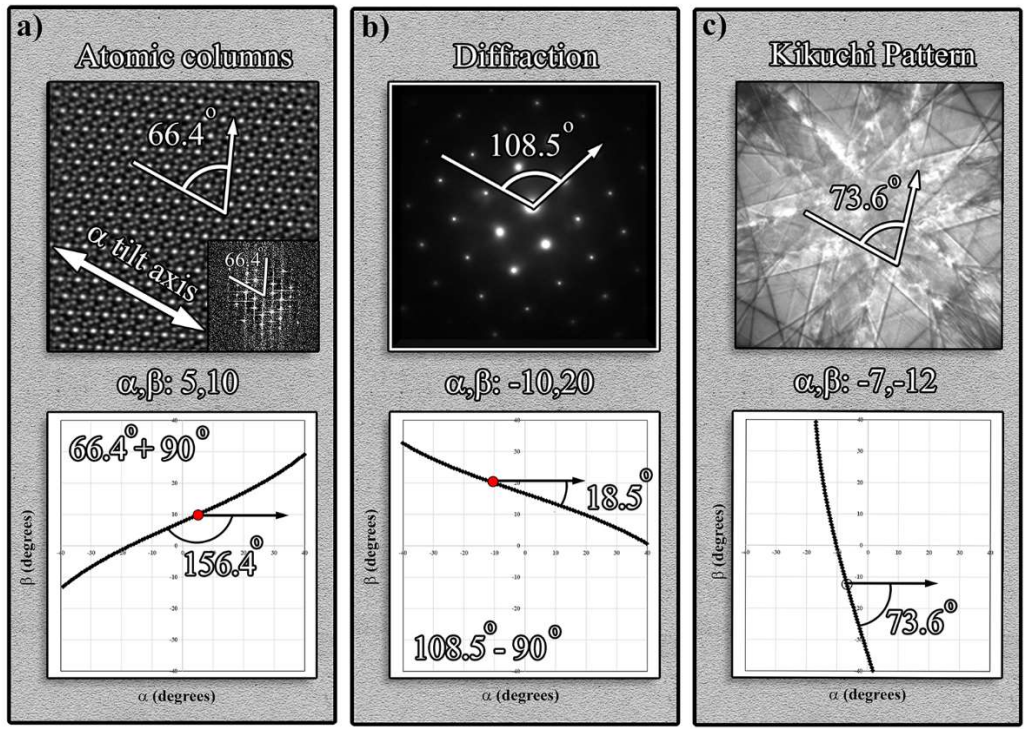

*Figure 24:* Mapping crystallographic planes using (S)TEM images and tilt conditions in a double tilt stage using either a) atomic columns, b) diffraction patterns, c) Kikuchi lines.

This technique can be utilized in a variety of manners to solve for unknowns, from crystals on the order of tens of nanometers to single crystals within a polycrystalline matrix. As a theoretical example, given one g-vector at specific tip/tilt coordinates (red dot, α,β:5,10, with the plane measured at 135° to the α tilt axis), the trace of that plane of atoms within the tilt range of the stage can be calculated (*Figure 25*). This would be similar to tracing a plane of atoms by any of the three means in *Figure 24*. By tilting along that g-vector, additional g-vectors associated with that crystal will eventually come into a diffracting condition (e.g., blue dot, α,β:-19.7,-25.2). Note that as this is a theoretical example, a random point along the line was chosen.

At these new tip/tilt coordinates the directions for any additional planes can be traced, thereby creating a map of that specific crystal (blue dot, *Figure 25*b, 90° at α,β:-19.7,-25.2). Once two poles have been identified and mapped (second pole, green dot, at α,β:-13.3,-23.5), the specific directions for at least one additional pole will be provided through the intersection of the trace of the planes of the individual poles. Without ever having tilted to that position with the stage (third pole, pink spot, α,β:22.9,15.7), the knowledge that there is a pole present provides assurance to the microscopist that additional poles can be quickly determined. As illustrated in *Figure 25*b, the weight of each trace can be scaled according to the apparent indexing of the planes (i.e., low index planes will appear darker due to multiple planes expressed due to the structure factor). This can further be used as a fingerprinting technique for analyzing unknowns. To illustrate how this would work in practice, a movie highlighting the identification of an austenitic stainless steel grain is shown in *Figure 25*c.

Lastly, once a number of poles have been identified and planes traced, additional crystallographic information can be garnered from the map. For example, the pure tilt angles between poles can be calculated (Eqn. 52), poles not attainable in the tilt stage will be apparent, and finally symmetry of the system can be observed and tested. If a mirror plane is suspected, tilting to specific angles to either side of the plane can be performed to observe whether identical diffraction patterns are exhibited. The collection of this data can then be used to identify a large number of poles within a sample to compare to crystallographic mapping programs such as Crystal Maker (Palmer, 2015) or JEMS (Stadelmann, 1987).

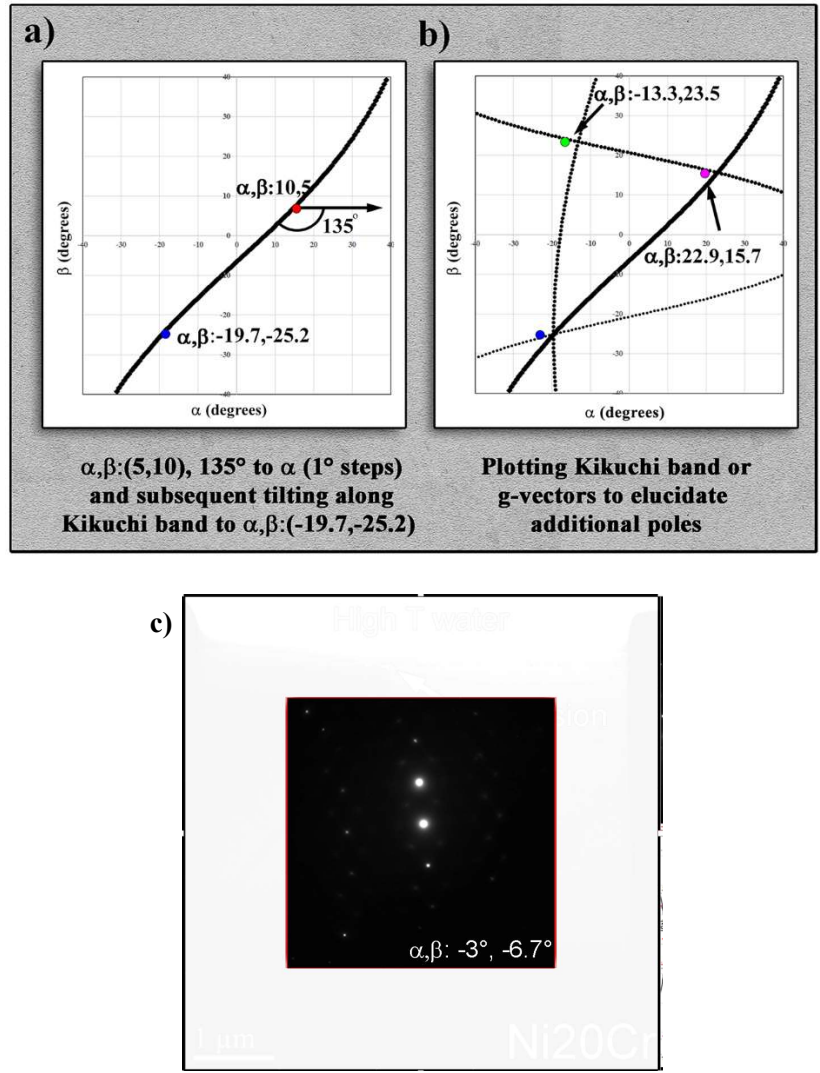

*Figure 25:* Unknown calculator illustrating how to tilt along g-vectors or Kikuchi lines to build unknown crystal maps using a) a single line, b) additional lines. c) The provided movie shows how this works in practice.

# 4. Practical Applications of Nanocartography

## 4.1 Calibration of the Double Tilt Stage

The premise of utilizing nanocartography for electron microscopic analysis centers on the knowledge of the α and β axes of a double tilt stage (note: as described in section 2, the α,β convention is equivalent to X and Y tilt axes). A similar procedure can be performed for a single tilt stage, but it precludes much of the information that can be collected (i.e., the ability to tilt within a crystal and stage motion are limited to one axis). The power of utilizing these techniques is that once the spatial relation of the α and β tilt axes are known for a specific microscope (note that the β can be simply deduced as orthogonal to the α once it is determined), any information gathered from a sample can be translated in any fashion for use on another microscope as long as the tilt axes have been similarly calibrated. This information allows any microscopist to perform routine pre-screening of data regardless of time or peripherals on any microscope before either reanalyzing the sample in the same microscope, or transferring it to a different facility or scope with more advanced capabilities.

The simplest manner in which to calibrate the α axis in imaging modes is through the use of a carbon contaminated sample. Once loaded, a region of interest that appears to be relatively uniform and flat should be selected. Carbon mounds on both the top and the bottom of the sample can be produced by condensing the beam to a fine point over the region (*Figure 26*). After a set amount of time, the probe can be defocused to detect whether carbon mounds were produced. Once a sufficient amount of cracked carbon has been deposited, an image at α,β:0,0 can be collected for reference (*Figure 26*a). The sample should then be tilted approximately both positive and negative 10° in the α, at which tilt point images should be collected. From the three images collected, the circular projection of the carbon mound at α,β:0,0 should become elliptical at both positive and negative α tilts. The long axis of the elongation is the α tilt axis. For the β axis, one can either draw a line orthogonal to the α or repeat the procedure of starting at α,β:0,0 and tilting to positive and negative 10° in the β to ensure the correct location of the axis. The calibration of the α axis needs to be performed at tilt condition α,β:0,0, from which the subsequent calibration across tilt space can be applied (e.g., at tilt conditions farther from 0,0 the actual α will be some mathematical rotation from the calibration at 0,0 depending on location).

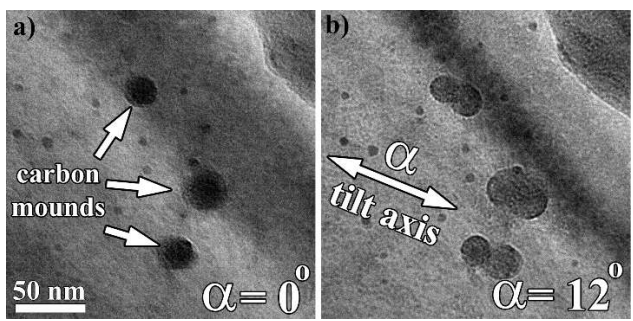

*Figure 26:* TEM (BF) images showing carbon contamination (mounds) deposited on a sample (a), followed by a pure α tilt to highlight orientation of the α tilt axis (b). For all other tilt calculations in TEM (BF) mode the angle to the α axis will be measured from this line.

As previously demonstrated in section 3.1, the calibration of TEM diffraction and STEM (Ronchigram) modes can be performed with any crystalline sample, preferably one with larger, single crystals as the calibration will be performed utilizing the position of diffraction discs and Kikuchi lines emanating from a single source (*Figure 27*).  In either TEM diffraction mode or Ronchigram mode, a grain should be selected that is relatively close to a zone axis (ZA) when the stage is at α,β:0,0.  The grain should be uniformly flat and set at the eucentric height.  This procedure works best if a grain oriented close to a higher index ZA grain is selected since the systematic reflections intersecting at the ZA will provide a good fiduciary marker (too low an index could skew the visibility of the center of the ZA).  Using a large condenser aperture (preferably in CBED mode if the microscope allows it), the probe should be condensed to a point followed by subsequent analysis in diffraction mode.  A center fiduciary image can then be taken with the oriented ZA as the nominal zero position (*Figure 27*a and c).  The sample should then be tilted approximately ±5˚ in the α at which time an image should be taken (*Figure 27*b and d). From each image collected a trace of the center of the ZA can be drawn, hence providing a measure of the α tilt axis.  As with the calibration of the TEM bright field and STEM imaging conditions, the β tilt can be ascertained as orthogonal to the α tilt axis, or the aforementioned protocol can be repeated using the β tilt in place of the pure α tilts.

Depending on the sample thickness, the observation of the Kikuchi lines intersecting at a high index ZA are the easiest to observe.  The same procedure can be used in parallel beam diffraction mode, except the center of the ZA needs to be identified by tracing a circle intersecting the Ewald sphere and determining the center of the circle to pinpoint the ZA.  This approach can be more difficult and is less favorable than using Kikuchi lines.  In diffraction mode the orientation of the α tilt axis can corroborated by defocusing the diffraction pattern either over or under focus (*Figure 27*e and f, respectively) to observe the physical relation of the diffraction pattern to the sample.

Lastly, a large, relatively flat single crystal can be used to calibrate the α tilt axis using two ZA and the methodology formulated to solve unknown crystals.  *Figure 28* illustrates selected area diffraction (SAD) patterns of two ZA ((a)<110> and (b)<112>) of a CuNi alloy, of which the details of the alloy are provided in subsequent sections and in the Supplemental.  The tilt coordinates of each of the found ZA is shown in *Figure 28*c.  The angle necessary to connect these two ZA will correspond to a specific g-vector in the diffraction pattern, in this case the {111} vector connects the <110> and <112> and the angle from the α tilt axis is 6°.  Therefore, in the SAD patterns an angle orthogonal to the measured angle in *Figure 28*c (i.e., angle +/- 90° due to the Kikuchi band being orthogonal to the g-vector) will illustrate where a fiduciary marker for the α tilt axis can be measured for all diffraction patterns.  As the projection of the lattice planes will rotate slightly at oblique angles, the use of the unknown calculator will accurately account for the measuring and plotting traces between ZA.  This is the exact reasoning for why diffraction patterns collected across the entire tilt stage require slight rotations by which to keep the g-vectors aligned.  The S-curves illustrated in section 2 represent the angles by which the planes rotate.

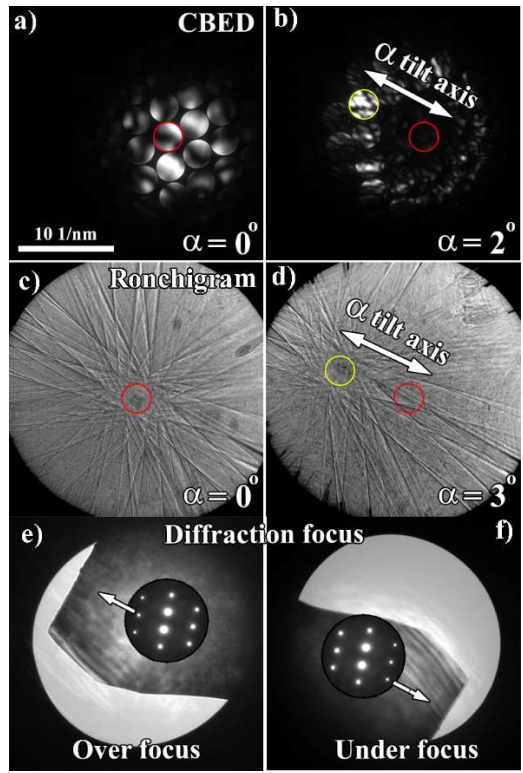

*Figure 27:* Calibration of the orientation of the α tilt axis in TEM diffraction and STEM (Ronchigram mode) using a single crystalline sample. CBED patterns (a-b) illustrate the crystal on zone (a) and tilted ~2° in the α tilt (b). Ronchigram mode showing the crystal on zone (c) and ~3° in the α tilt (d). Over (e) and under (f) focus images of the central diffracted beam showing the orientation of the sample with respect to the diffraction pattern.

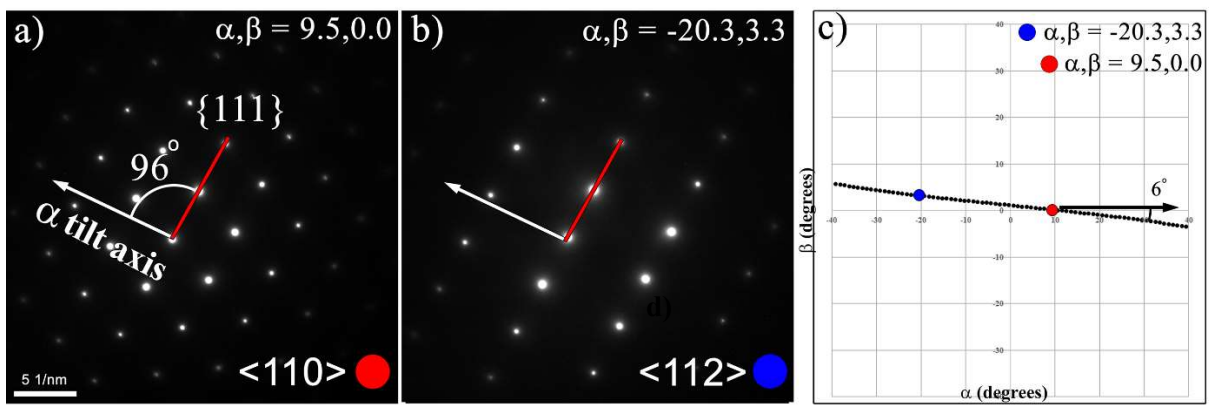

*Figure 28:* Calibration of the alpha tilt axis using a single, locally flat sample. SAD patterns (a,b) were collected at various tilt conditions and then plotted (c) to obtain an approximate measurement of the α tilt axis.

Once each mode is calibrated for the location of the tilt axes, it is suggested that digital templates be created for future analysis by which to overlay and measure data currently being collected. The most important is the orientation of the α tilt axis because every measurement moving forward in these protocols are measurements to this axis. The β could similarly be used except that the default is to the α because it is typically the most eucentric of the two axes. In order to best be able to utilize these protocols, it is necessary to be able to measure the radial angle between the α axis and a given fiduciary, whether it be a diffraction spot, Kikuchi line, or interface. As noted in section 3.1, these calibrations can be incorporated into an algorithm utilizing digital capture in order to use the computer screen to perform small angle tilting of the sample.

## 4.2 Tilt Stage Limit

In order to understand the limitations of the tilt map produced for any number of operations, it is necessary to calculate and map out the limit of the stage motions. This can be achieved by assuming symmetry of the stage motion and tilting to a set number of α tilts and then determining the corresponding limit of the β tilt. When mapped on the JEOL ARM200CF the standard double tilt holder has the shape of a "squircle" or superellipse. The superellipse is defined by:

*Equation 85*

$$\alpha = a \cos \theta^{2/r}$$

*Equation 86*

$$\beta = b \sin \theta^{2/r}$$

Where α and β are the tilt stage readouts, a and b are the α and β tilt limits, respectively, and r is a shape factor. ***Figure 29*** is a plot illustrating stage limits of a = 36 and b = 32 for a number of shape factors with an r value of ~3.3 used a good description of the JEOL double tilt holder for the ARM200CF. While these tilt limits can be programmed as look up tables to curtail outputs (e.g., when plotting specific crystals or for calculating tilt series calculations) the ability to trace Kikuchi bands outside of the tilt limit can be beneficial to understanding crystallographic data of an unknown sample.

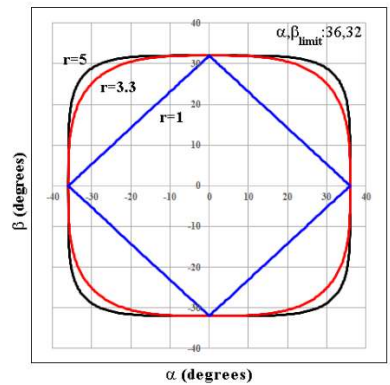

***Figure 29:*** *Double tilt stage limits as calculated through a squircle (superellipse) estimation using various r values.*

## 4.3 Crystallographic Orientation

The calculations and derivations of formulae surrounding crystallographic orientation using a double tilt stage have already been presented, but a more practical discussion of how to apply these techniques can be discussed. First, it is necessary to have a general understanding the movement of the double tilt stage in comparison to a pole figure or stereographic projection (see ***Figure 4***). Whereas the planes of traces of atoms in a stereographic projection appear to move in a straight or arced path, in the tilt map the planes move in an S-curve fashion due to the α,β relationship. For instance, with a cubic crystal positioned with the [001] at the α,β:0,0 tilt coordinates and the [100] along the α or β orientation, the tilt conditions for the [111] will be observed by first tilting alpha to 35.3° and then beta to 45°, but the reverse order (beta to 35.3° and then alpha to 45°) will not because of the dependency of the order to the stage motion. Utilizing this allows orientation of any crystal using the following protocols.

With experience and familiarity with a single system, microscopists can eventually identify diffraction or Kikuchi patterns from memory, and eventually identification of each plane within the diffraction pattern becomes second nature. However, for inexperienced users identifying these crystallographic waypoints can be a challenging if not daunting experience. It should be noted that the work Xie et al. has automated this in an open source manner (Xie and Zhang, 2020), but for completeness the protocols will be provided here.

In order to optimize crystallographic solutions with regard to the sample and stage, at least one known or identified pole must be determined as a proposed starting point. The stereotypical, six-fold symmetry of a <111> cubic crystal is presented at stage positions α,β:10.9,5.0 (*Figure 30*). Depending on the level of confidence with the found pole, the next step is to either tilt to a second known pole somewhere within the tilt stage to set a second waypoint or to follow orient to a known g-vector.

Since the preliminary calculations of the crystallographic solution are based upon the first known pole (e.g., <111>) the remainder of the crystal will rotate concentrically about that pole. Given the crystallographic system has already been identified, the crystal can be rotated about the first pole until a second proposed pole lines up with the tilt conditions of the second pole. For instance if a second pole were discovered at α,β:13.3,-14.8 the crystal could be rotated until the <112> type pole would line up. Depending on the confidence of the user in the identification of the two known poles and the symmetry of the crystal system, a third and final pole could be identified on the map and confirmed by tilting to that point (e.g., the <110> type at -16.6, -17.3). For mapping and predictive tilting of the sample, the symmetry of the given crystal system is not as important in setting this rotation correction. When considering the orientation relationship between two grains (i.e., misorientation angle) it is necessary that the directionality of each adjacent grain be considered.

For more experienced microscopists, measuring the angle of a known plane to the α tilt axis angle can be utilized in combination with the first known pole. In *Figure 30* the {110} is measured ~20º to the α tilt axis and can then be adjusted accordingly in the tilt map to get the projection of the {110} correct. Again, depending upon the symmetry of the crystal system, other poles can be located and tilted to in order to confirm the orientation. Once enough poles have been identified to satisfy the correct orientation, the crystallographic solution for that crystal has been mapped and can be utilized for future tilt experiments and the determination of vector normals to interfaces and adjacent grains. This is extremely important for inexperienced users to gain familiarity with a crystal system, as a large number of poles can be quickly visited thereby building up experience with that crystal type. For this reason alone, for less experienced users this is more beneficial than an automated system of which it may be tempting to blindly trust the output. Automation can be a powerful, but without a proper understanding of the underlying principles, it can be improperly used.

As shown in *Figure 30*, the collection of a CBED or Kikuchi map makes this exercise trivial because the measured angle to the α tilt axis can be quickly, efficiently, and accurately measured. This does not preclude microscopists without the ability to collect such pattern from utilizing these methods. Once two known poles have been identified, the tilt positions of the poles can be used to calibrate the crystallographic solution similar to the aforementioned protocol. The poles/planes of the entire crystal can then be traced and visited.

Mapping of crystals in this manner can be extremely useful in confirming suspected crystals as well as rapid tilting for collecting diffraction or atomic column images. Depending on the crystal, one pole may have a unique projection of atomic planes that helps identify specific atomic positions. For instance, in dislocation contrast imaging (DCI) the <110> is a preferential pole in FCC materials to indicate a large basis set of {111} planes (Zhu et al., 2018, Phillips et al., 2011, Sun et al., 2019). Understanding the exact location of the <110> within a specific crystal could suggest that the sample is too thick (i.e., a large tilt condition) to achieve clear dislocation images. It would allow a microscopist to then either find an additional crystal or tilt to a less desirable pole. Lastly, when presented with opportunity to utilize a crystallographic solution, microscopists could more confidently confirm the specific crystal and avoid mis-identification by rapidly tilting to multiple poles. This type of analysis would not only be beneficial for assuaging scientific reviewers, but would again be most beneficial to inexperienced users who are not accustomed to rapid identification of specific poles through fingerprinting.

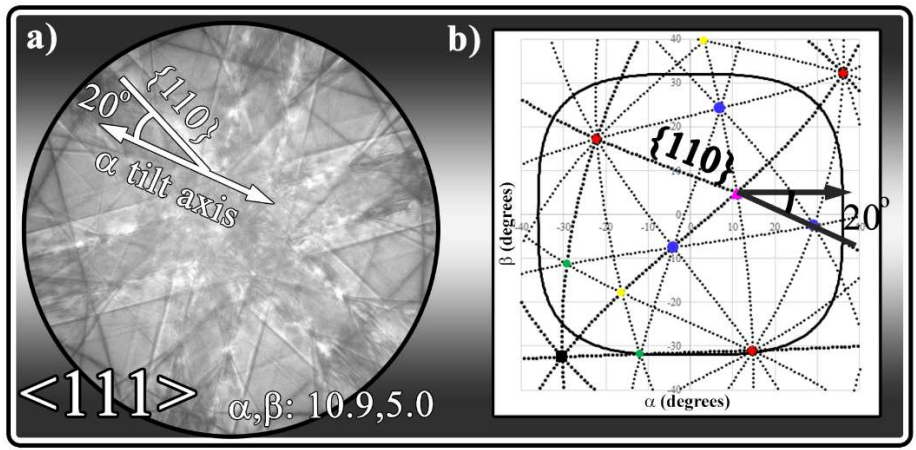

*Figure 30:* Determination of orientation of a known cubic crystal by measuring angles with relationship to the α tilt axis using a Kikuchi pattern (a) and plotting out the pattern with respect to the α tilt axis (b).

## 4.4 Grain Boundary Characterization

The notion of crystals as physical objects using vectors was considered in the development of the mathematics for orientating crystalline material with the double tilt stage to more logically tie it to the sample as a physical object. Microstructural analysis is highly dependent upon objects that, while being often crystallographically related, are not themselves crystalline. Objects such as grain boundaries, surfaces, and interfaces are all important to microscopic analysis, and as such, the description of their motion is of great interest to microscopists. In section 3, an interface calculator was derived which allows full control of finding a boundary on edge, and how to tilt along the long axis of the boundary. This is discussed through a variety of different techniques in these papers, and there are limitless protocols for which having this control could be utilized.

One example of how this enables nanocartography is through the analysis of nanocrystalline phases on a grain boundary. The difficulty of nanocrystalline phase identification is often related to the size limitation of selected area diffraction apertures. Once an aperture becomes too small, it can start to influence the scattering of the probe, thereby skewing the crystallographic information within the sample (Thomas, 1962, Carter et al., 1996). With advanced instrumentation, nanobeam analysis techniques were developed to overcome this limitation, but they themselves suffer from being too site specific. That is to say, in nanobeam mode it becomes difficult to position the site-specific region with respect to the rest of the sample given that the beam width is on the order to 10-50 nm. An appropriate analogy is the use of the Ronchigram to align aberration probe corrected beams because they are on the order of a fraction of an angstrom (and hence cannot be observed on a physical detector). Having full control and knowledge of not only the crystalline components of the sample but as well, the control over orientation of interfaces can serve to overcome much of these limitations.

Most often nanocrystalline phases on interfaces and boundaries are crystallographically oriented to one of the adjacent crystalline phases, and therefore knowing the orientation of the adjacent matrix will make phase identification of the nanocrystalline phases more accessible (Bhadeshia, 1987). Yet, if the interface is not oriented correctly (i.e., not viewed edge on), parasitic reflections from an adjacent grain could skew identification. Therefore, a protocol could be utilized by which to first solve the crystal orientations of both crystals (*Figure 31*), and then subsequently find the interface on edge condition. In *Figure 31* (with the sample at α,β:0,0) the grain boundary is observed ~10° from an edge on condition and 38° to the α tilt axis. By overlaying the directions of the interface lines over top of the crystalline solutions, a map of the optimum orientation for crystallographic analysis could be achieved. There will be only one tip/tilt position on the tilt directions normal to the interface where the boundary would be oriented edge on (in this condition 10° from α,β:0,0 is α,β:-6.1,-7.9), and once at that orientation the boundary could then be tilted along the its long axis to keep the boundary edge on (along the blue line). By tilting along the boundary in the edge on condition to a position where either adjacent grain is near a low index pole/ZA, selected area diffraction (SAD) could then be achieved without parasitic reflections from either adjacent grain (e.g., G1 oriented to [121] at α,β:17.4,-26.8, with no major pole present in G2). Any systematic reflections (albeit weak due to the small volume of material) could then be compared against the low index orientation of either grain, and additionally since the orientation of the opposite grain would already be known, any reflections from that grain could easily be identified and temporarily ignored. Additionally, if a tilt condition exists where both grains are preferentially oriented (e.g., α,β:-28.8, 11.9 where G1 is oriented to the [12-1] and G2 is on the [110]) when the grain boundary is edge on, HAADF atomic column imaging can clearly elucidate the boundary conditions.

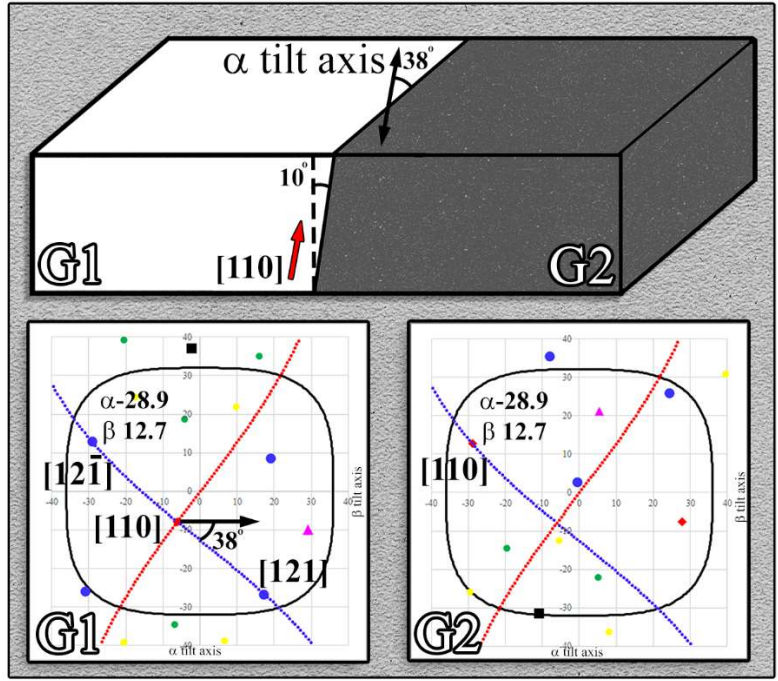

*Figure 31*: Schematics illustrating a protocol for determining the grain boundary physical orientation relationship to the adjacent crystalline grains. At α,β:0,0 the boundary is ~10° from an edge on condition with the boundary oriented 38° from the α tilt axis. The crystallographic solution for both grains is presented with an overlay of the tilt orientations for tilting the boundary along (blue) and against (red) the long axis of the boundary.

## 4.5 Multiple Session Sample Analysis

While the impetus for developing nanocartography as a technique was to rapidly and accurately perform tiling along known planes and between known poles/ZA (similar to the impetus for the development of ALPHABETA (Cautaerts et al., 2018)), it soon became more evident as to the real power behind utilizing vector calculations. While any number of publications have outlined the geometry and mathematics to navigate a double tilt stage as well as various crystals, none have taken into consideration the ability to re-orient the sample for additional analyses. Through the inclusion of the $R_{stage}$ rotation matrix, which utilizes improper rotations as reflections, sample analysis can be done over multiple sessions on one microscope without losing the previously mapped crystallographic or interface positioning. As previously mentioned, this opens the ability to take full advantage of smaller and smaller analysis windows in an era where crowd sourcing of the instrument places microscope time at a premium.

More importantly, with the ability to map out samples it makes collaborations between institutions more attractive. *Figure 32* illustrates how this might work between an institution such as Colorado School of Mines (CSM) and Pacific Northwest National Laboratory (PNNL). Regardless of what peripherals, detectors, or systems are on a single microscope, the ability to measure the α tilt axis on a double tilt stage, provides a manner to communicate the data collected from one scope to another. The provided movie illustrates how diffraction patterns would rotate with sample rotation in the cradle.

At CSM a sample could be loaded, and two fiduciary markers would be logged. The first, a global fiduciary marker in relation to the holder (which will later be accounted for in the rotation matrix as $R_{horz}$ and $R_{vert}$). Once loaded, a conspicuous local fiduciary marker on the sample itself is also measured. Mapping of any of the crystals on the sample surface could be performed, with each pole/ZA being logged. The sample and sample map could subsequently be shipped to PNNL for further analysis. This could be done for any reason; lack of a specific technique, higher end instrumentation, or collaborative analysis. This could save a great deal of time, money, and effort in requiring staff from each institution to travel.

At PNNL, regardless of reason, the sample could be loaded in a horizontally flipped orientation. The reason for a change in orientation could vary from a simple mistake to a geometric necessity (e.g., a double tilt holder which can only load the sample with the crescent normal to the long axis of the stage to account for the position of EDS detectors), but because the mathematics derived in this work, this can be taken into account and the original crystallographic data can be translated. The identical fiduciary marker initially recorded is measured along with the horizontal flip of the sample, and the conversion of the previously noted poles/ZA are quickly achieved. The movie included in *Figure 32* demonstrates how diffraction patterns would rotate commensurate with sample rotation in the cradle.

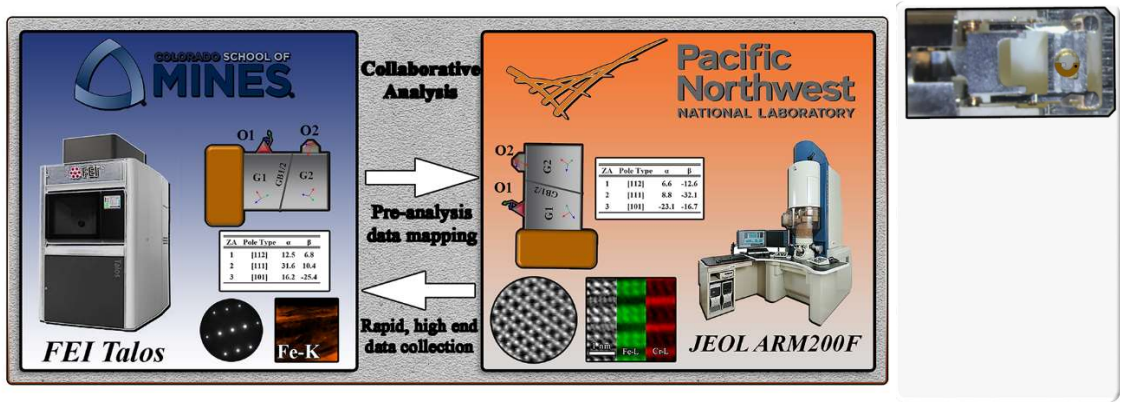

*Figure 32: Collaborative sample analysis at multiple institutions through predicative tilting and mapping and movie illustrating sample rotation.*

## 5. Discussion

The ability to accurately control the stage motion with respect to crystallography has long been a desirable function in any microscopist's toolbox. While Desktop Microscopist was one of the first programs to accurately predict diffraction patterns and measure relative stage positions given a specific crystallographic information, others have done similar plotting programs for a variety of applications. ALPHABETA a was designed to accurately determine two beam locations for proper dislocation and microstructural analysis of stainless steels (Cautaerts et al., 2018). A number of papers have detailed the study of either the double tilt stage and/or its relation to crystallographic analysis, but the overwhelming majority have approached it from an a priori standpoint of having specific crystallographic information (Liu, 1994, Liu, 1995, Qing et al., 1989, Qing, 1989, Cautaerts et al., 2018, Xie and Zhang, 2020). These approaches are applicable for navigating known crystals, but the challenge for materials science is the unknown, especially at the nanoscale. This necessitates an approach by which to consider not only the stage motion for known crystals but unknown structures as well, whether they be crystals or non-crystalline physical constructs important to materials science (e.g., grain boundaries and interfaces).

The approaches detailed in this paper focus on breaking down the complex nature of materials analysis using a double stage tilt mechanism into distinct parts for ease of understanding and explanation; 1) a simple vector analysis approach to explain the physical nature of a crystal as opposed to defining it through diffraction, 2) the stage motion for crystallographic vectors and for physical constructs such as interfaces and boundaries, 3) the introduction of crystallographic analysis converting the physical description of a crystal to reciprocal space, and finally, 4) the use of the structure factor as a discrimination filter to be applied on top of the physical description of the crystal to illuminate the allowable planes/pole for a pre-defined crystal.

Taking each of these components as individual parts assists in developing tools by which it becomes easier to create and develop maps of the crystallographic orientations within a sample in addition to its overall relationship to the sample. These tools can then be utilized in a variety of manners to quickly and efficiently manipulate the sample, and also plan for future analyses.

## 5.1 Crystals as Physical Objects

In order to deconvolute what is often described as one of the more difficult subjects in the conversion of materials science to electron microscopy, namely crystallography in reciprocal space, the derivations in this research focus on delineating the description of crystals as real space objects from their description in k-space. These derivations serve a dual function in that movement of a crystal in real space is more intuitive than in reciprocal space, and it also assists in relating the crystal to additional non-crystalline physical objects (e.g., grain boundaries) and their motion using a double tilt stage. Much of the published research regarding the use of a double tilt stage for TEM analysis has focused solely on the motion of the stage, and very few have combined it with its relation to crystals. As noted prior, (Cautaerts et al., 2018), Cautaerts et al. approached the use of a double tilt stage to calculate the motion of a cubic crystal for use in determining the optimal tilt conditions for two beam analysis. While this provided more information than simple stage motion, it still did not fully concern all crystal types.

The explanation of both the description of vectors of non-cubic crystals and their motion is often confusing to inexperienced microscopists along with a larger portion of materials scientists. The understanding of low index poles and vectors is typically the extent to which most curricula extend. Even in many prominent electron microscopy textbooks, the measurement of angles between planes is provided as a series of equations confined to a given system. This further serves to shroud the explanation of systems more complex than cubic.

By first showing the conversion of all systems to cubic and then demonstrating that vectors are transformed in real space provides a clearer, more manageable pathway towards more complex analysis in reciprocal space.

## 5.2 Stage motion of Crystallographic Vectors and Non-Crystallographic Structures

This research illustrates a logical method by which to convert all crystallographic systems to a cubic system, and then demonstrate systematically how the motion of any vector occurs through a double tilt stage. Building upon these mathematical operations, the motion or pathways between vectors could also then be calculated. Finally, using similar mathematics of plotting planes of atoms, the motion of non-crystalline physical objects such as grain boundaries, surfaces, or matrix/precipitate interfaces could quickly be defined. The power and flexibility of considering crystals and their motion in real space is demonstrated when they could then be related back to non-crystallographic objects within the sample.

Grain boundary and interface analysis are key components in assessing the nanoscale properties of a material to further explain bulk properties (e.g., micromechanical, thermoelectrical, electronic).

Therefore, correct orientation (i.e., on-edge) of grain boundaries for chemical analysis becomes imperative for proper investigation. Equally important, although often overlooked, is the crystallographic relationship of crystals to these interfaces. This is often relegated to more automatic detection/analysis approached such as EBSD in SEM because of its ease of use (Wilkinson and Britton, 2012, Alam et al., 1954, Venables and Harland, 1973, Harland et al., 1981). Unfortunately, the deeper understanding of these programs more so than the general meaning of inverse pole figures (IPFs) is taken for granted. Having a manner by which to orient interfaces and then quickly relate their orientation to adjacent crystallographic objects provides for more thorough analysis opportunities during a session. Newer technologies such as precession electron diffraction and 4-D STEM ((Ophus, 2019, Ghamarian et al., 2014)) will speed up these analyses, but similar to EBSD, they lack the ability to take into account the physical description of surrounding non-crystalline objects such as grain boundaries.

Additionally, the ability to describe the full stage motion with respect to these objects also provides a pathway to understand their three-dimensional structure within the confines of the foil. To date, electron tomography plus atom probe tomography provides the highest three-dimensional spatial resolution (even considering a multitude of artifacts) of materials within the spectrum of analytical materials science analysis tools. For all of the advantages these techniques provide, the major drawback is their extremely localized view of the sample. This tradeoff can be detrimental to a more representative analysis of the sample as a whole, but as well can be costly and time consuming. Using the double tilt stage to tilt samples about any given axis allows microscopists to tell a more complete story of the sample, even within a 50-100 nm volume without having to remove the sample and orient to a logical axis (e.g., α tilt). This motion, when combined with the knowledge of the surrounding crystal, becomes even more important.

Lastly, having the orientation solution of any possible grain within a sample allows for a number of discriminatory actions by a microscopist. If the local orientation of any neighboring ZA or pole is mapped, it can then be compared to the tilt stage limit. Depending on the desired crystallographic orientation, it may not be possible to achieve said orientation within that tilt range. As has been previously demonstrated, with the location of a few zone axes, the coordinates of the principle axes can be accurately calculated (Liu, 1995). The comparison of the principle axes of two neighboring grains can then be utilized) to calculate the location misorientation and the axis of misorientation that then leads to grain boundary type. Again, future development of scanning diffraction techniques will eventually automate this analysis, but the basic knowledge and understanding of this technique will assist in demystifying the often black box approach to their use.

## 5.3 Crystallographic Analysis/Conversion to Reciprocal Space

The description and understanding of crystals as real space objects defined by vectors and vector motion is the first step in developing a more logical pathway for materials scientists and electron microscopists. The additional step, which can then be finally utilized to develop a mapping filter, is to describe crystals within reciprocal space. The latent introduction to reciprocal space is often performed through the preliminary description of Miller indices to describe planes of atoms. This is typically accomplished through utilization of inverse nomenclature, but then drawn as a real space object. This can lead to confusion as to the crystallographic conversion to reciprocal space, especially for non-cubic crystals where the normal to the plane (a real space description) is the same as the Miller indices description (reciprocal space). As previously noted, the motion of a crystal is predicated on the real space description, but in electron microscopy analysis, the majority of terminology deals with reciprocal space.

Therefore, decoupling these two explanations as first described in the section above through vector math and then subsequently the conversion to reciprocal space is necessary for better understanding.

This additional conversion takes advantage of the previous conversion of non-cubic systems to cubic. In doing so, the normal to any plane can quickly be calculated given that the descriptions of the normals and the Miller indices are the same. Utilizing these normals to describe the plane motion in a double tilt stage provides a full description of any plane. Given the understanding of dislocation imaging along specific planes, the ability to accurately map along the trace of a given plane is imperative for accurate imaging of defects and other crystallographically dependent structures with a sample.

## 5.4 Structure Factor as a Filter

A priori knowledge of the allowable planes and poles within a given crystal based on the atomic arrangement in a specific crystal system is the most frequent approach for the analysis of crystals within the microscope. Programs such as Desktop Microscopist and K-space Navigator relied on this a priori knowledge of any given crystal to output diffraction and stereographic projections of said crystal. While these programs were extremely well written descriptions of crystallographic motion and were imperative to the development of this research, their use was confined to known crystals. This utilization of this crystallographic knowledge was based upon the structure factor. By deconvoluting each step and finishing with the structure factor, it can be shown that the structure factor simply acts a filter to be placed upon the results of the previous calculations.

Describing it in this fashion, the position and motion of all possible vectors within a crystal (regardless of system) can be detailed. Similar methodologies can be applied in the conversion of planes from reciprocal space to real space to describe their motion in a double tilt stage. The structure factor can be viewed as which of those planes, and as well ZA/poles, are possible thereby dropping out a large fraction that need to be plotted. Granted, as has been described herein, what is being plotted is not diffraction or Kikuchi lines, but the pathways between poles. Whether tilt stages will ever be developed that provide the precision to accurately discriminate between the (200) or (400) Kikuchi lines is beyond the scope of this paper, but similar methodologies could be developed if and when this ever becomes reality. Even so, given the tendency for local sample foiling and misorientations, even with the most accurate stages this type of tilting may never be practical.

## 5.5 General Applications of Nanocartography

Whereas previous research has been confined to crystallographic analysis and through the motion of a double tilt stage, it will become evident that only by decoupling real space and reciprocal space can strategies for solving unknowns and mapping of samples as real space objects be attained. This takes the extensive amount of previous research on this subject into a new realm that changes the dynamic of how sample analysis in the electron microscope is conducted. The limitations of quickly performing single analyses are removed, and most importantly, it creates a pathway by which rapid utilization of analytical tools at various institutions for a single sample can be performed with ease.

The necessity to accurately and rapidly calculate crystallographic orientations using computer-aided programming is an idea as old as the personal computer itself. With the advent of crystallographic programs to assist in understanding and comparing diffraction data to simulated patterns, it has assisted microscopists in more precisely describing samples. Programs such as Desktop Microscopist even provided the ability to input stage conditions to further predict additional tilt protocols. Unfortunately, due to the delayed data analysis due to film capture, immediate reaction to these directions was difficult if not time prohibitive. Additionally, many of these programs consider only a priori knowledge of crystallographic samples. Much like current EBSD analysis on modern SEMs, their analyses are beholden to input of candidate crystals for optimal results. By considering crystals (and their motion) as physical objects rather than through diffraction and reciprocal space, it allowed and easier transition to considering the motion of non-crystalline objects such as grain boundaries and interfaces. More importantly, it opens the avenue to mapping samples that can then be later scrutinized for planning of subsequent microscopic analysis or even for others to rapidly repeat experiments.

Crystallographic analysis using TEM (diffraction) as well as STEM (atomic column imaging) provides highly localized, site-specific identification at the nanoscale of both known and unknown phases. Having full control of the stage both in terms of guiding crystallographic analysis and the orientation of non-crystallographic features can assist in a more complete description of any sample analysis. Combining these together then becomes the ultimate tool for microstructural sample analysis. Describing the sample as a solid object, which can be manipulated similar to crystallographic directionality, allows a greater sense of flexibility. Accurate tilting of an interface or a surface to an edge on condition can mean the difference between measuring a diffusion profile of a few nanometers as compared to tens of nanometers. Similarly, the ability to tilt a boundary or a structure along or against a logical axis can elucidate a wide variety of latent microstructures. Two such examples are accurate tilting a boundary in combination with crystallographic knowledge of adjacent grains and rapid development of tilt series.

Grain boundaries are an extremely important subject in all of materials science analysis due their excess free energy that provides a wide array of phenomena to occur within a microstructure. Rapid diffusion of chromium in stainless steels provides the means for a thin protective layer of chromia to form both on free surfaces and at crack tips to arrest stress corrosion cracking (SCC) (Bruemmer et al., 2017, Olszta et al., 2014). Gallium can decimate the structure of an aluminum body as it quickly diffuses along grain boundaries, unzipping the entire structure and leaving behind individual grains (Rajagopalan et al., 2014). Therefore, the study of how elements diffuse and segregate along grain boundaries is extremely valuable, especially at the nanoscale. At this scale, phase analysis can be difficult because crystallographic information from adjacent grains can obfuscate proper analysis of the desired boundary phase. Given the crystallographic solution of each adjacent grain in combination with the motion of the boundary to an on-edge condition can assist in deduction of the unknown phase. With a boundary edge on, tilting the boundary along the plane can be directed to an adjacent ZA of either grain which might then provide for low index planes to be expressed (Carter et al., 1996). More importantly, understanding of parasitic reflections from adjacent grains can be used to discriminate the possible orientation of the unknown phase to either grain.

While the description of the precipitation along grain boundaries is most relevant, understanding morphology and density can also be an effective means in describing more global bulk properties. Whereas the typical goal for most effective sample preparation techniques is to achieve the thinnest possible sample, here it is posited that even without extremely high accelerating voltages, preliminary analysis of slightly thicker samples (100-200 nm) can be just as informative as to the data garnered from subsequent thinning and high-resolution analysis. Within the volume of a 100-200 nm thick sample the density and distribution of grain boundary precipitates can provide a more representative picture of the sample being analyzed. This can be accomplished through simple logical tilt series that takes advantage of having the ability to tilt against or along a given interface. Tomography and APT will always be a more accurate description of the three-dimensional volume, but they both suffer from being locally destructive techniques in addition to only providing an extremely narrow view of the sample volume. Creating rapid tilt series at any given step size using the protocols, all interfaces within a sample, regardless of orientation, may be transformed into a digital movie that allows for more informative data presentation. Since the step sizes between tilts are minimized for a more accurate description, non-eucentric tilting of non-orthogonally oriented interfaces is not as drastic, which in turn allows for quicker data collection. Lastly, while tilt series of dislocations have been demonstrated in the literature (Liu and Robertson, 2011, Hata et al., 2020, Yamasaki et al., 2015), if the tilt map for any given crystal has been solved, tilt series directions for any plane can quickly be calculated. Instead of following the trace of the plane systematically, if the directions for the trace of the plane are calculated it provides for easier data collection.

As with many of the subjects described herein, relating adjacent crystals to one another is an important topic in material science analysis and has been discussed in a variety of different manners. Qui et al. demonstrated how knowledge of crystallographic poles of two cubic crystals could assist in solving the local misorientation angle between them (Liu, 1994, Liu, 1995), and Jeong et al. (Jeong et al., 2010) attempted the use of a triangulation method in solving the same problem. In section 2 it was demonstrated how one could the orientation of cubic crystals be solved, but all crystal systems as well. The research herein takes a similar approach to Qui to demonstrate how the calculation of the unit vectors for any crystal can be calculated from the solution of the crystal and then be compared to an adjacent crystal through a misorientation matrix to achieve similar results (Qing, 1989). In the derivation of these formulae an important distinction must be considered in that the rotation about an arbitrary rotation axis to move a known pole to the [001] beam orientation must be performed instead of two successive rotations about the α and β axes. In solving the crystallographic orientation of any crystal, the two procedures provide identical results because the known vector is rotated to the [001] position, but in comparing the location of the unit vectors of two adjacent crystals the final misorientation angles does not yield a unique solution. Tilting in the α then subsequently in the β will yield a different result than first β then α (as shown the Supplemental *Figure S4*). Additionally, through these calculations it has been determined that the triangulation method is not sufficient in accurately describing the local misorientation (Jeong et al., 2010). Vectors chosen closest to the [001] beam direction will provide a differing result than vectors farther away from the [001]. This is because the triangulation method does not consider the dependency of the β tilt on the first α tilt, and therefore errors can be compounded for vectors farther from the [001] beam direction.

Finally, whereas the capture of data on film, either by diffraction or imaging, was tedious or time consuming (and often erroneous), digital capture has provided microscopists with the ability to optimize and improve data collection. The improved accuracy can be utilized in two distinct manners, first through the tilt of the stage through small angles, and secondly by calibration of probe deflection. Observation of a large field of view in k-space in larger crystals provides a sense of ease because microscopists can immediately observe the motion of the crystal much like traveling along an open highway. With the advent of probe corrected instruments, the ability to observe even small volumes in Ronchigram mode has allowed microscopists to observe small regions of k-space down to the order to 10s of nanometers. Yet, the ability to tilt within this small area can be difficult, and therefore being able to directly provide directions through self-identified regions on a screen (i.e., a mouse click) is highly desirable. The calculations provided in this research expand upon the tilt motion of the stage in combination with the calibration of k-space within digital capture. Conversely, dark field imaging in TEM mode, a widely useful technique in its own right, is dependent upon the microscopist's knowledge of accurately deflecting the beam using condenser lens deflectors. Most often, this is performed by eye, but calibrating the digital capture with respect to this deflection allows for more difficult deflection protocols. For instance, blind tilting can could be achieved by simply pointing to a region on the screen and having the computer read out the necessary deflections. Once a diffraction pattern is collected, the beam could be shuttered and a complete array of deflections could be planned out without introducing additional dose to the sample. Finally, even more complex schemes by which the a darkfield map exploring the entire k-space of an FCC versus BCC crystal could be programmed to best discriminate within a field of suspected dissimilar phases. Digital manipulation and control of either small stage tilts or beam deflections provides a clear advantage for rapid and accurate data collection.

## 6. Conclusions

The ability to precisely navigate nanoscopic volumes will become increasingly important to the field of materials science and electron microscopy. While there have been a number of researchers who have developed methodologies to predict both the stage motion of a double tilt stage and crystallographic motion and diffraction analysis, these have all typically been approached from an a priori stance where the crystallographic information is known. The research herein focuses on the approach of stage motion and crystal analysis initially where the crystal and sample are treated as physical objects with no regard for the physics of diffraction. Consideration of the vector motion, regardless of crystal type, and the structure factor as a filter to determine allowable planes/poles provides a more logical approach to mapping for both known crystalline solids as well as unknown phases. Additionally, it provides a direct relation of the crystal motion to physical constructs (such as grain boundaries) in the microscope.

With the rising cost of high-end microscopes and a push for nanoscale research, the cost-sharing model of financing purchases amongst a wide user base has put a premium on microscope time similar to that of a beamline user facility. Currently, crowd sourcing of instrumentation has become the norm due to a variety of reasons ranging from peripheral detectors approaching the price of base microscopes to ever-expanding service contracts. The more sources that pool financial resources, the higher the demand for instrumentation time. To overcome this, microscopists will be required to not only become faster and more accurate, but as well optimize every protocol available. Coupled with the increased ease of site-specific sample preparation of FIB and plasma FIB technology where a large number of samples can be prepared from all sample types, there has is a need to perform better and more efficient microscopy in order to take full advantage of collaborations between institutions. Nanocartography, mapping of TEM samples from the global scale down to the atomic column level, is a path forward to provide all researchers a manner to achieve optimal results and rapidly share information.

The necessity of nanocartography will be imperative toward the development of automated TEM. Regardless of how well automated image analysis may become, the projection aspect of TEM analysis will still require human directed analysis to a point. On the road to more automated data, collection and analysis there must be a transition by which microscopists are provided full control to map samples. More importantly, the cost of instrumentation is an issue that will only become more complicated with automated microscopy.

Mapping of samples and optimization of data collection for each sample will be necessary for overcoming these challenges. Nanocartography is a means by which to achieve this. Data from a preliminary session be transferred to additional sessions, and samples can be shared between instruments without each microscopist having to re-map the sample. Preplanning of data analysis in between sessions could also assist in targeting desired information.

# Acknowledgements

The authors acknowledge financial support the U.S. Department of Energy (DOE) Office of Science, Basic Energy Sciences, Materials Sciences and Engineering Division. The authors also thank Drs. Larry Thomas, Steven Spurgeon, and Libor Kovarik for their fruitful discussions on electron microscopy and crystallography. Additionally, Jacob Haag's contribution of providing soundproofing and revisions are duly noted. PNNL is a multiprogram national laboratory operated by Battelle for the U.S. DOE under contract DEAC05-76RL01830.

# Declaration of Competing Interest

The authors declare that they have no known competing personal relationships or financial interests that could have appeared to influence the work in this paper.

# Supplemental for Nanocartography: Planning for success in analytical electron microscopy

Olszta, M.J.[1*], Fiedler, K.R.[2]

[1] Pacific Northwest National Laboratory, Richland, WA 99354, USA

[2] Washington State University, Tri-Cities, Richland, WA, 99354, USA

Phone: (509) 371-7217

Fax: (509) 375-3033

[*]Corresponding author email: matthew.olszta@pnnl.gov

## S1. Conversion to Cubic

In order to perform math on any crystal system it is first necessary to transform each system such that it can be described as a cubic cell. The simplest manner in which to accomplish this is to first constrain one of the new vectors onto the z-axis. For simplicity, and to match the convention that the beam axis is along the z-axis, c is chosen to align with the z-axis (**Figure S1**). As mentioned in the main body of the text, other equivalent starting choices could be made but they do not inhibit the overall conversion. The next step is to constrain the second of the new vectors in one plane. In this case, a is constrained to the xz plane, and then decomposed into the components of that plane using the angle β to the z-axis. Finally, the third vector is decomposed onto each of the x, y and z axes using the angles α, γ, and δ. The angle α is the angle from b to the z-axis. The angle γ is the angle between the vectors a and b. The last angle, δ, is an auxiliary quantity that is the angle between the projection of the b vector onto the xy plane and the y axis. Typically, a crystallographic system is specified with a, b, c, α, β, and γ. The auxiliary angle, δ, is then expressed in terms of these known quantities.

To understand how δ is related to the known quantities, a set of geometric objects is shown in **Figure S2** to break down the calculation into smaller steps. Using the law of cosines on the triangle in the xy plane gives a relationship between the lengths of the three sides and the angle δ.

$$D^2 = (a \sin \beta)^2 + (b \sin \alpha)^2 - 2(a \sin \beta)(b \sin \alpha) \cos \delta \qquad Eqn. S1$$

This has introduced another quantity, $D$, which is the length of the side of the triangle opposite the angle δ. Solving for δ yields:

$$\delta = \cos^{-1}\left(\frac{(a \sin \beta)^2 + (b \sin \alpha)^2 - D^2}{2ab \sin \alpha \sin \beta}\right) \qquad Eqn. S2$$

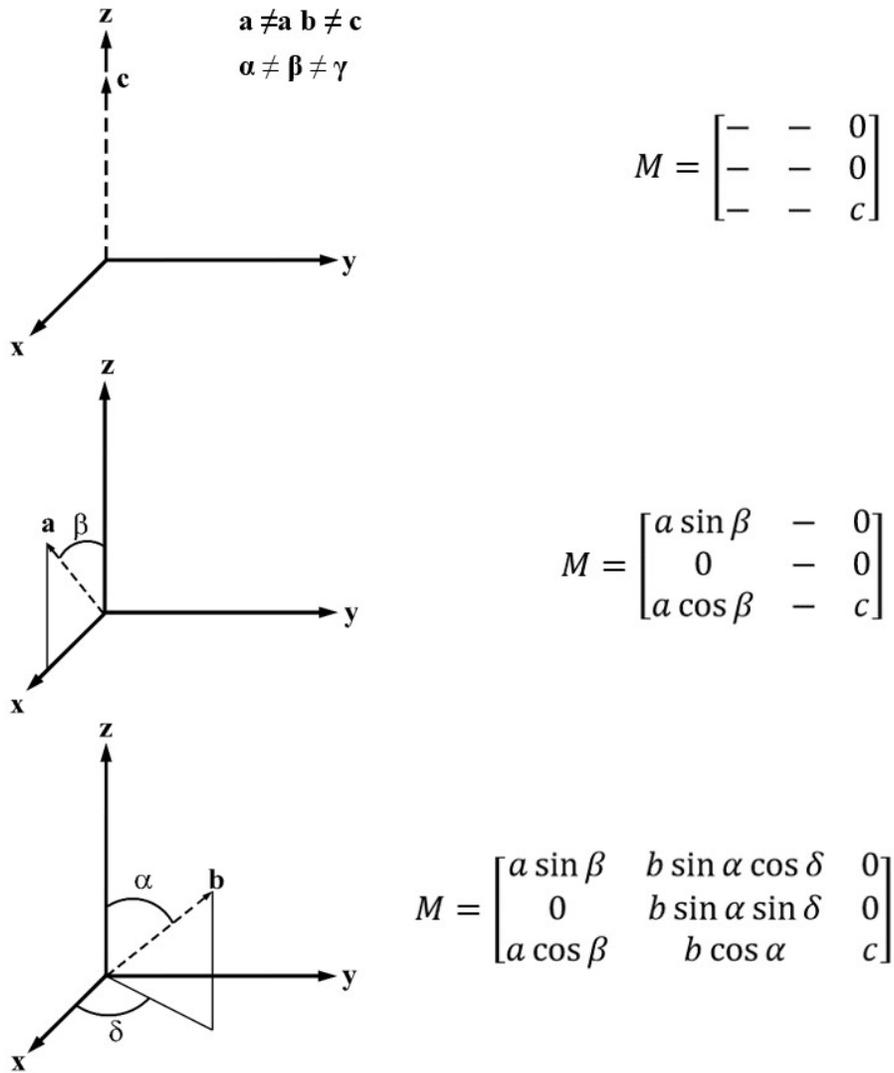

*Figure S1:* Schematics illustrating step-by-step derivation of the non-cubic to cubic conversion matrix.

To find the value of D, the Pythagorean Theorem is used on the right triangle with D as the base.

$$D^2 + (a\cos\beta - b\cos\alpha)^2 = G^2 \qquad Eqn.\,S3$$

In this equation, G, is the side opposite the angle $\gamma$ in the triangle which has legs a and b. The law of cosines is also applied here to find:

$$G^2 = a^2 + b^2 - 2ab\cos\gamma \qquad Eqn.\,S4$$

Eliminating $G^2$ to find $D^2$ and then substituting back into the equation for δ and simplifying the trigonometric functions yields a simplified expression:

$$\delta = \cos^{-1}\left(\frac{\cos\gamma + \cos\alpha\cos\beta}{\sin\alpha\sin\beta}\right) \qquad Eqn.\,S5$$

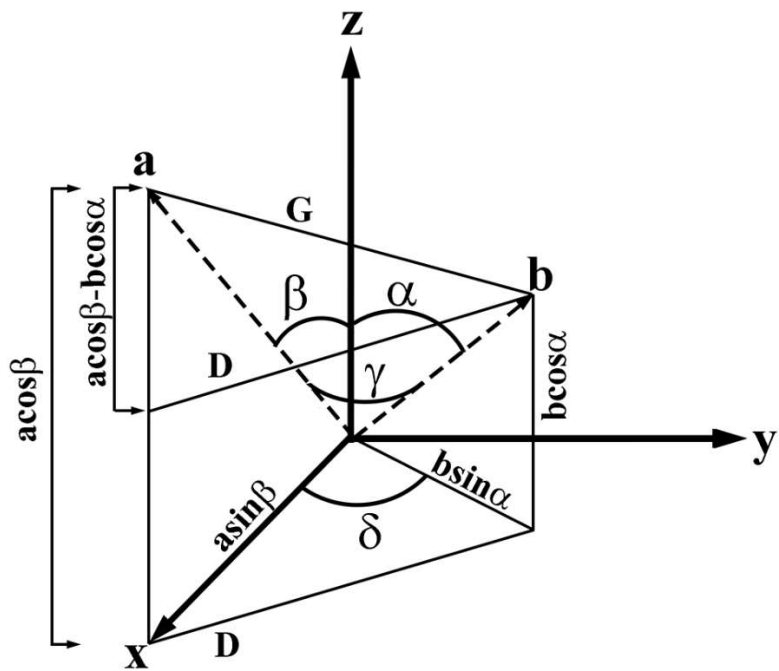

*Figure S2: Schematic illustrating the calculation of the angle δ that accounts for non-orthogonal angles between axes in monoclinic and triclinic systems in the conversion matrix.*

With the angle $\delta$ known, the full expression for the conversion matrix is what was presented above in the text as Eqn. 1:

$$M = \begin{bmatrix} a\sin\beta & b\sin\alpha\cos\delta & 0 \\ 0 & b\sin\alpha\sin\delta & 0 \\ a\cos\beta & b\cos\alpha & c \end{bmatrix} \qquad Eqn.\,S6$$

This matrix will convert a vector from a non-cubic system to a cubic system after matrix multiplication. The reverse process (going from cubic to a non-cubic system) is accomplished by multiplying by the inverse matrix. The inverse matrix for M is:

$$M^{-1} = \begin{bmatrix} \dfrac{1}{a\sin\beta} & \dfrac{-\cos\delta}{a\sin\beta\sin\delta} & 0 \\ 0 & \dfrac{1}{b\sin\alpha\sin\delta} & 0 \\ \dfrac{-\cos\beta}{c\sin\beta} & \dfrac{\cos\beta\sin\alpha\cos\delta - \sin\beta\cos\alpha}{c\sin\alpha\sin\beta\sin\delta} & \dfrac{1}{c} \end{bmatrix} \qquad Eqn.\,S7$$

## S2. Hexagonal to Cubic

Whereas *Figure 1* provided a simplistic example of the need to convert non-cubic systems to cubic before performing mathematical calculations, *Figure S3* illustrates a more complicated crystal system. A number of low index vectors in the hexagonal unit cell provided in *Figure S3*, which are described in the native nomenclature and are invariable despite the c/a ratio (*Figure 2*). A schematic presentation of the cubic matrix conversion can be envisioned by placing a cube with the x and z axes commensurate with the hexagonal x and z axes and observing the traces of the vectors in the hexagonal system overlapping with the inserted cube. After a mental transcription of the vectors onto the cube, the hexagonal vectors can be removed, thereby leaving the vectors in the cubic cell. The positions of the vectors within the cube can then be observed. Mathematical operations, such as the angle between poles, can then be performed.

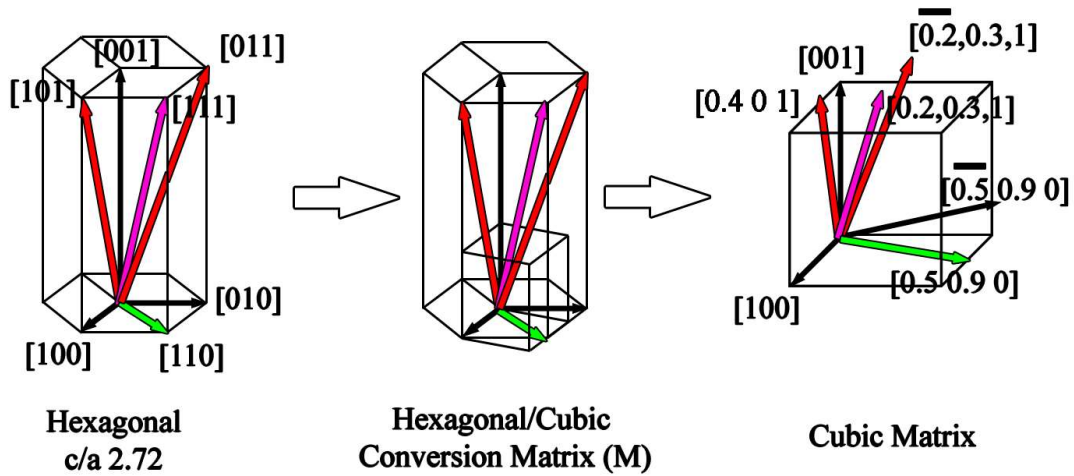

*Figure S3:* Illustration showing the conversion of a hexagonal matrix to cubic form.

## S3. Rotation about an Arbitrary Axis

Rotations about the principal axes are simple to state, but are of limited use when considering general crystal orientations. Instead, it is useful to rotate about an arbitrary axis in space for any angle, not just rotation about one of the principal axes (Eqns. 6-8). As previously stated, this rotation is not necessary to create useful tip/tilt maps, but it is necessary when comparing two crystals in a misorientation matrix. The misalignment that arises in the misorientation matrix as compared to the tip/tilt maps is that while the vector moves to the correct position in both derivations (e.g., the [111] to the [001]), the location of the unit vectors (i.e., [001], [100], and [010]) will vary depending on pathway. This will be derived subsequently, but is illustrated schematically in *Figure S4*.

The vector [uvw] is rotated to the [001] orientation about [xy0] through angle θ. The schematic details two pathways by which to achieve this rotation; path 1 which is about a single axis, and path 2 which is through two concurrent tilts about the additional unit vector axes (in this case x and y). A movie is provided that indicates the motion of path 2. In both cases the [111] vector is rotated to [001], but the location of the unit axes (as shown by the projection of a cube down the [111]) will differ by some angle (Δ) as compared to the rotation about the arbitrary axis. Note that the illustrated angle is exaggerated for effect, and is only approximately a few degrees. Depending on the order of the pathway, the angle (Δ) would also be different, hence providing two distinct projections of the unit vectors for misorientation calculations.

The example described above and illustrated in *Figure S4* pertains directly to the formulation of a tip/tilt map for a double tilt stage (i.e., the rotation axis is always calculated normal to the [001] orientation), but a more general formulation of the tilt about an arbitrary axis needs be derived to account for non-crystallographic motions. Therefore, the full derivation is provided below.

To describe this general rotation, a sequence of steps can be used to decompose any rotation down into rotations about the principal axes. To accomplish this the desired axis of rotation needs to be rotated to coincide with any one of the principal axes through two rotations about the other axes (in the tip/tilt map calculation the z or probe axis is chosen). The desired rotation about the arbitrary axis is accomplished with the system aligned along a principal axis (e.g., $\varphi_c$). Finally, the arbitrary axis is returned to its original location by performing the inverse of the first two rotations. For more details of this process, consult (Ian R., 2015).

The axis of rotation is a unit vector, $\hat{\mathbf{u}} = (u_x, u_y, u_z)$ and the desired angle of rotation about this axis is $\theta$. Without loss of generality, the arbitrary axis will be rotated to the x-axis where the rotation by $\theta$ occurs. First, $\hat{\mathbf{u}}$ rotates into the x-y plane by rotating about the y-axis by an angle $\varphi_1$. This angle is computed by the requirement that the resulting vector, $\widehat{\mathbf{u_1}} = (u_{1,x}, u_{1,y}, 0)$ has no y-component. The angle $\varphi_1$ which accomplishes this rotation is $\varphi_1 = \tan^{-1}\left(\frac{u_z}{u_x}\right)$ and can be found from the z-component (*Figure S5*).

$$\begin{bmatrix} \cos\varphi_1 & 0 & \sin\varphi_1 \\ 0 & 1 & 0 \\ -\sin\varphi_1 & 0 & \cos\varphi_1 \end{bmatrix} \begin{bmatrix} u_x \\ u_y \\ u_z \end{bmatrix} = \begin{bmatrix} u_x \cos\varphi_1 + u_z \sin\varphi_1 \\ u_y \\ -u_x \sin\varphi_1 + u_z \cos\varphi_1 \end{bmatrix} = \begin{bmatrix} u_{1,x} \\ u_{1,y} \\ 0 \end{bmatrix} \qquad Eqn. S8$$

In particular, since the y-component is unchanged ($u_y = u_{1,y}$) and the magnitude of the vector is still 1, this intermediate vector can be written as $\widehat{\mathbf{u_1}} = \left(\sqrt{1 - u_y^2}, u_y, 0\right)$. Next, the vector $\widehat{\mathbf{u_1}}$ is rotated to point along the x-axis. This rotation is by an angle $\varphi_2$ about the z-axis and the angle is determined by the requirement that the resulting vector should have no x-component. This angle is $\varphi_2 = \tan^{-1}\left(\frac{-u_y}{\sqrt{1-u_y^2}}\right)$ as can be seen by equating the y-component of the vector to 0.

$$\begin{bmatrix} \cos\varphi_2 & -\sin\varphi_2 & 0 \\ \sin\varphi_2 & \cos\varphi_2 & 0 \\ 0 & 0 & 1 \end{bmatrix} \begin{bmatrix} \sqrt{1-u_y^2} \\ u_y \\ 0 \end{bmatrix} = \begin{bmatrix} \left(\sqrt{1-u_y^2}\right)\cos\varphi_2 - u_y\sin\varphi_2 \\ \left(\sqrt{1-u_y^2}\right)\sin\varphi_2 + u_y\cos\varphi_2 \\ 0 \end{bmatrix} = \begin{bmatrix} 1 \\ 0 \\ 0 \end{bmatrix} \qquad Eqn. S9$$

These two angles move the vector to the x-axis, where it is rotated by an angle $\theta$ and then rotated back to the original position by rotating the negative of $\varphi_2$ and $\varphi_1$. Combining all of these rotations into one yields the overall rotation matrix, $R_{\hat{u},\theta}$:

$$R_{\hat{u},\theta} = R_{-\varphi_1,y} R_{-\varphi_2,z} R_{\theta,x} R_{\varphi_2,z} R_{\varphi_1,y} \qquad Eqn. S10$$

In this equation, each rotation matrix is defined by the angle and the axis about which the rotation occurs. Note that the negative angle rotations are undoing the initial rotations to bring the vector to the x-axis and are the inverses of the first two rotations which can be found by taking the transpose of the $R_{\varphi_1,x}$ and $R_{\varphi_2,y}$ since for any rotation matrix, $R^{-1} = R^T$. If we express these matrices in terms of the matrices and simplify the trigonometric expressions in terms of the components of $\hat{u}$, then the matrix product is:

$$R_{\hat{u},\theta} = \begin{bmatrix} \frac{u_x}{\sqrt{u_x^2+u_z^2}} & 0 & \frac{-u_z}{\sqrt{u_x^2+u_z^2}} \\ 0 & 1 & 0 \\ \frac{u_z}{\sqrt{u_x^2+u_z^2}} & 0 & \frac{u_x}{\sqrt{u_x^2+u_z^2}} \end{bmatrix} \begin{bmatrix} \sqrt{1-u_y^2} & -u_y & 0 \\ u_y & \sqrt{1-u_y^2} & 0 \\ 0 & 0 & 1 \end{bmatrix} \begin{bmatrix} 1 & 0 & 0 \\ 0 & \cos\theta & -\sin\theta \\ 0 & \sin\theta & \cos\theta \end{bmatrix} \begin{bmatrix} \sqrt{1-u_y^2} & u_y & 0 \\ -u_y & \sqrt{1-u_y^2} & 0 \\ 0 & 0 & 1 \end{bmatrix} \begin{bmatrix} \frac{u_x}{\sqrt{u_x^2+u_z^2}} & 0 & \frac{u_z}{\sqrt{u_x^2+u_z^2}} \\ 0 & 1 & 0 \\ \frac{-u_z}{\sqrt{u_x^2+u_z^2}} & 0 & \frac{u_x}{\sqrt{u_x^2+u_z^2}} \end{bmatrix} \quad Eqn. S11$$

After matrix multiplication and simplification, recalling that the magnitude of $\hat{u}$ is one ($u_x^2 + u_y^2 + u_z^2 = 1$), the full matrix for rotation of angle $\theta$ about an axis of rotation $\hat{u}$ is:

$$R_{\hat{r},\theta} = \begin{bmatrix} u_x^2 + (u_y^2 + u_z^2)\cos\theta & u_x u_y(1-\cos\theta) - u_z \sin\theta & u_x u_z(1-\cos\theta) + u_y \sin\theta \\ u_x u_y(1-\cos\theta) + u_z \sin\theta & u_y^2 + (u_x^2 + u_z^2)\cos\theta & u_y u_z(1-\cos\theta) - u_x \sin\theta \\ u_x u_z(1-\cos\theta) - u_y \sin\theta & u_y u_z(1-\cos\theta) + u_x \sin\theta & u_z^2 + (u_x^2 + u_y^2)\cos\theta \end{bmatrix} Eqn. S12$$

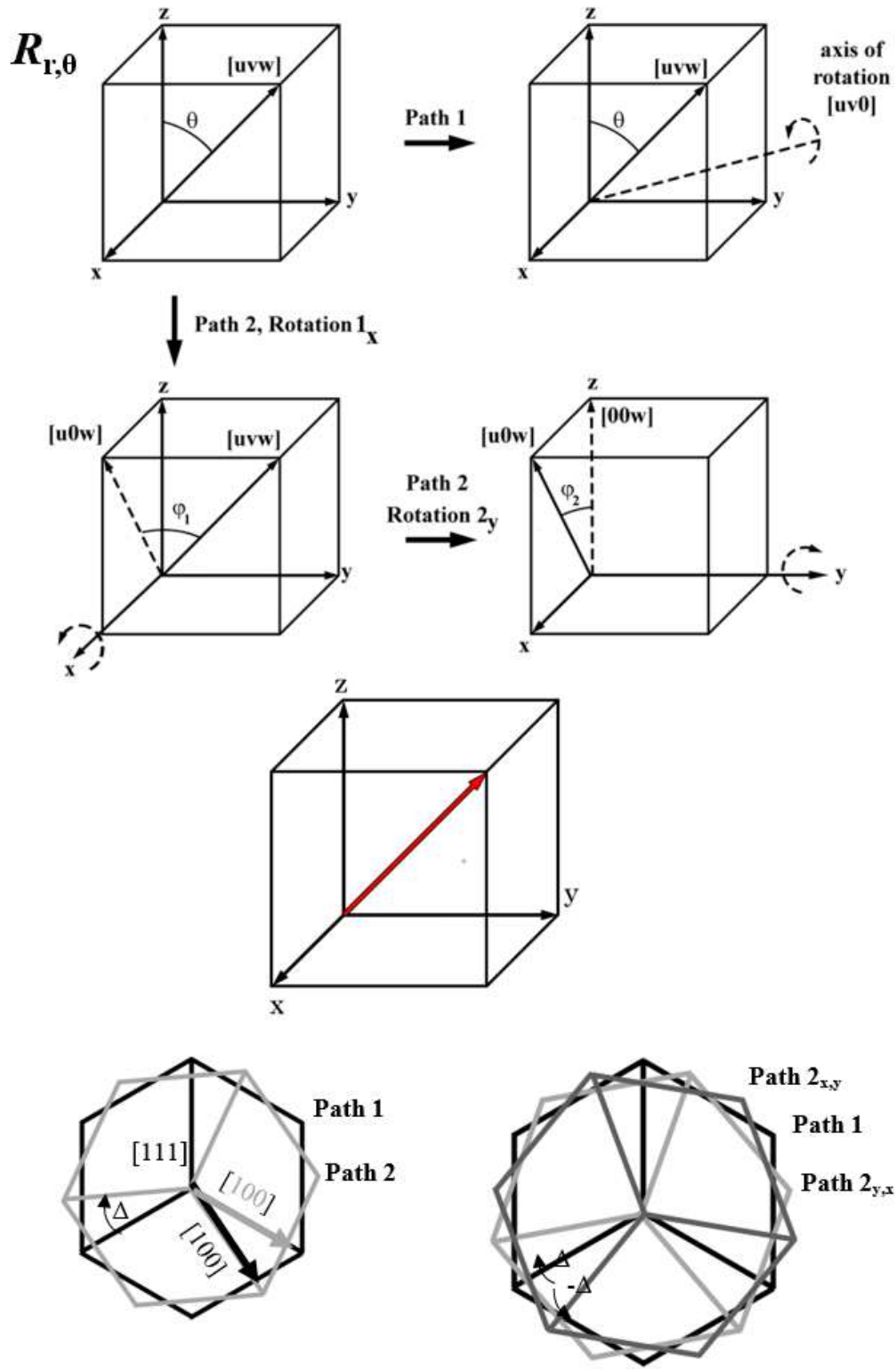

*Figure S4:* *Direct rotation about an arbitrary axis as compared to two rotations about unit vector axes.*

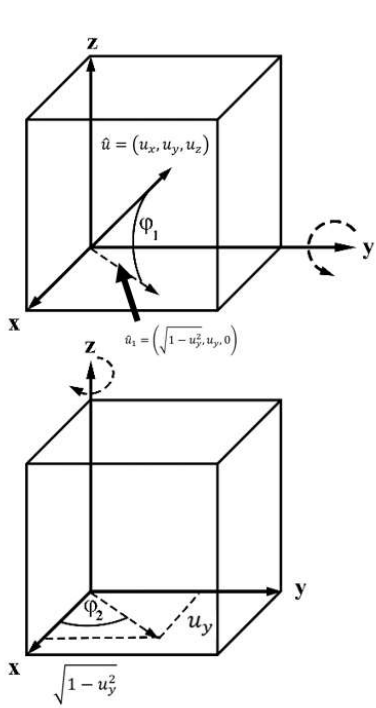

$$\hat{u} = \left[ \frac{u}{\sqrt{u^2+v^2+w^2}} \quad \frac{v}{\sqrt{u^2+v^2+w^2}} \quad \frac{w}{\sqrt{u^2+v^2+w^2}} \right]$$

$$\begin{bmatrix} \cos\varphi_1 & 0 & \sin\varphi_1 \\ 0 & 1 & 0 \\ -\sin\varphi_1 & 0 & \cos\varphi_1 \end{bmatrix} \begin{bmatrix} u_x \\ u_y \\ u_z \end{bmatrix} = \begin{bmatrix} u_x\cos\varphi_1 + u_z\sin\varphi_1 \\ u_y \\ -u_x\sin\varphi_1 + u_z\cos\varphi_1 \end{bmatrix}$$

$$-u_x \sin\varphi_1 + u_z \cos\varphi_1 = 0$$

$$\varphi_1 = \tan^{-1}\left(\frac{u_z}{u_x}\right)$$

$$\begin{bmatrix} \cos\varphi_2 & -\sin\varphi_2 & 0 \\ \sin\varphi_2 & \cos\varphi_2 & 0 \\ 0 & 0 & 1 \end{bmatrix} \begin{bmatrix} \sqrt{1-u_y^2} \\ u_y \\ 0 \end{bmatrix} = \begin{bmatrix} \left(\sqrt{1-u_y^2}\right)\cos\varphi_2 - u_y\sin\varphi_2 \\ \left(\sqrt{1-u_y^2}\right)\sin\varphi_2 + u_y\cos\varphi_2 \\ 0 \end{bmatrix} = \begin{bmatrix} 1 \\ 0 \\ 0 \end{bmatrix}$$

$$\varphi_2 = \tan^{-1}\left(\frac{-u_y}{\sqrt{1-u_y^2}}\right)$$

*Figure S5:* Schematics demonstrating the angles necessary to rotate around an arbitrary vector.

# S4. Examples of Tip/Tilt Maps in Lower Symmetry Systems

As demonstrated in *Figure 7,* tip tilt maps for cubic and hexagonal crystal systems can be procured in any number of orientations. The same can be done for other non-cubic systems such as tetragonal and orthorhombic (*Figure S6*a-c). In the [001] orientation a tetragonal crystal with a c:a ratio of 2 (as illustrated in *Figures 1* and *10* ) there are more accessible poles within an ~40° range as compared to the cubic form (i.e., *Figure 7*a). To demonstrate the asymmetry of the tetragonal system the [010] is plotted in *Figure S6*b. When the third unit axis vector is changed (i.e., orthorhombic), the larger asymmetry can be observed in *Figure S6*c. In order to best demonstrate how the tip/tilt maps can vary between a high symmetry system (i.e., cubic) and lower symmetry systems (e.g., monoclinic and triclinic) the [112] of the three systems is plotted in *Figure S6*d-f with the orientation of the [001] pole in each system ~15° from the α tilt axis. The various tilt positions of the [001] between the three systems as well as other surrounding poles highlights the large discrepancy when the unit axes and angles between axes varies. While each of these crystal system parameters were utilized from real crystals, it should be noted that the plotting of poles is in a general form to highlight the variability and does not fully represent whether each of the poles are expressed in those systems. Additionally, while the family of poles (e.g., the <112>) are plotted with the same format (e.g., blue dots) practically they are different families due to the variation in axes and angle between axes (e.g., the [112] is not in the same family of the [121] for a triclinic system).

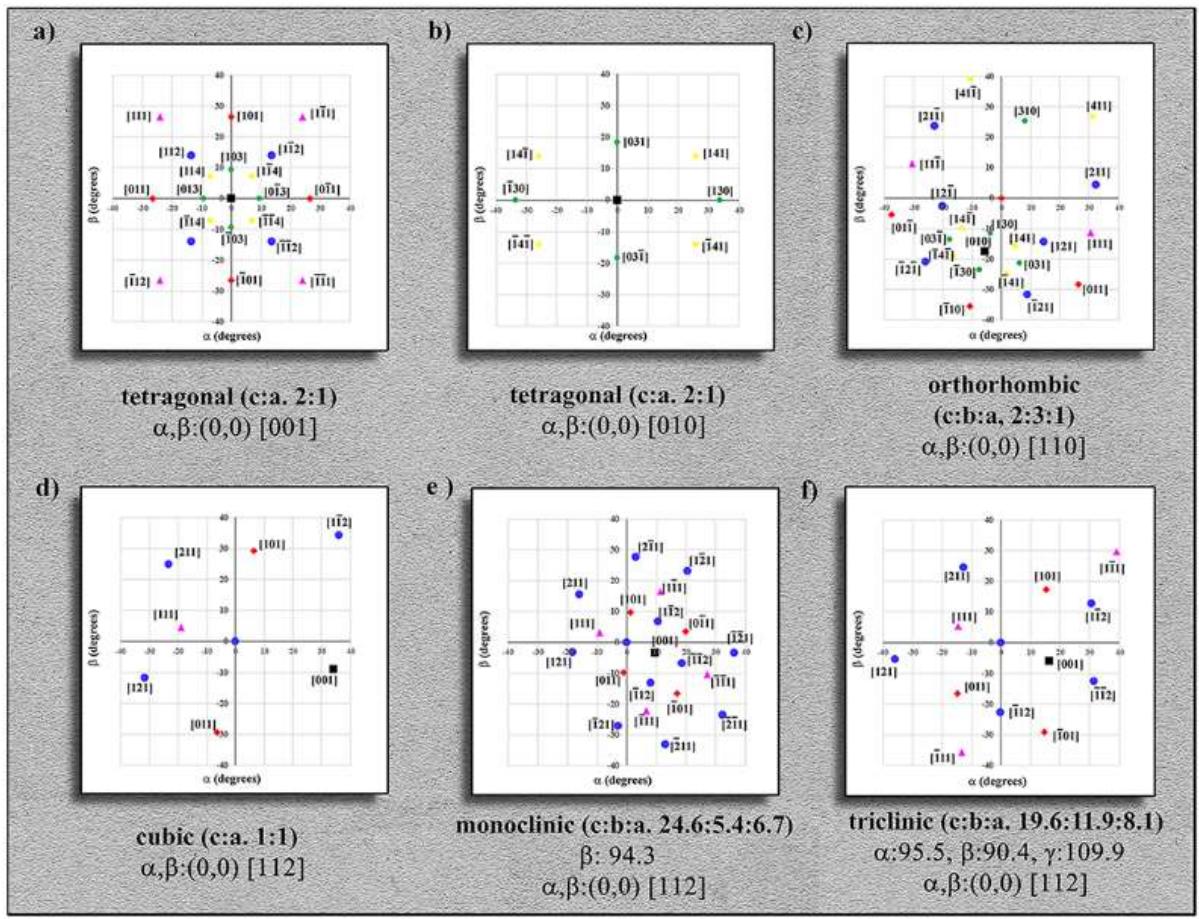

*Figure S6*: Tip/Tilt plots of various crystal systems.

## S5. Plotting Kikuchi Bands

Plotting the traces of planes of atoms was accomplished by using the normal to the plane as an arbitrary axis by which to rotate any vector in the plane. This approach is not appropriate for tracing Kickuchi bands that are located at an angular distance equaling the Bragg angle from the trace of the plane of atoms. If these planes were to be plotted on a tip tilt map (*Figure S7*) it is unlikely that the precision of current double tilt stages would be able to discretely tilt to these conditions. In *Figure S7* the [111] orientation of an FCC crystal is shown plotted on a tip tilt map with a corresponding Kikuchi pattern. The strong (4-40) bands shown in the Kikuchi pattern are the approximately the smallest d-spacings observed in all materials systems, and when plotted in the tip/tilt map it becomes apparent that all larger d-spacings would be bound within these lines (hence not within the precision of current double tilt holders). The projection of the large condenser aperture (~103 mrad) shown in the Kikuchi pattern is overlaid on the tip/tilt map for scale.

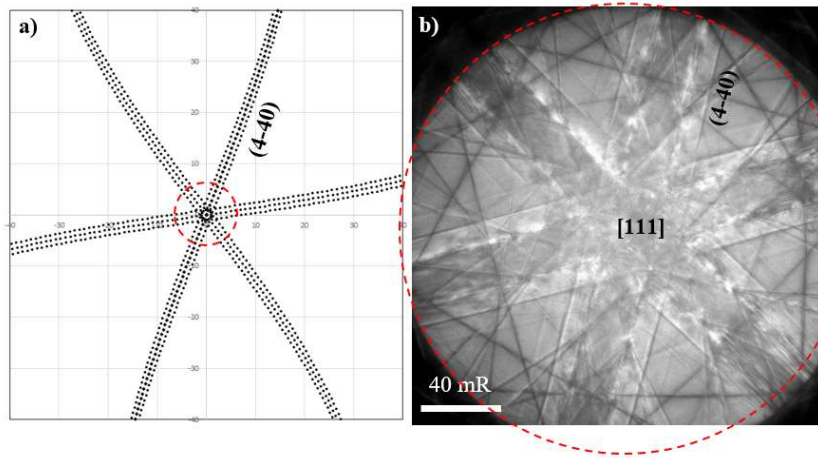

*Figure S7:* Plot of tilt map for [111] FCC austenitic stainless steel in the [111] orientation with the {440} planes expressed (a) and a CBED pattern in the same orientation (b).

## S6. Vector of Equations

The key to solving the system of equations for the coordinate vectors in Eqns. 57-59 is to rewrite them as vector equations for the elements of $\vec{p}, \vec{q},$ and $\vec{t}$. Writing out all nine equations gives:

$$u_{A1}p_{Ax} + v_{A1}q_{Ax} + w_{A1}t_{Ax} = x_{A1} \qquad \text{Eqn. S13}$$

$$u_{A1}p_{Ay} + v_{A1}q_{Ay} + w_{A1}t_{Ay} = y_{A1} \qquad \text{Eqn. S14}$$

$$u_{A1}p_{Az} + v_{A1}q_{Az} + w_{A1}t_{Az} = z_{A1} \qquad \text{Eqn. S15}$$

$$u_{A2}p_{Ax} + v_{A2}q_{Ax} + w_{A2}t_{Ax} = x_{A2} \qquad \text{Eqn. S16}$$

$$u_{A2}p_{Ay} + v_{A2}q_{Ay} + w_{A2}t_{Ay} = y_{A2} \qquad \text{Eqn. S17}$$

$$u_{A2}p_{Az} + v_{A2}q_{Az} + w_{A2}t_{Az} = z_{A2} \qquad \text{Eqn. S18}$$

$$u_{A3}p_{Ax} + v_{A3}q_{Ax} + w_{A3}t_{Ax} = x_{A3} \quad \text{Eqn. S19}$$

$$u_{A3}p_{Ay} + v_{A3}q_{Ay} + w_{A3}t_{Ay} = y_{A3} \quad \text{Eqn. S20}$$

$$u_{A3}p_{Az} + v_{A3}q_{Az} + w_{A3}t_{Az} = z_{A3} \quad \text{Eqn. S21}$$

In this order, the patterns in these equations are not clear, so instead, they are grouped as Eqns. {S13, S16, S19}, Eqns. {S14, S17, S20}, and Eqns. {S15, S18, S21}. These groupings yield:

$$u_{A1}p_{Ax} + v_{A1}q_{Ax} + w_{A1}t_{Ax} = x_{A1} \quad \text{Eqn. S13}$$

$$u_{A2}p_{Ax} + v_{A2}q_{Ax} + w_{A2}t_{Ax} = x_{A2} \quad \text{Eqn. S16}$$

$$u_{A3}p_{Ax} + v_{A3}q_{Ax} + w_{A3}t_{Ax} = x_{A3} \quad \text{Eqn. S19}$$

$$u_{A1}p_{Ay} + v_{A1}q_{Ay} + w_{A1}t_{Ay} = y_{A1} \quad \text{Eqn. S14}$$

$$u_{A2}p_{Ay} + v_{A2}q_{Ay} + w_{A2}t_{Ay} = y_{A2} \quad \text{Eqn. S17}$$

$$u_{A3}p_{Ay} + v_{A3}q_{Ay} + w_{A3}t_{Ay} = y_{A3} \quad \text{Eqn. S20}$$

$$u_{A1}p_{Az} + v_{A1}q_{Az} + w_{A1}t_{Az} = z_{A1} \quad \text{Eqn. S15}$$

$$u_{A2}p_{Az} + v_{A2}q_{Az} + w_{A2}t_{Az} = z_{A2} \quad \text{Eqn. S18}$$

$$u_{A3}p_{Az} + v_{A3}q_{Az} + w_{A3}t_{Az} = z_{A3} \quad \text{Eqn. S21}$$

In this grouping, these systems become much more recognizable as matrix equations:

$$\begin{bmatrix} u_{A1} & v_{A1} & w_{A1} \\ u_{A2} & v_{A2} & w_{A2} \\ u_{A3} & v_{A3} & w_{A3} \end{bmatrix} \begin{bmatrix} p_{Ax} \\ q_{Ax} \\ t_{Ax} \end{bmatrix} = \begin{bmatrix} x_{A1} \\ x_{A2} \\ x_{A3} \end{bmatrix} \quad \text{Eqn. S22}$$

$$\begin{bmatrix} u_{A1} & v_{A1} & w_{A1} \\ u_{A2} & v_{A2} & w_{A2} \\ u_{A3} & v_{A3} & w_{A3} \end{bmatrix} \begin{bmatrix} p_{Ay} \\ q_{Ay} \\ t_{Ay} \end{bmatrix} = \begin{bmatrix} y_{A1} \\ y_{A2} \\ y \end{bmatrix} \quad \text{Eqn. S23}$$

$$\begin{bmatrix} u_{A1} & v_{A1} & w_{A1} \\ u_{A2} & v_{A2} & w_{A2} \\ u_{A3} & v_{A3} & w_{A3} \end{bmatrix} \begin{bmatrix} p_{Az} \\ q_{Az} \\ t_{Az} \end{bmatrix} = \begin{bmatrix} z_{A1} \\ z_{A2} \\ z_{A3} \end{bmatrix} \quad \text{Eqn. S24}$$

In particular, the 3x3 matrices in each grouping are the same, which means that the same set of row operations will row reduce all of them simultaneously. As such, they are combined into a single augmented matrix that is Eqn. 60 in the text.

$$\begin{bmatrix} u_{A1} & v_{A1} & w_{A1} & | & x_{A1} & y_{A1} & z_{A1} \\ u_{A2} & v_{A2} & w_{A2} & | & x_{A2} & y_{A2} & z_{A2} \\ u_{A3} & v_{A3} & w_{A3} & | & x_{A3} & y_{A3} & z_{A3} \end{bmatrix} \quad \text{Eqn. S25}$$